\documentclass[smallcondensed]{svjour3}     
\smartqed  
\usepackage{graphicx}
\usepackage[utf8x]{inputenc}
\usepackage{epstopdf}
\usepackage{graphicx}
\usepackage[normal]{subfigure}
\usepackage{overpic}
\usepackage{float}
\usepackage{parskip}
\usepackage{natbib}
\usepackage{float}
\usepackage{dblfloatfix}
\usepackage{fixltx2e}
\usepackage{color}
\usepackage{geometry}
\usepackage{amsmath}
\usepackage{amssymb}
\usepackage{array}
\usepackage{lineno}
\usepackage[american]{babel}
\usepackage{times}
\usepackage{mathtools}
\usepackage[labelformat=empty]{caption}

\usepackage[amssymb]{SIunits}
\newcommand{\vphi}{\mbox{{\boldmath{$\phi$}}}}

\bibpunct[; ]{(}{)}{;}{a}{}{;}
\journalname{Contributions to Mineralogy and Petrology}
\begin{document}
\title{Phase-field study of grain boundary tracking behavior in crack-seal microstructures}

\titlerunning{Phase-field study of grain boundary tracking behavior in crack-seal microstructures}        

\author{Kumar Ankit\and Britta Nestler\and Michael Selzer\and Mathias Reichardt}


\institute{Kumar Ankit\and Britta Nestler\and Michael Selzer\and Mathias Reichardt\at
             Karlsruhe Institute of Technology (KIT), IAM-ZBS, Haid-und-Neu-Str. 7, D-76131 Karlsruhe, Germany \\
              Tel.: +49 721 608-45022\\
              Fax: +49 721 925-2348\\
              \email{kumar.ankit@hs-karlsruhe.de}
}
\date{Received: date / Accepted: date}
\maketitle
\begin{abstract}
In order to address the growth of crystals 
in veins, a multiphase-field model is used 
to capture the dynamics of crystals 
precipitating from a super-saturated
solution. To gain a detailed understanding
{of the polycrystal growth phenomena in veins}, 
we investigate the influence of various boundary
conditions on crystal growth. In particular, we 
analyze the formation of vein microstructures 
resulting from the free growth of crystals as 
well as crack-sealing processes. {We define the 
crystal symmetry by considering the anisotropy 
in surface energy to simulate crystals 
with flat facets and sharp corners.
The resulting growth competition of crystals 
with different orientations is studied to 
deduce a consistent orientation
selection rule in the free-growth regime.}
 {Using} crack-sealing simulations, we correlate 
the grain boundary tracking behavior depending on the 
relative rate of crack opening, {opening} trajectory, 
initial grain size and wall roughness. Further, we illustrate
how these parameters induce the microstructural 
transition between blocky (crystals growing 
anisotropically) to fibrous morphology (isotropic)
and formation of grain boundaries. 
The phase-field simulations of crystals in {the} free-growth 
regime (in 2D and 3D) indicate that the 
growth or consumption of a crystal is dependent on 
the orientation difference with neighboring crystals. 
The crack-sealing simulation results (in 2D and 3D) 
reveal that crystals grow isotropically and
grain boundaries track the opening trajectory 
if the wall roughness is high, opening increments 
are small and crystals touch the wall before the 
next crack increment starts. Further, we find that
within the complete crack-seal regime, anisotropy in surface 
energy results in the formation of curved/oscillating grain
boundaries (instead of straight) when the crack opening 
velocity is increased and wall roughness is not sufficiently high.
{Additionally, the overall capability of phase-field method 
to simulate large-scale polycrystal growth in veins (in 3D)
is demonstrated enumerating the main advantages 
of adopting the novel approach.}

\end{abstract}

\section{Introduction}
\label{sec:Introduction}
Veins are sub-planar discontinuities 
in the earth’s crust,
containing minerals which precipitate from a super-saturated
solution in a fracture \citep{Urai:1991ec}.
They exhibit a wide range of crystal habits from 
dendritic, acicular to crystalline or fibrous structures depending 
on the boundary conditions 
{\citep{Durney:1973fr,Ramsay:1983mb,Fisher:1992kb,Bons:1997qa}}. 
{\citet{Hilgers:2002bv} provide an extensive list of
such boundary condition: relative rate of crystal
growth with respect to fracture opening 
{velocity}, degree of crack surface roughness,
fluid properties (pressure, temperature, flow 
velocity and supersaturation)
\citep{Cox:1987qq,Knipe:1994lh} or
transport mechanisms (diffusion
and advection) \citep*{Durney:1976xw, McCaig:1988ly,
Yardley:1997mi}. \citet{Durney:1972na} and \citet{Durney:1976xw}
also correlate fibrous veins to pressure solution and diffusion.}
The morphology of crystals growing 
in freely flowing fluid differs
significantly from the crystals growing in a constrained
environment such as present in crack-sealing conditions. 
Among the numerous distinguishable 
crystal morphologies that develop due to a combination 
of boundary conditions, an interesting case
is the formation of curved crystals in veins.
{
\citet{Durney:1973fr} propose that
syntectonic fibres grow along the opening trajectory
{of fracture walls} with
the transport mechanism being diffusion into the dilational
sites. {The crystal growth is assumed to be continuous because
of which growth competition is not observed.} As an alternative to diffusional
and continuous accretion, \citet{Ramsay:1980jl} discusses the
discontinuous crack-seal mechanism, a repeated process of
crack opening and sealing, as indicated by the characteristic
inclusion bands/trails and stair-stepped grain boundaries. In
recent papers based on {natural microstructural 
features, laboratory experiments} and theoretical modeling
\citep{Fisher:1992kb,Bons:1997qa}, it is suggested that
{crystal fibers in veins} can only form by a diffusional accretion process.}

\citet{Williams:1989fk} elucidate that the primary 
reason responsible for such deviations from the
equilibrium crystal shape is mechanical coupling 
of wall rock and evolving crystals. Further, 
\citet{Hilgers:2002bv} amend that mechanical 
coupling of vein and wall is not a necessary 
condition for the formation of curved fibrous
crystals, if the crack increments are smaller 
than about 10 $\mu$m.

\citet{Cox:1983fk} and \citet{Urai:1991ec} propose a kinematic
model for crystal growth in crack seal veins and show the
importance of the wall-rock morphology for the resulting vein
microstructure. According to this model, if the crystals have
already sealed the space available before the next crack event,
the facets are lost and they assume the morphology of the rough
vein wall interface. If opening increments are sufficiently
small, the crystals cannot develop crystal facets and therefore,
grow isotropically.  An important
conclusion of this model is that the crystal growth kinetics is
effectively isotropic if the crack surface is sufficiently rough
and crack opening rate is {smaller as compared to 
rate of crystal growth front.} As a step to implement
crystal anisotropy in 2D, an efficient numerical program
\textit{Vein Growth} is developed by \citet{Bons:2001ve} 
which is further utilized to simulate the
anisotropic growth of crystals under complex boundary
conditions. Simulations with \textit{Vein Growth} produce fibrous
crystals with the potential to track the opening
trajectory of the crack when the wall morphology is rough and
the average opening velocity is smaller than the growth velocity
of the crystals \citep{Koehn:2000fk,Hilgers:2001rw}. 
A 2D simulation program FACET is separately developed by 
\citet{Zhang:2002fk} to study the growth of polycrystal based 
on deposition flux of atoms.
\citet{NOLLET:2006oz} use the algorithms \textit{Vein Growth} 
and FACET to study crystal growth 
competition leading to orientation selection and transition from blocky to 
fibrous morphology during crack-seal growth. 

The numerical programs \textit{Vein Growth} and FACET suffer from
geometric restrictions and are vulnerable to inaccuracies at
triple/quadruple crystal junctions. The artifacts of the 
algorithm \textit{Vein Growth} namely `Crystal terminations' 
and `Long-distance effects' cause the euhedral angles 
between facets at crystal terminations to depart 
from {an} angle corresponding to  
{the} equilibrium shape and 
{the} effect of non-neighboring crystals
on growth process respectively. Further, the method
advocates the switching of the numerical program to FACET in
order to produce \textit{crystallographically correct} facets 
that develop during free growth of crystals. 
However, the switching of numerical program
induces new {complications}; the complete sealing of
{the} crack
can no longer be simulated correctly. Besides, the 
algorithm FACET is not capable of generating the 
crystal facets from a randomly shaped nucleus 
and implicitly assumes the presence of 
\textit{crystallographically correct} facets 
prior to {the} growth process. 
{These aspects are 
addressed in great detail by \citet{Nollet:2005qo}.}
We recognise that since neither of the methods used in the past
are able to completely
describe the crystal growth in veins and 
{restricted to two dimensions}, it is 
important to develop better algorithms to 
enhance the knowledge concerning the
vein growth process.

{The phase-field method, long established 
in the material science community, is a 
stand-out approach to describe 
microstructural evolution during 
phase transitions for e.g. 
solidification, spinodal 
decomposition etc.
\citep{Chen:2002rv,Thornton:2003uq,Singer-Loginova:2008gb,Nestler:2011fk} 
The phase-field method's 
popularity in modeling 
material processes is 
due to the elegance with 
which it treats moving 
boundary problems by 
obviating the necessity 
to explicitly track the 
interfaces. Thus it} can 
overcome many of the 
problems suffered by earlier
models such as the \textit{Vein Growth} and FACET. 
The model equations are derived 
on the basis of general thermodynamic
and kinetic principles and contain 
a number of phenomenological
parameters related to the physical 
properties of the material.
These parameters are determined 
based on experimental and
theoretical information. 
Different thermodynamic 
driving forces for microstructure 
evolution, such as chemical bulk 
free energy, interfacial energy, 
elastic strain energy and
different transport processes, 
such as mass diffusion and
advection, can be coupled and 
the effect on the over all evolution 
process can be studied simultaneously.

{In the present article, we use a thermodynamically consistent
multiphase-field model \citep{Stinner:2004uq,Nestler:2005ye} to
extend the 2D numerical studies of vein growth to 
3D. We present the specific formulation 
of the multiphase-field model for
crystall growth by considering anisotropy of the surface energy
to produce flat facets and sharp corners
\citep{Stinner:2004uq,Nestler:2005ye}. 
We also briefly discuss about the computational
optimization technique used to 
make large-scale studies feasible in
the present simulations.
The phase-field model is used to simulate
and analyze the kinematic properties of the
anisotropic growth competition in polycrystal, 
arising due to mis-orientation with respect to most preferred 
growth direction. In this regard, we also make a strict 
check for any model artifacts (if present).
Further, we modify the 
boundary condition and parameters to 
present a systematic study of crack-sealing,
e.g.
crack surface roughness, crack 
opening velocity, opening trajectory and number 
of crystal nuclei. 
Finally, we present a detailed 
discussion of 2D and 3D phase-field
simulation results, benefits of using the new 
approach and possible extensions of the
current numerical model.}

\section{Phase-field model}
\label{sec:phase_field}

We consider a set of phase-field parameters, denoted by
$\vphi\left(\vec{x},t\right) =
\left(\phi_{1}\left(\vec{x},t\right)\cdots\phi_{N}\left(\vec{x},t\right)\right)$ 
where each component of the vector $\phi_{\alpha}\left(\vec{x},t\right)$
varies smoothly from 1 inside
0 outside the crystal $\alpha$ over a small finite
distance $\varepsilon$ (diffuse interface). The  location of the crystal-liquid or
crystal-crystal interface is defined by a level set at $\phi_{\alpha}\left(\vec{x},t\right) = 0.5$
which is determined mathematically by solving the evolution
equation. The external fields like temperature or concentration
can be coupled to the phase-field parameter $\vphi$, hence no
external boundary conditions needs to be applied at the
interfaces.

In this article, we use the phase-field equations derived from
non-equilibrium thermodynamics guaranteeing a locally
positive entropy
production. The model is well suited to describe a
polycrystalline system by accounting for an arbitrary number of
phase-field parameters \citep{Nestler:2005ye}. In the considered
applications to crystal growth in veins, we restrict model definition
to an isothermal problem. The Helmholtz free energy functional
can be formulated as

\begin{eqnarray}
{\cal F}\left(\vphi\right)=\int_{\Omega}\left(f\left(\vphi\right)+\varepsilon
a\left(\vphi,\nabla\vphi\right) + \frac{1}{\varepsilon}
w\left(\vphi\right)\right)dx,
 \label{eq:free_energy_func}
\end{eqnarray}

where $f\left(\vphi\right)$ is the bulk free energy density,
$\varepsilon$ is the small length scale parameter related to the
interface width, $a\left(\vphi,\nabla\vphi\right)$ is a gradient
type and $w\left(\vphi\right)$, a potential type energy density. 
We discuss each of these terms in detail.

The phase-field parameter $\vphi\left(\vec{x},t\right) =
\left(\phi_{1}\left(\vec{x},t\right)\cdots\phi_{N}\left(\vec{x},t\right)\right)$
describes the location of `N' crystals with different
orientation. The value of each phase field paramter
$\phi_{\alpha}$ lies in the interval $\left[0,1\right]$ and
fulfils the constraint $\displaystyle\sum\limits_{\alpha=1}^N
\phi_{\alpha}=1$. The integral shown in functional
$\left(\mathrm{\ref{eq:free_energy_func}}\right)$ extends 
over the entire domain of consideration.

The gradient energy density $a\left(\vphi,\nabla\vphi\right)$ is
given by

\begin{eqnarray}
a\left(\vphi,\nabla\vphi\right) =
\displaystyle\sum_{\substack{\alpha,\beta=1 \\
\left(\alpha<\beta\right)}}^{N,N}\gamma_{\alpha \beta}
a_{\alpha\beta}^{2}\left(\vphi,\nabla\vphi\right)\vert
\vec{q}_{\alpha\beta}\vert^{2}
 \label{eq:grad_energy}
\end{eqnarray}

where $a_{\alpha\beta}\left(\vphi,\nabla\vphi\right)$ defines the form
of the surface energy anisotropy of the evolving phase boundary and
$\gamma_{\alpha\beta}$ is the surface free energy per unit area of 
the $\alpha-\beta$ boundary which may additionally depend on 
the relative orientation of the interface.  
The vector quantity $\vec{q}_{\alpha\beta}=\phi_{\alpha}\nabla\phi_{\beta}-\phi_{\beta}\nabla\phi_{\alpha}$
is a generalized gradient vector normal to the $\alpha-\beta$ interface.
To assign an isotropic surface energy to the $\alpha-\beta$
phase boundary, $a_{\alpha\beta}=1$ is chosen. For including
a strongly anisotropic surface energy, so that
crystals develop flat facets and sharp corners according
to directions of the crystal symmetry, a piecewise
defined function is used
\begin{eqnarray}
a_{\alpha\beta}\left(\vphi,\nabla\vphi\right)= \displaystyle
\max_{1\leq k\leq
\eta_{\alpha\beta}}\left\lbrace\dfrac{\vec{q}_{\alpha\beta}}{\vert
\vec{q}_{\alpha\beta}\vert}\cdot\vec{\eta_{\alpha\beta}}\right\rbrace
 \label{eq:facet_anis}
\end{eqnarray}
\newpage
\begin{figure}[!htbp]
\centering
\subfigure{\includegraphics[scale=0.25]{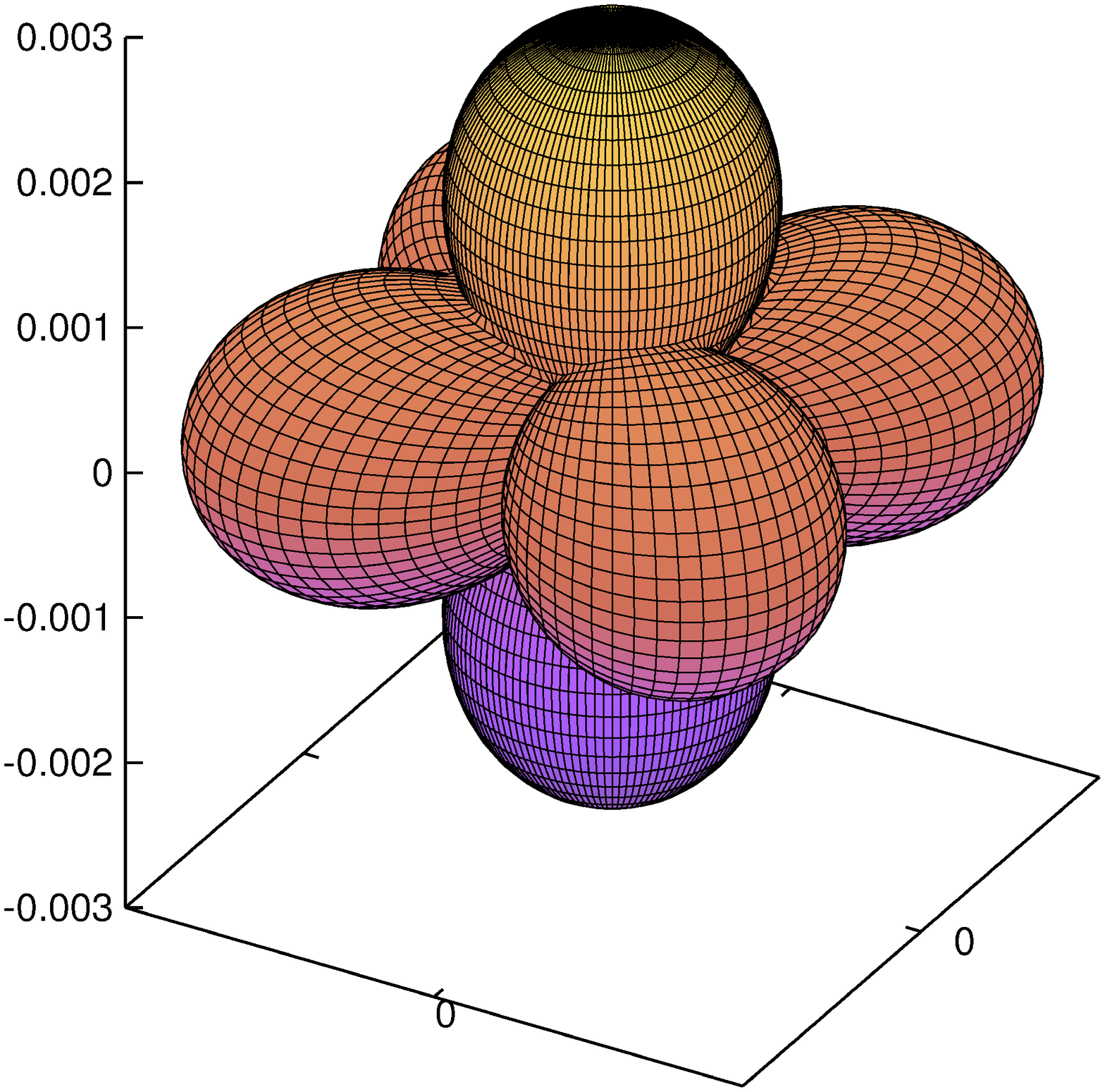}\label{fig:gammaplot_octa}}
\put(-5,48){\includegraphics[scale=0.06]{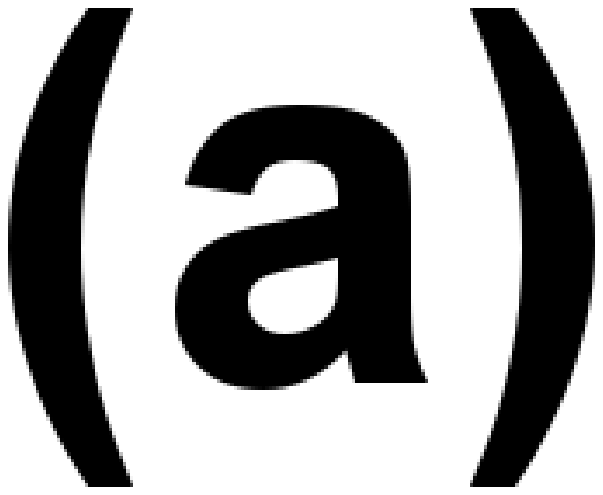}}\qquad\qquad
\subfigure{\includegraphics[scale=0.25]{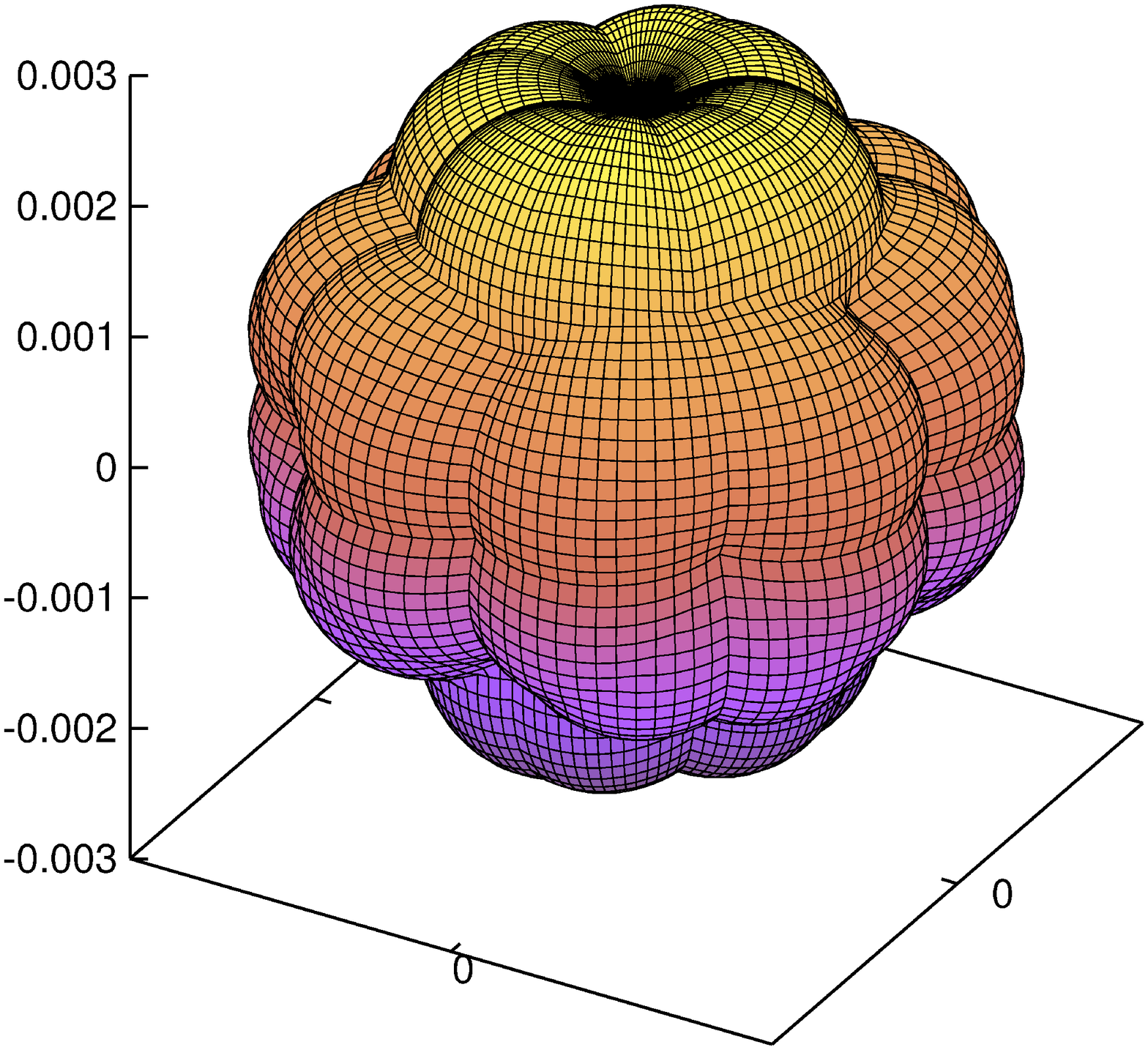}\label{fig:gammaplot_alum}} 
\put(-5,48){\includegraphics[scale=0.06]{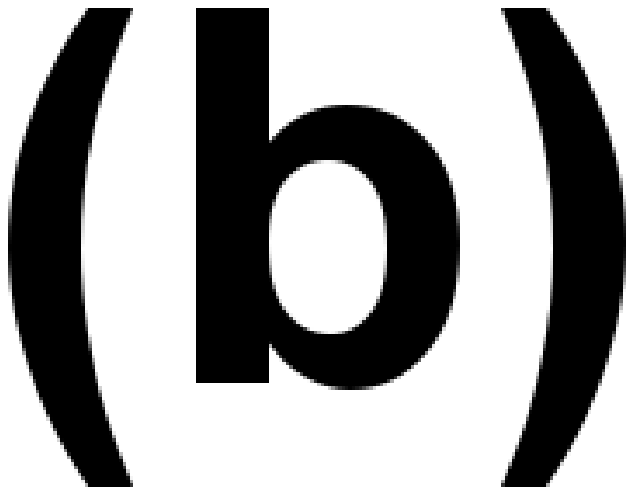}}\\
\subfigure{\includegraphics[scale=0.14]{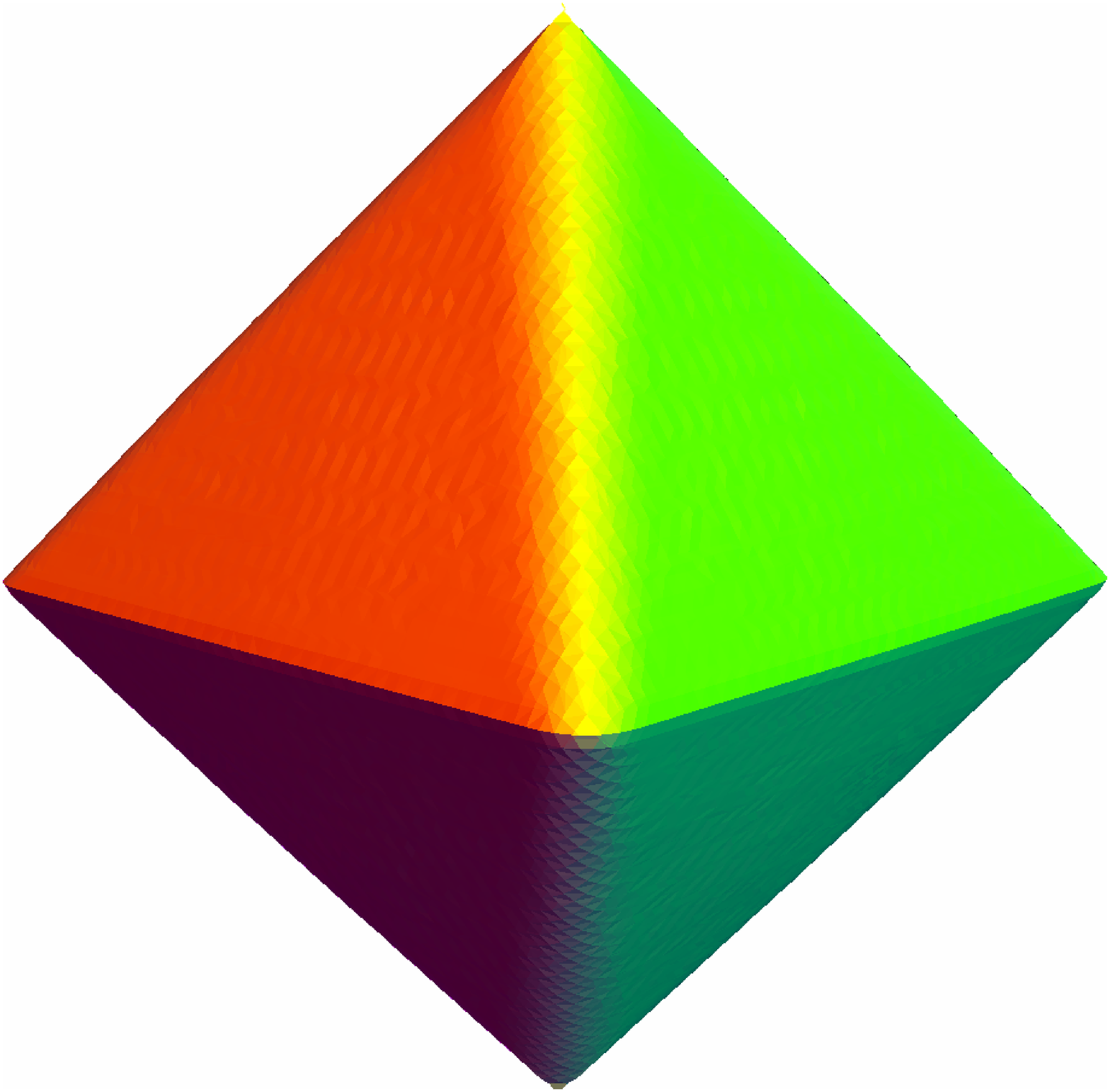}\label{fig:ECS_octa}}
\put(-5,40){\includegraphics[scale=0.06]{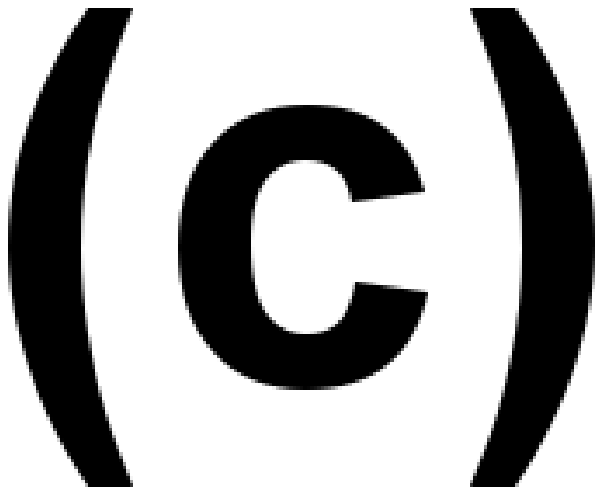}}\qquad\qquad
\subfigure{\includegraphics[scale=0.2]{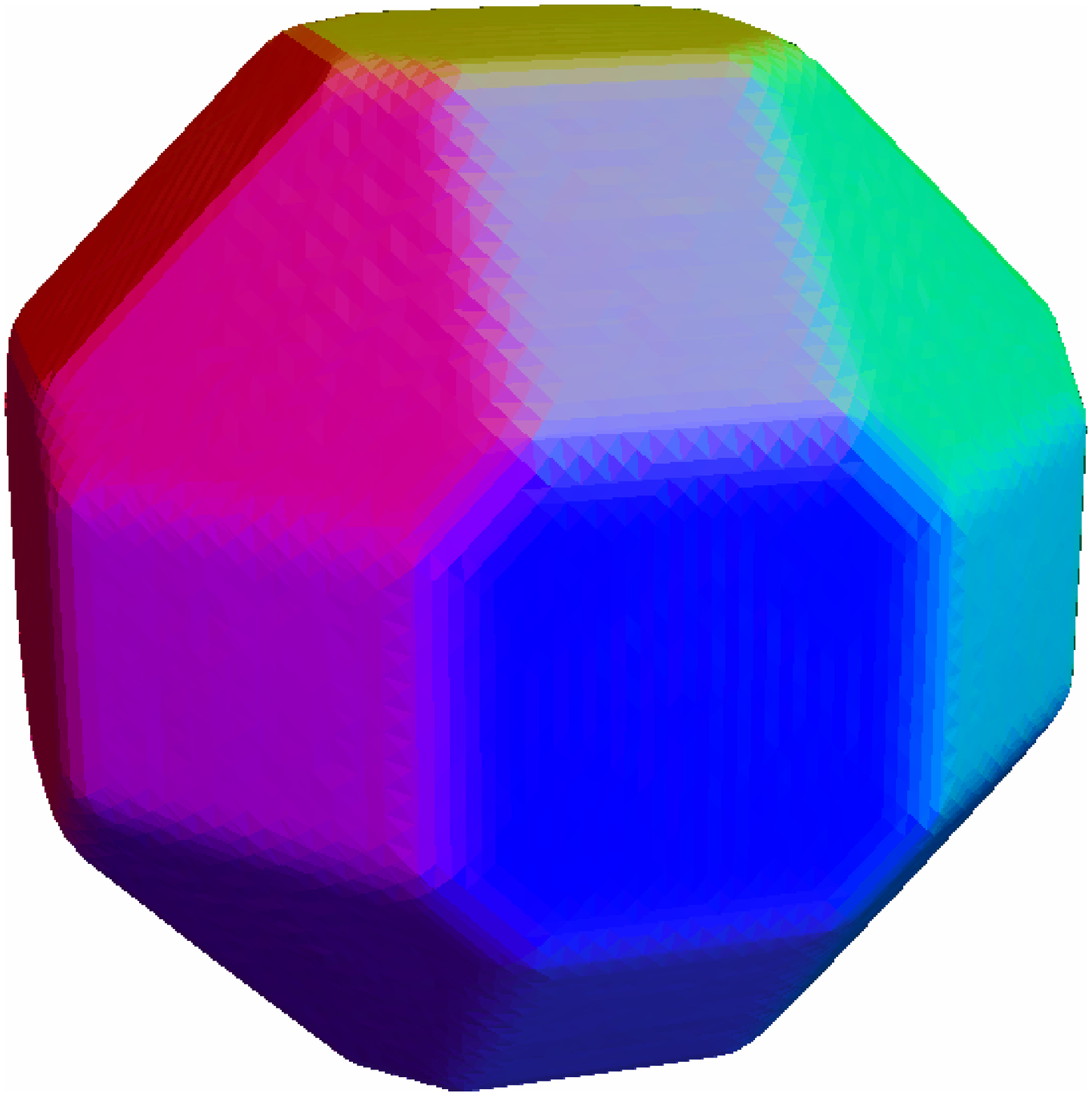}\label{fig:ECS_alum}}
\put(-5,40){\includegraphics[scale=0.06]{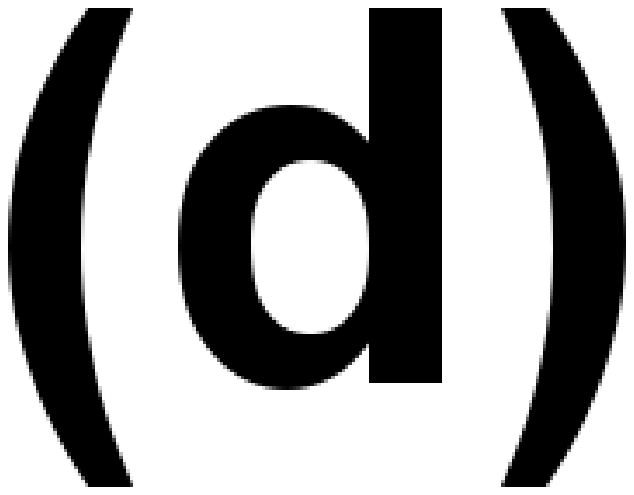}}
\caption{\textbf{Fig. 1 }Polar plot of function $\left(\mathrm{\ref{eq:facet_anis}}\right)$ for
(a) cubic and (b) alum (truncated octahedral symmetry). These
anisotropic functions are incorporated in phase-field model  
to simulate the equilibrium crystal shapes for (c) an octahedral
crystal and (d) an alum in free growth
conditions. The colors in Figs. 
\ref{fig:ECS_octa} and \ref{fig:ECS_alum} 
differentiate among crystal facets.} \label{fig:gam_plot}
\end{figure}
\clearpage
where $\lbrace\vec{\eta}_{k,\alpha\beta}\vert
k=1,\cdots,\eta_{\alpha\beta}\rbrace$ for $\eta_{\alpha\beta}\in
M$ denotes the complete set of vertex vectors of the
corresponding Wulff shape of a crystal $\alpha$ embedded in the
bulk phase $\beta$ and $M$ represents the number of edges. 
In Figs. \ref{fig:gammaplot_octa} and 
\ref{fig:gammaplot_alum}, a polar plot of
function $\left(\mathrm{\ref{eq:facet_anis}}\right)$ for a cubic symmetry and for a
truncated octahedral shape are shown. The
corresponding equilibrium crystal shapes are displayed in Figs.
\ref{fig:ECS_octa} and \ref{fig:ECS_alum}, respectively.

In the free energy functional $\left(\mathrm{\ref{eq:free_energy_func}}\right)$,
we choose the function $w\left(\vphi\right)$ to represent a multi-obstacle potential in
the form of

\begin{align}
w\left(\vphi\right) = \begin{cases}
\dfrac{16}{\pi^{2}}\displaystyle\sum_{\substack{\alpha,\beta=1
\\
\left(\alpha<\beta\right)}}^{N,N}\gamma_{\alpha\beta}\phi_{\alpha}\phi_{\beta}+
\displaystyle\sum_{\substack{\alpha,\beta,\delta=1 \\
\left(\alpha<\beta<\delta\right)}}^{N,N,N}\gamma_{\alpha\beta\delta}\phi_{\alpha}\phi_{\beta}\phi_{\delta}&
\text{if\hspace{0.5cm}$\vphi\in\sum$} \\
\infty & \text{elsewhere}
\end{cases}
\label{eq:surface_potential}
\end{align}

where $\displaystyle\sum =
\left\lbrace\vphi\;\vert\;\displaystyle\sum\nolimits_{\substack{\alpha=1}}^{N}\phi_{\alpha}
= 1 \;\mathrm{and}\; \phi_{\alpha}\geq 0\right\rbrace$. The
higher order term proportional to 
$\phi_{\alpha}\phi_{\beta}\phi_{\delta}$ 
in function $\left(\mathrm{\ref{eq:surface_potential}}\right)$
is added to reduce the presence of {an}
unwanted third or higher-order phase at binary interfaces.

The term $f\left(\vphi\right)$ represents the interface driving
force due to the occurrence of different bulk phases. A general
formulation of $f\left(\vphi\right)$ can be given as an
interpolation of different free energy densities $f_{\alpha}$
of the bulk phases, 

\begin{align}
f\left({\vphi}\right) =
\displaystyle\sum_{\substack{\alpha}}f_{\alpha}h\left(\phi_{\alpha}\right)
\label{eq:bulk_energy_interpol}
\end{align}

For studying the kinematics of crystal growth, we apply 
the interpolation function
$h\left(\phi_{\alpha}\right) = \phi_{\alpha}^{3}\left(6\phi_{\alpha}^{2}-15\phi_{\alpha}+10\right)$
as a suitable interpolaton function and a constant value for
$f_{\alpha}$ for the bulk free energies. 

The evolution equations for the phase fields can be
derived from the free energy functional 
$\left(\mathrm{\ref{eq:free_energy_func}}\right)$ 
by relating the temporal change of the order parameter, i.e. 
$\dfrac{\partial\phi_{\alpha}}{\partial t}$ to the 
variational derivative of the functional $\cal{F}$.
Applying the Euler Lagrange formalism yields:

\begin{align}
\tau\varepsilon\dfrac{\partial\phi_{\alpha}}{\partial t}=
\varepsilon\left(\nabla\cdot a_{,\nabla\phi_{\alpha}}\left(\vphi,\nabla\vphi\right)-
a_{,\phi_{\alpha}}\left(\vphi,\nabla\vphi\right)\right)-
\dfrac{1}{\varepsilon}w_{,\phi_{\alpha}}\left(\vphi\right)-
f_{,\phi_{\alpha}}\left(\vphi\right)-\lambda
\label{eq:evolution_eqn}
\end{align}

\begin{align}
\lambda=\dfrac{1}{N}\displaystyle\sum_{\alpha}\varepsilon
\left(\nabla\cdot a_{,\nabla\phi_{\alpha}}\left(\vphi,\nabla\vphi\right)-
a_{,\phi_{\alpha}}\left(\vphi,\nabla\vphi\right)\right)-
\dfrac{1}{\varepsilon}w_{,\phi_{\alpha}}\left(\vphi\right)-
f_{,\phi_{\alpha}}\left(\vphi\right)
\label{eq:Lagrange}
\end{align}

where the comma separated subindices represent
derivatives with respect to $\phi_{\alpha}$ and 
gradient components $\dfrac{\partial\phi_{\alpha}}{\partial\chi_{i}}$.
The lagrange multiplier $\lambda$ guarantees the
summation constraint 
$\left(\displaystyle\sum_{\alpha=1}^{N}\;\phi_{\alpha}=1\right)$.
In the evolution equation $\left(\mathrm{\ref{eq:evolution_eqn}}\right)$, 
$\tau$ is the kinetic coefficient 
which establishes a relationship between the interface growth velocity 
and the driving force. In the following sections, we assume negligible
interface kinetic effect and choose the value of $\tau$ accordingly.
This assumption ensures that the rate at which a nuclei evolves 
into its equilibrium shape (fast or slow growth mode),
is not influenced by interface kinetics.

The phase-field evolution equation
$\left(\mathrm{\ref{eq:evolution_eqn}}\right)$ 
is solved numerically
using an explicit forward Euler scheme.
The spatial derivatives of the right hand side 
equation are discretized 
using a second order accurate scheme with a 
combination of forward and backward finite differences.
The phase-field solver is written in programming
language C and only solves the evolution equation 
next to the locally present interfaces. 
The implementation of such a locally 
reduced order parameter optimization
(LROP) facilitates a reduction in 
computation time from $O(N^3)$ to $O(1)$ and 
in memory consumption from $O(N)$ to $O(1)$ 
per cell in the domain, N being the number of
crystals in the system. Thus, the computation is 
independent of the number of phases making 
the large scale crystal growth studies feasible 
even in 3D
\citep{Kim:2006fk,Nestler:2008kx}.
Further, the simulation code is highly parallelized on
the basis of message passing interface (MPI) standard
including 3D domain decomposition and 
dynamic redistribution schemes. 
For the present article, the simulations are
performed on multiprocessor workstations 
as well as on a Linux high performance cluster.

\section{Free growth in an open cavity}
\label{sec:free_growth}
The dependence of growth rate on facet orientations 
for a single crystal  has been reported in the past
\citep{Mugge:1925fk,Mullin:2001uq}. However, when polycrystal
growth occurs, the neighboring crystals influence the growth
rate. Microstructures formed due to crystal growth in an open
cavity are characterized by an increase in grain size with more
favorably oriented facets out-growing the poorly oriented neighbors
\citep{Schmidegg:1928uq,Thijssen:1992kx}. 
The orientation selection rule responsible for the
growth competition in such systems 
is based {on mis-orientation} with 
respect to the most preferred growth 
direction. In the following section, we 
simulate grain growth competition 
in 2D and 3D and deduce
the orientation selection rule.
Since the scope of the current study is 
limited to kinematics 
of crystal growth, the driving 
force is assumed to be constant. 

One common mineral analogue that has 
been used to replicate free growth of crystals in 
laboratory experiments is Potash alum
$\left[\mathrm{KAl\left(SO_{4}\right)_{2}\cdot 12H_{2}O}\right]$
\citep{NOLLET:2006oz}. The crystal structure of alum is cubic
with eight $\left\lbrace111\right\rbrace$ facets. At room 
temperature it grows into an octahedral habit 
and also develops $\left\lbrace110\right\rbrace$ and 
$\left\lbrace100\right\rbrace$ facets.
However, the growth rate of the $\left\lbrace110\right\rbrace$
and $\left\lbrace100\right\rbrace$  facets is much higher 
as compared to $\left\lbrace111\right\rbrace$, hence 
they are much smaller in size as compared to the 
primary facet \citep{Bhat:1992uq}. \citet{Klapper:2002fk} 
report that temperature fluctuation 
increases the growth rate of $\left\lbrace110\right\rbrace$ 
and $\left\lbrace100\right\rbrace$ facets with respect to 
$\left\lbrace111\right\rbrace$. They also state that crystal 
dissolution and recrystallization can lead to formation of extra facets
in the early growth stages. In the present work, we limit 
the discussion to cubic symmetry since the kinetics of
precipitation of crystals from its salt solution is yet to be 
incorporated in the present phase-field model.

\begin{figure}[!htbp]
\centering
\subfigure{\includegraphics[width=0.25\textwidth,height=0.25\textwidth]{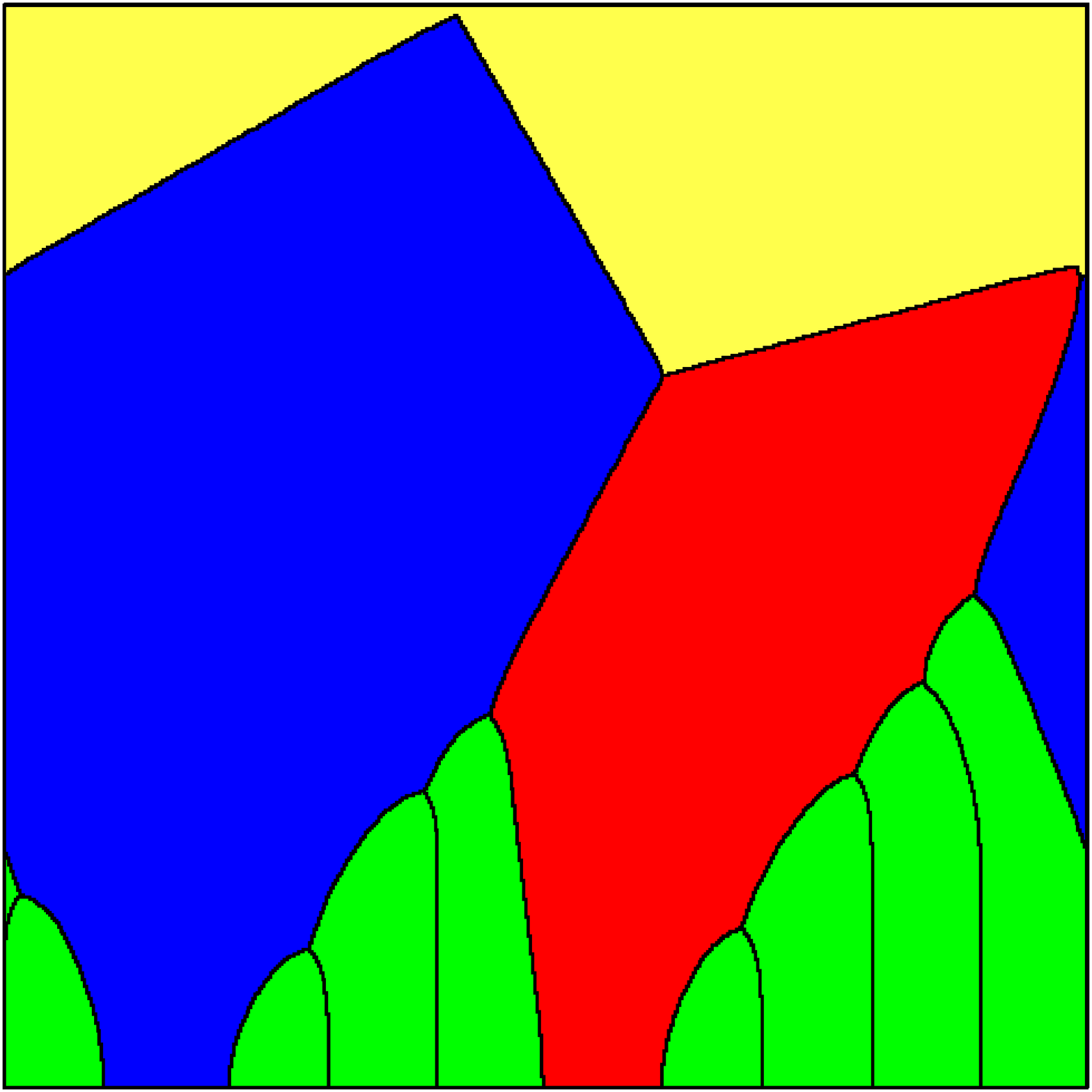}\label{fig:A_15_B_30_C_45}}
\put(-35,33){\includegraphics[scale=0.06]{a.eps}}\quad
\subfigure{\includegraphics[width=0.25\textwidth,height=0.25\textwidth]{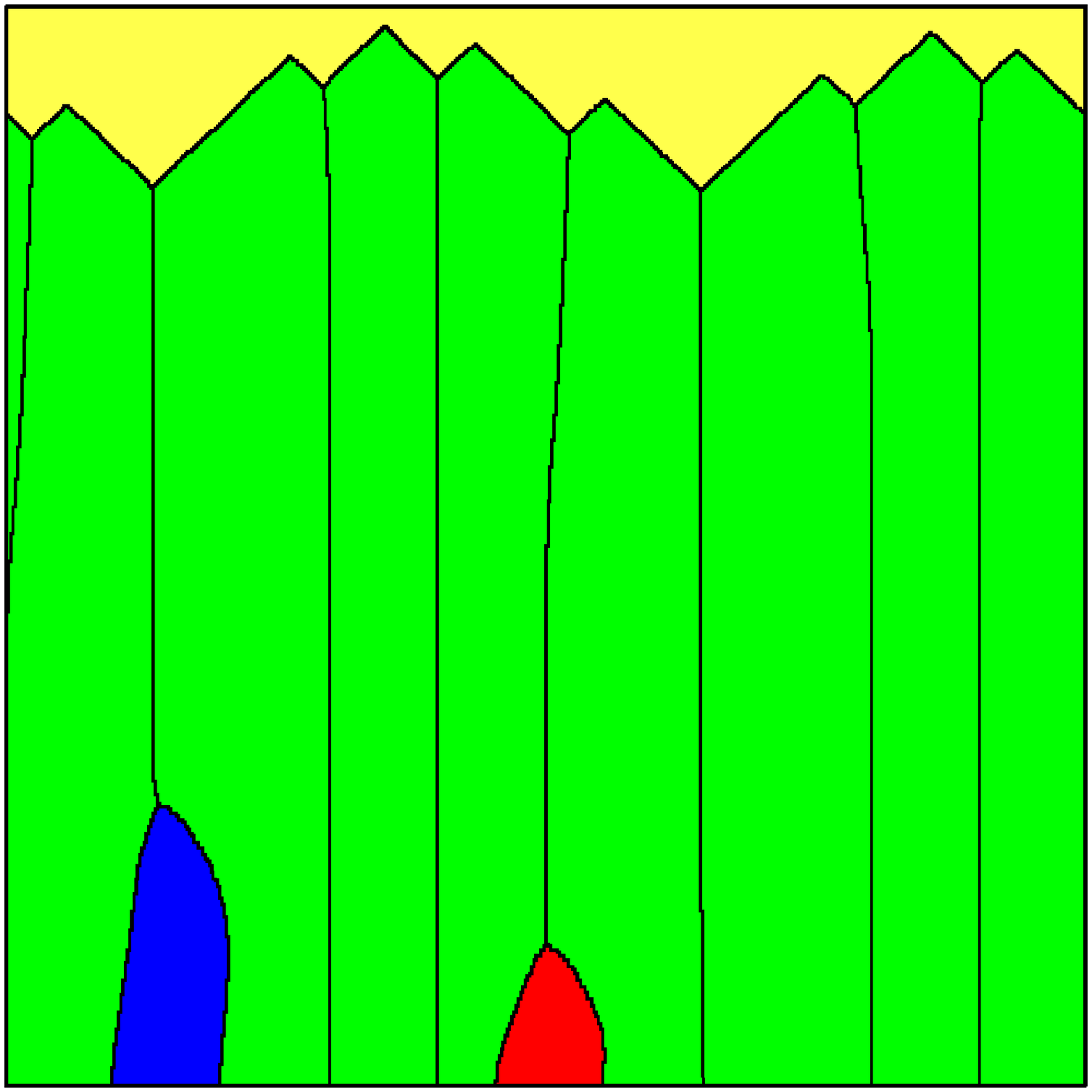}\label{fig:A_15_B_30_C_0}}
\put(-35,33){\includegraphics[scale=0.06]{b.eps}}\quad
\subfigure{\includegraphics[width=0.25\textwidth,height=0.25\textwidth]{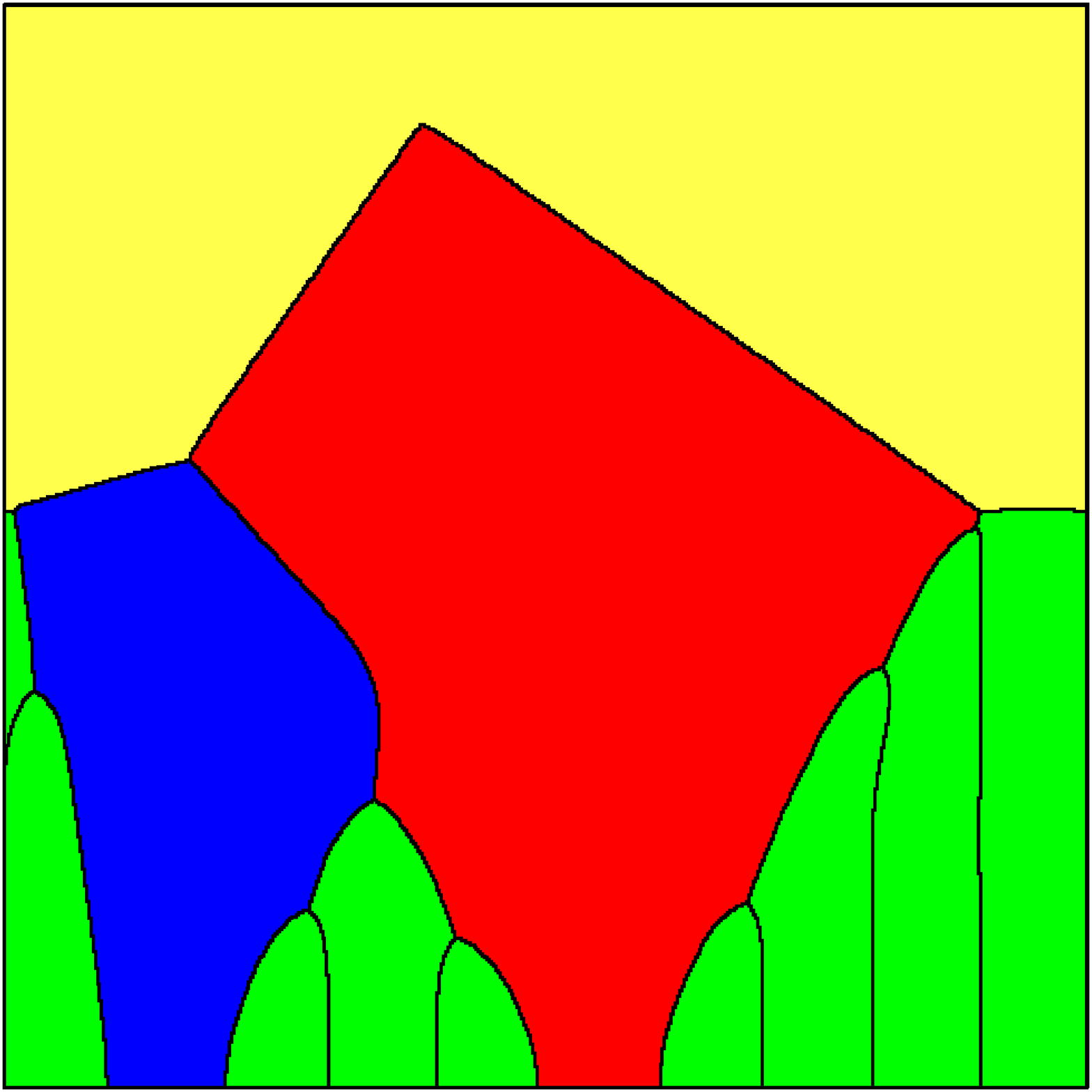}\label{fig:A_30_B_min5_C_45}}
\put(-35,33){\includegraphics[scale=0.06]{c.eps}}\\
\subfigure{\includegraphics[width=0.25\textwidth,height=0.25\textwidth]{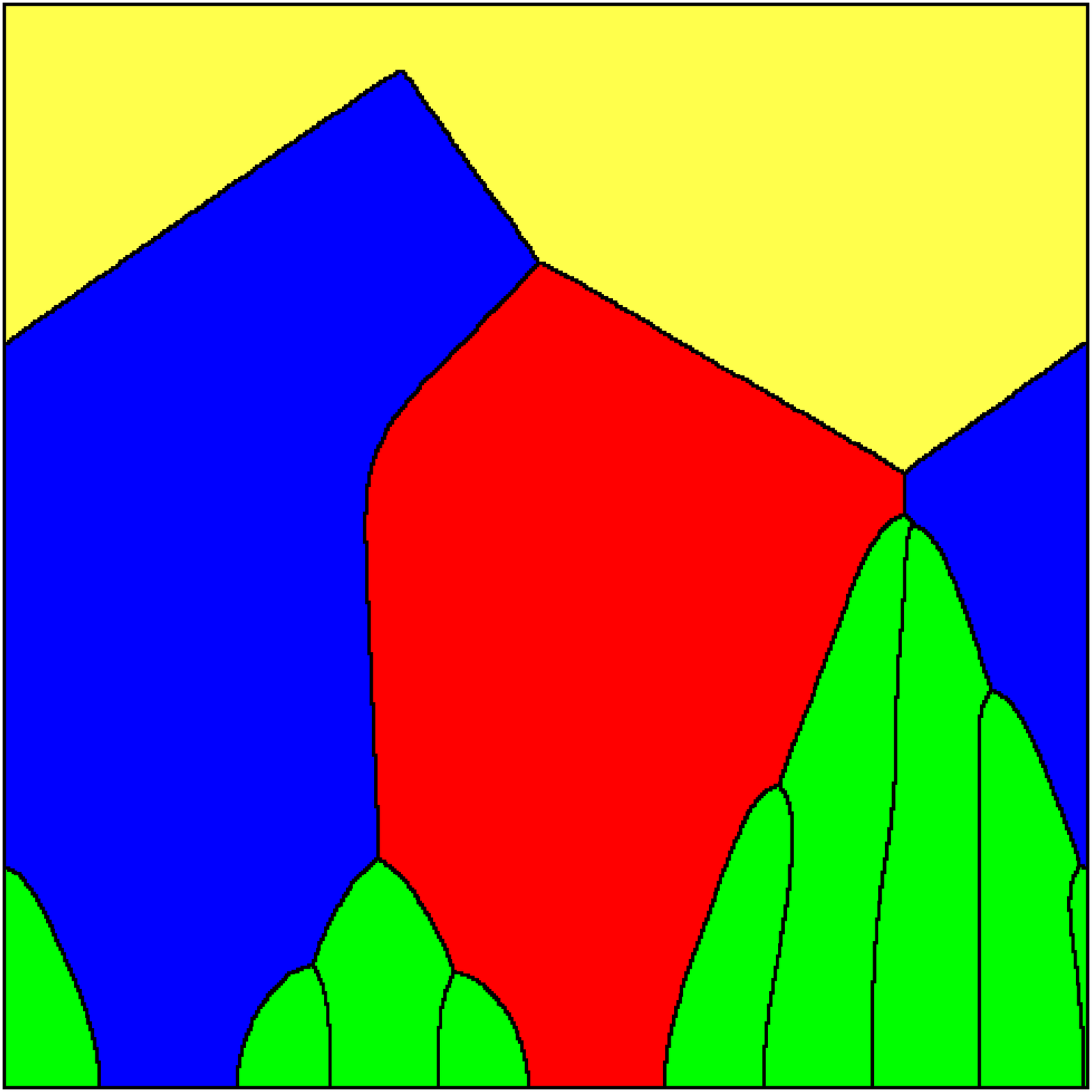}\label{fig:A_10_B_min15_C_45}}
\put(-35,33){\includegraphics[scale=0.06]{d.eps}}\quad
\subfigure{\includegraphics[width=0.25\textwidth,height=0.25\textwidth]{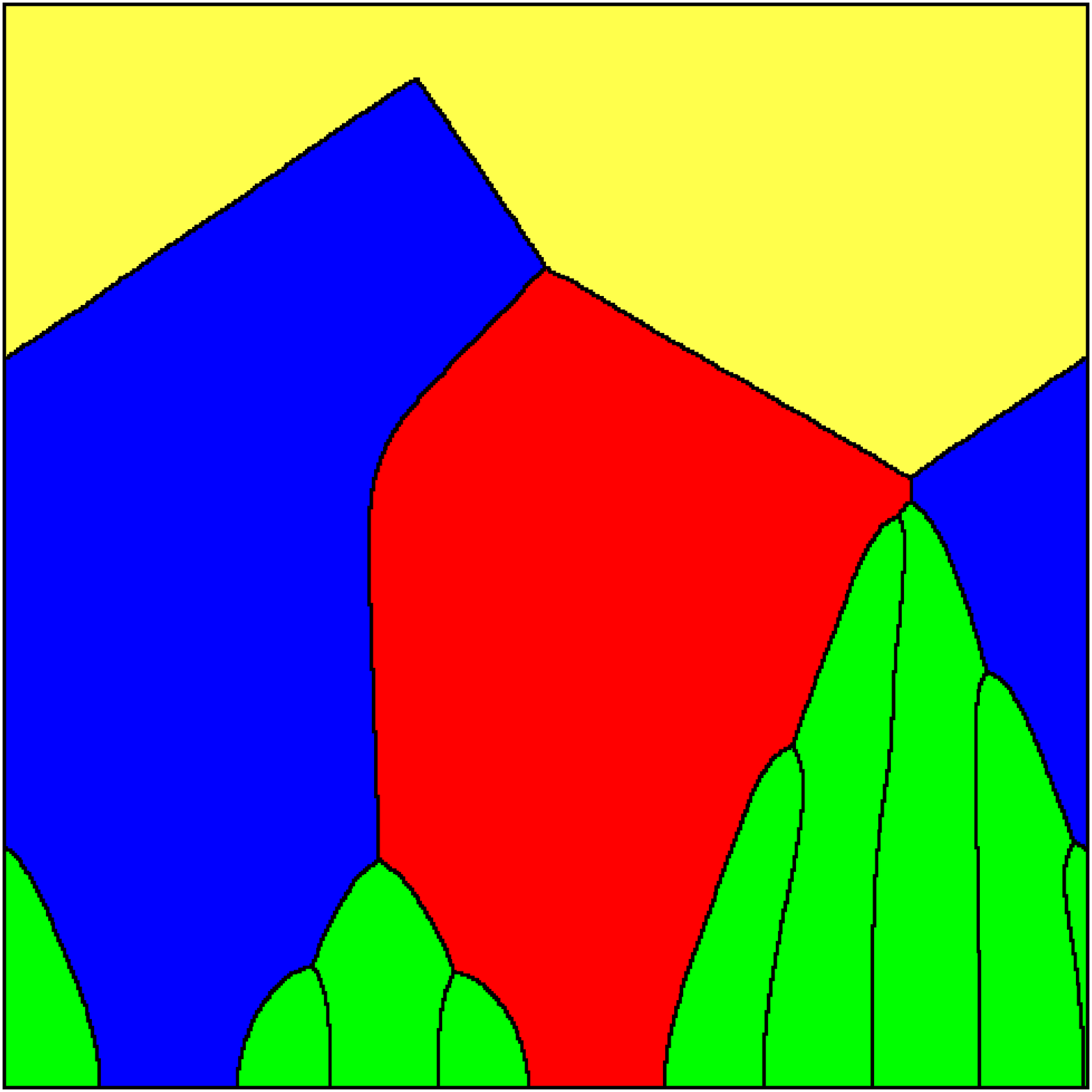}\label{fig:A_11_B_min15_C_45}}
\put(-35,33){\includegraphics[scale=0.06]{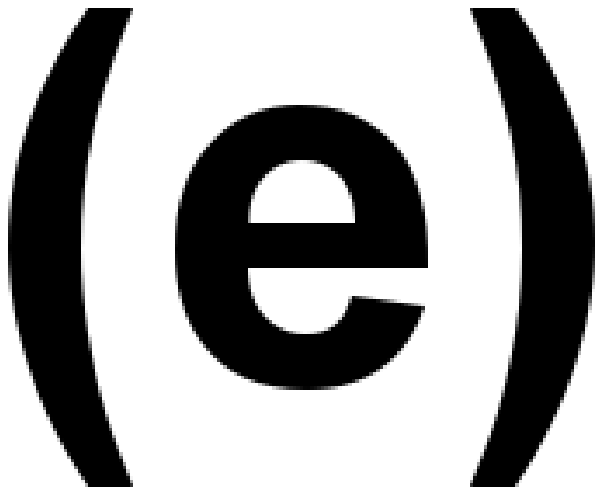}}\quad
\subfigure{\includegraphics[width=0.25\textwidth,height=0.25\textwidth]{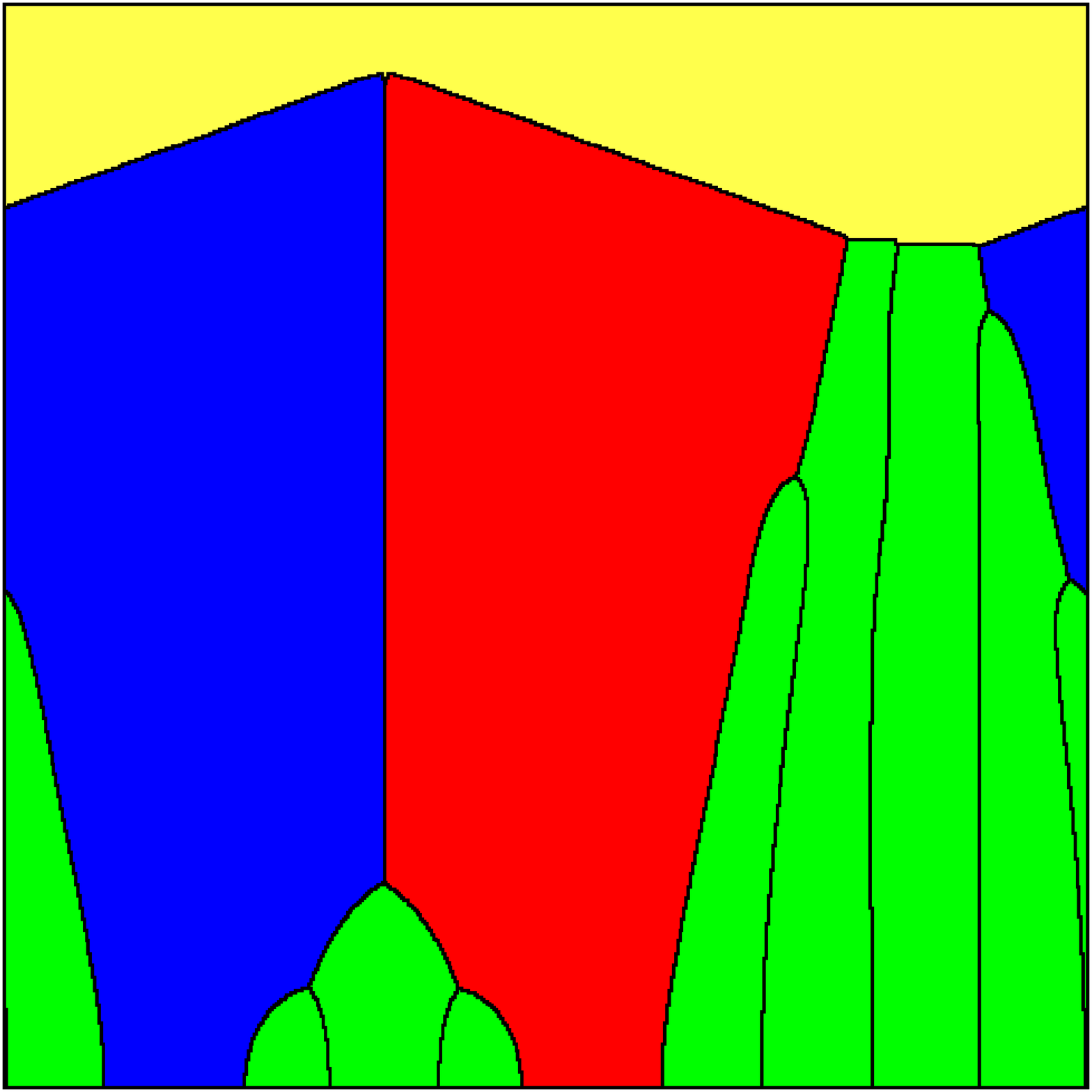}\label{fig:A_25_B_min25_C_45}}
\put(-35,33){\includegraphics[scale=0.06]{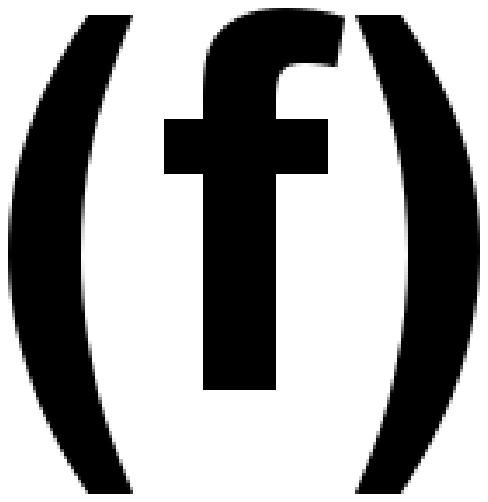}}\\
\subfigure{\includegraphics[scale=0.25]{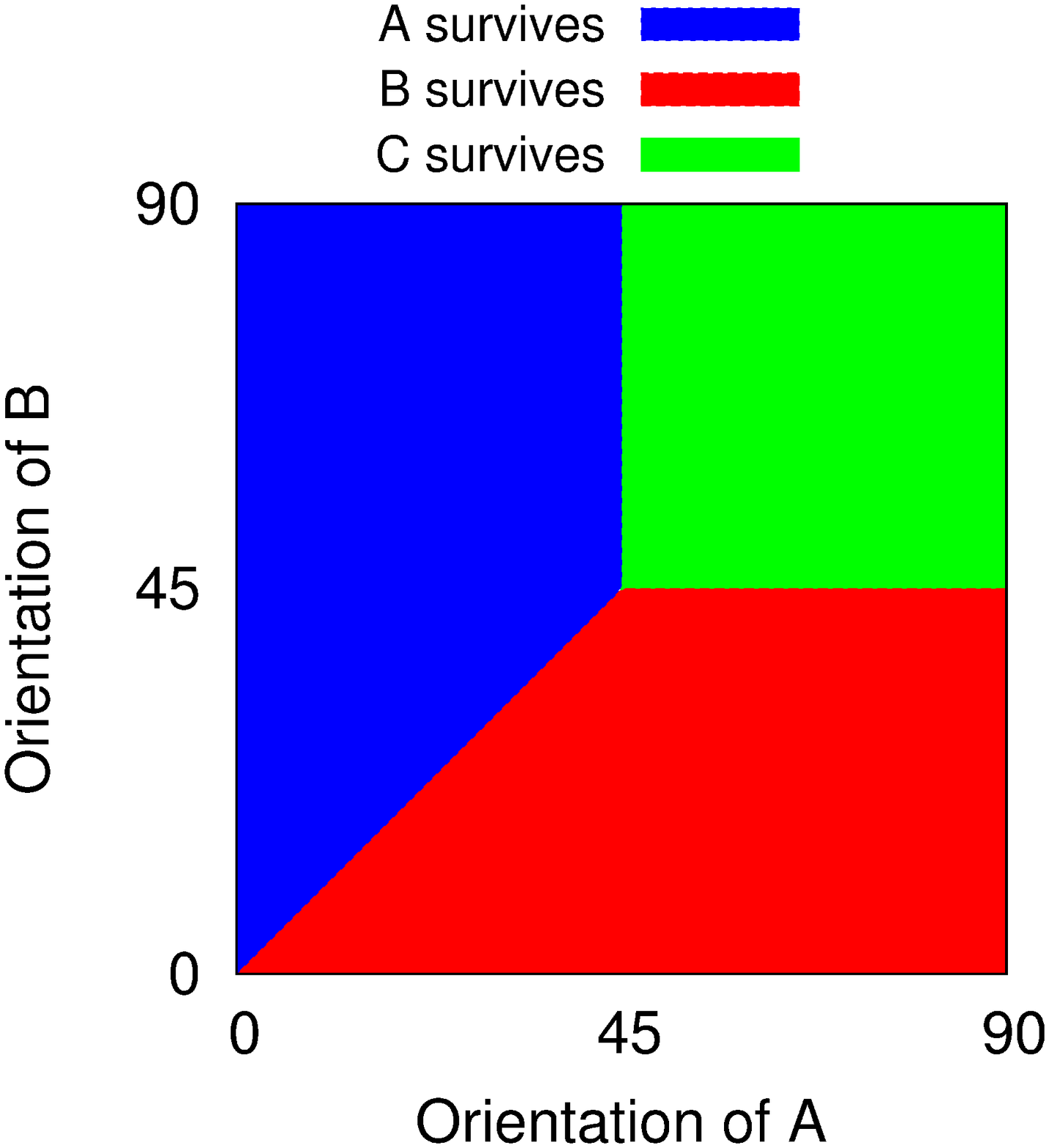}\label{orient_map}}
\put(-45,45){\includegraphics[scale=0.06]{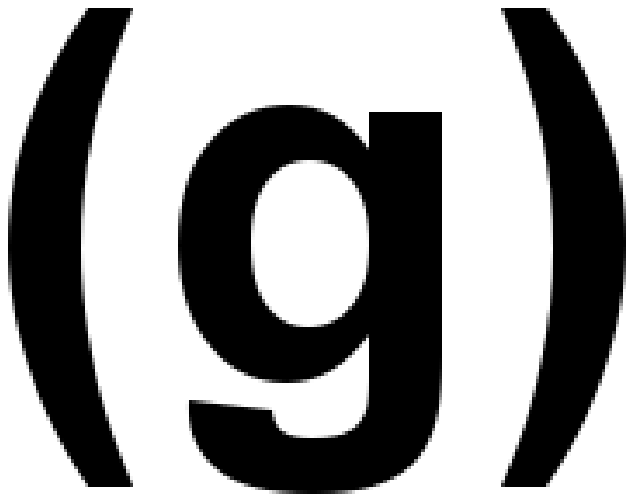}}\qquad
\subfigure{\includegraphics[scale=0.26]{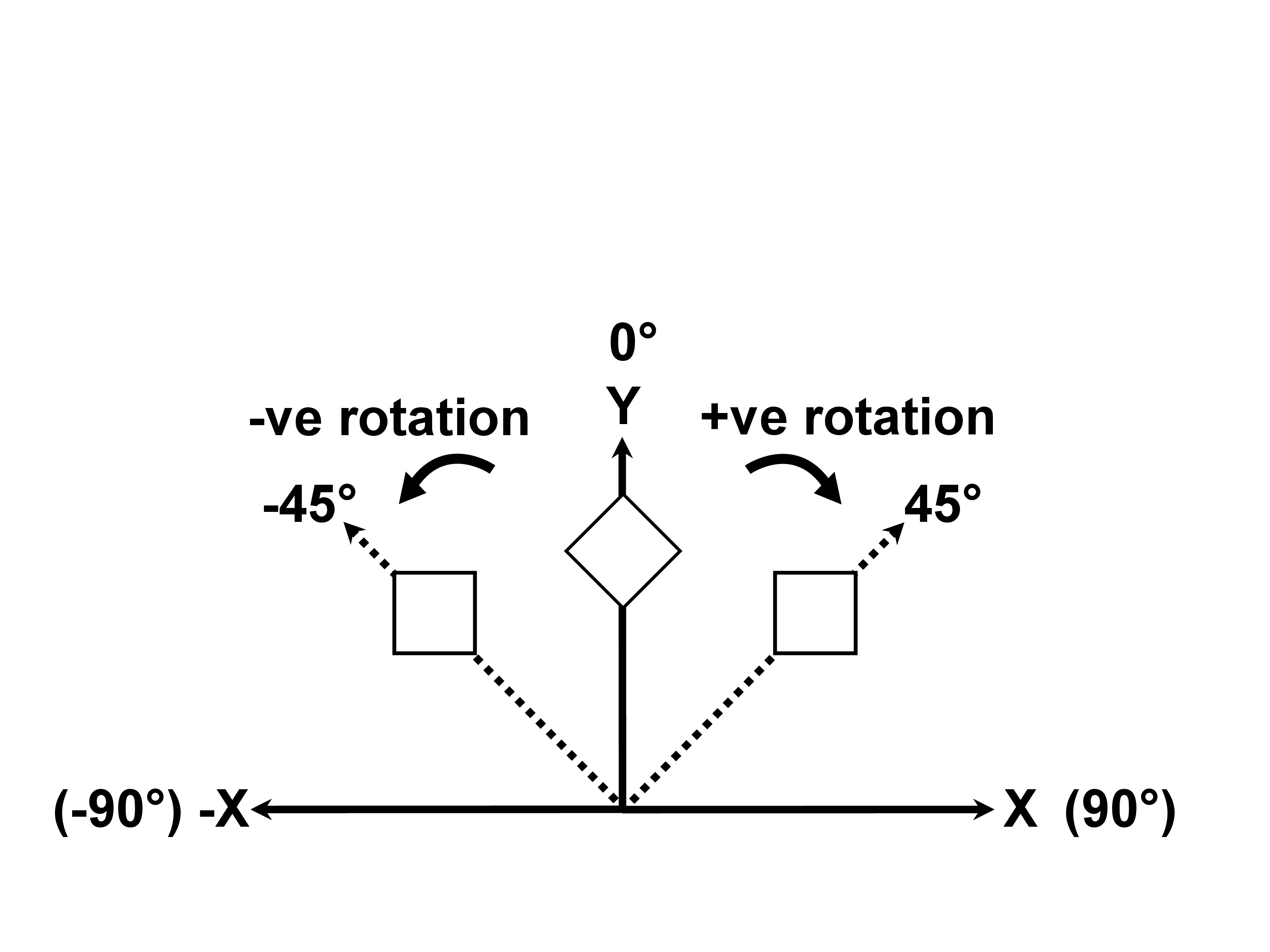}\label{fig:orient_def_2D_cubic}}
\put(-83,45){\includegraphics[scale=0.015]{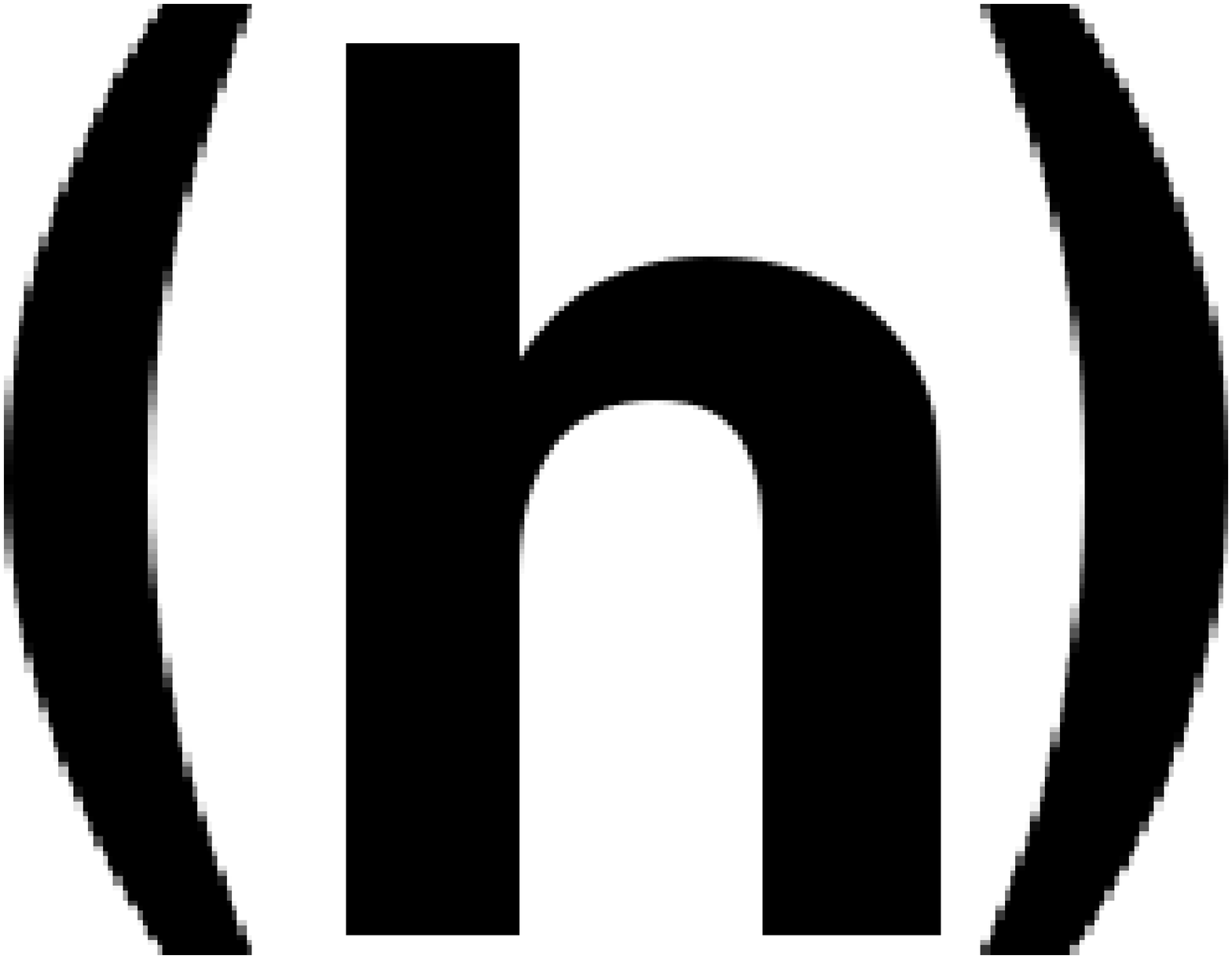}}
\caption{\textbf{Fig. 2 }Anisotropic growth competition of crystals with
orientation A (blue), B (red) and C (green).
(a) A = 15$\degree$, B = 30$\degree$ and C = 45$\degree$
{(b) A = 15$\degree$, B = 30$\degree$ and C = 0$\degree$}
(c) A = 30$\degree$, B = -5$\degree$ and C = 45$\degree$
(d) A = 10$\degree$, B = -15\degree and C = 45$\degree$
(e) A = 11$\degree$, B = -15\degree and C = 45$\degree$
(f) A = 25$\degree$, B = -25$\degree$ and C =45$\degree$
(g) Orientation map for C = 45\degree derived from phase-field simulation results
{(h) Orientation definition for 2D cubic crystal
growth simulations.}}\label{fig:Freely_growing_crystals}
\end{figure}

\begin{figure}[!htbp]
\centering
\subfigure{\includegraphics[scale=0.13]{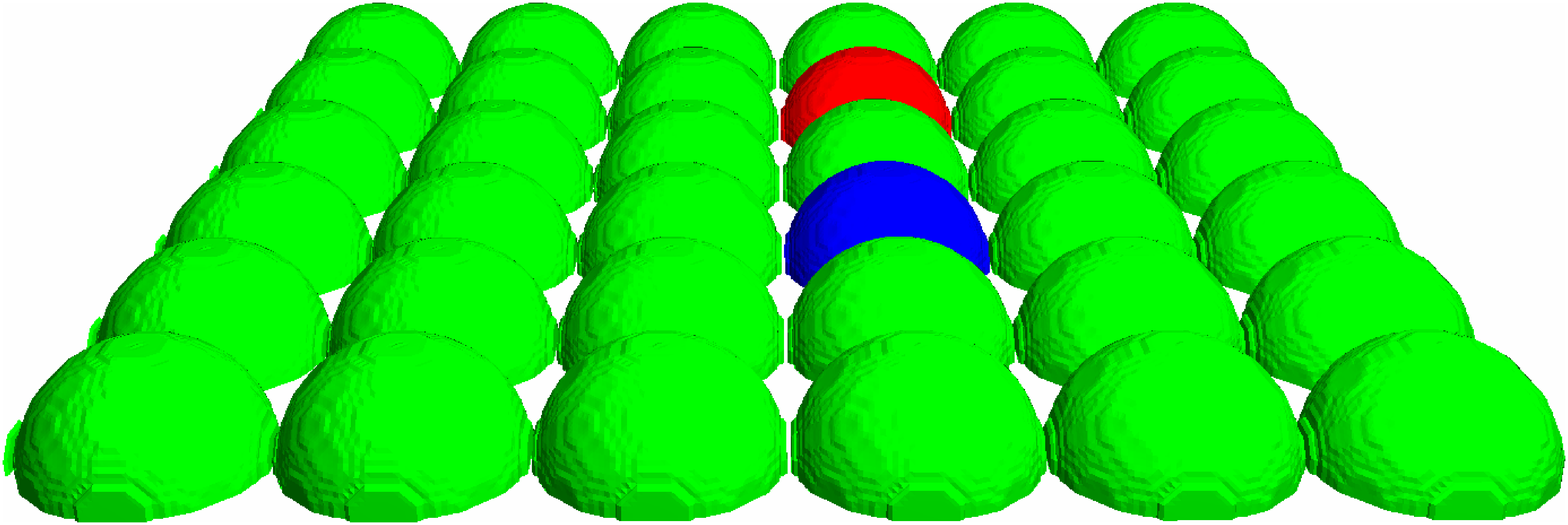}\label{fig:3d_T0}}
\put(-5,20){\includegraphics[scale=0.06]{a.eps}}\quad
\subfigure{\includegraphics[scale=0.10]{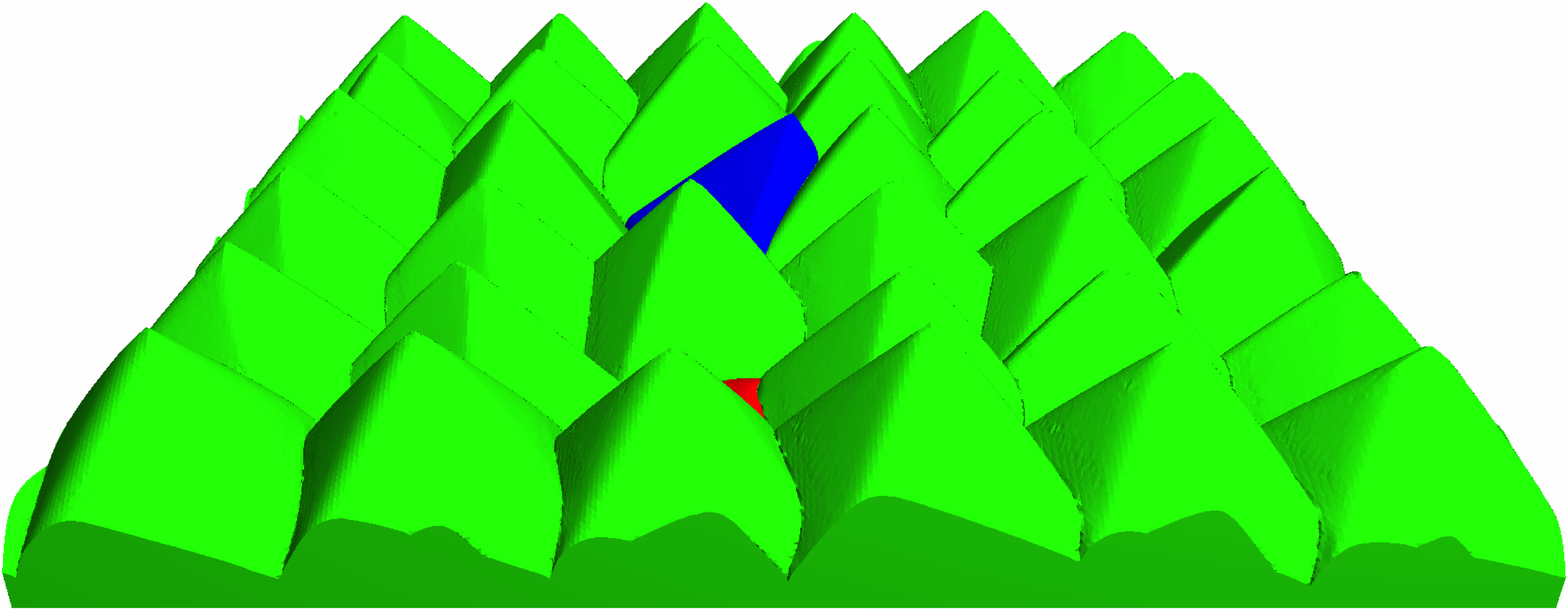}\label{fig:3d_A_15_B_30_C_0}}
\put(-5,20){\includegraphics[scale=0.06]{b.eps}}\\
\subfigure{\includegraphics[scale=0.11]{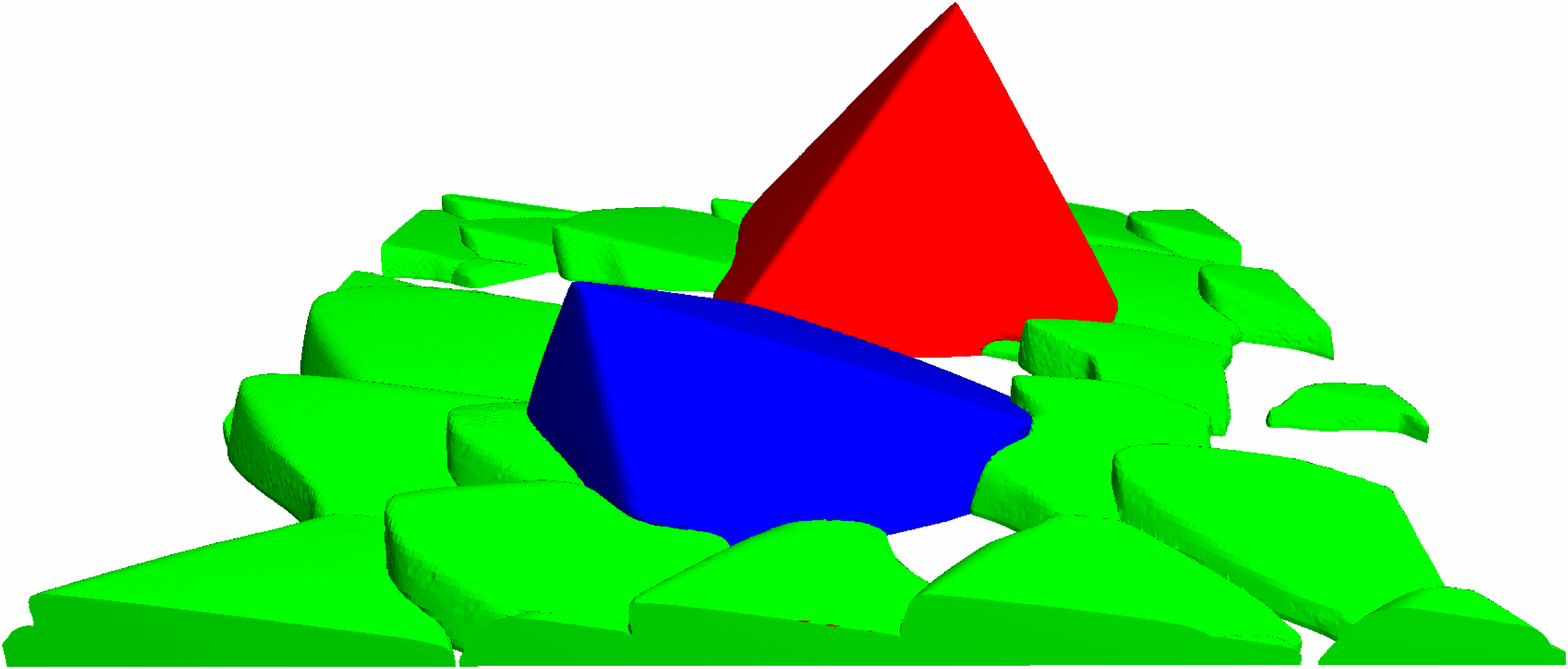}\label{fig:3d_A_min30_B_5_C_45}}
\put(-5,20){\includegraphics[scale=0.06]{c.eps}}\quad
\subfigure{\includegraphics[scale=0.17]{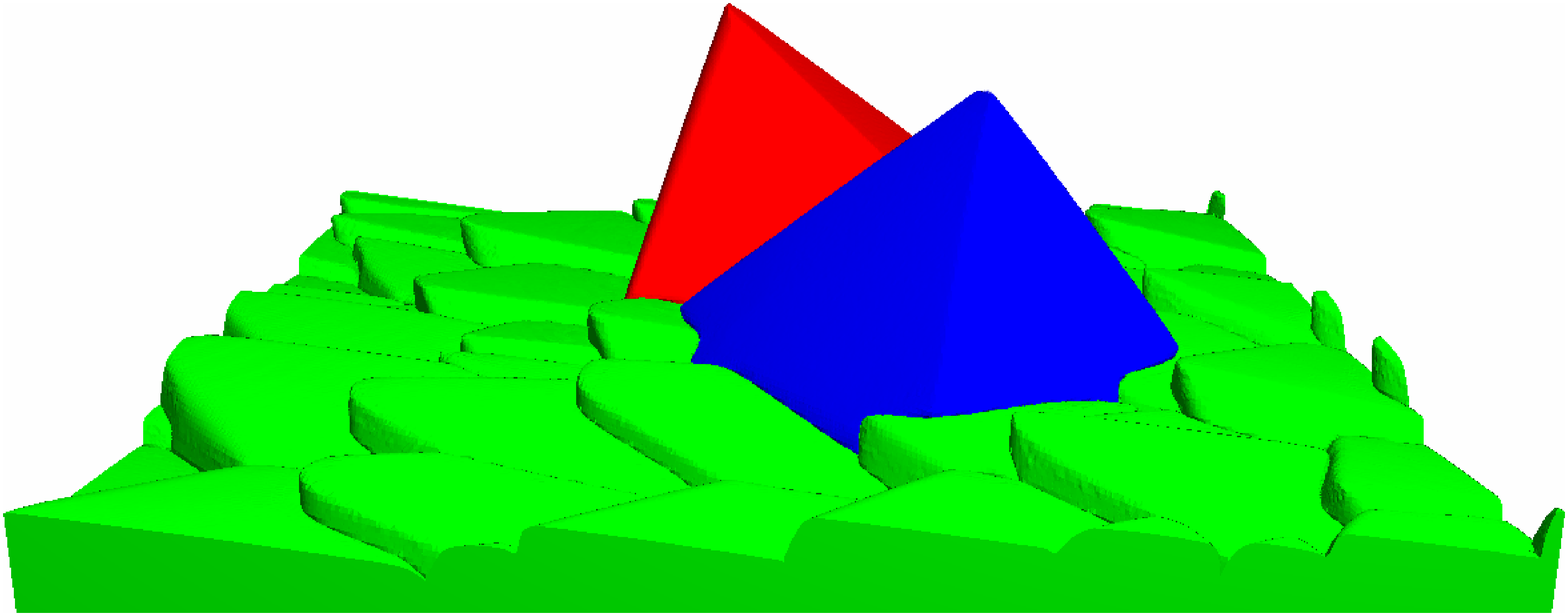}\label{fig:3d_A_15_B_min10_C_45}}
\put(-5,20){\includegraphics[scale=0.06]{d.eps}}\\
\subfigure{\includegraphics[scale=0.25]{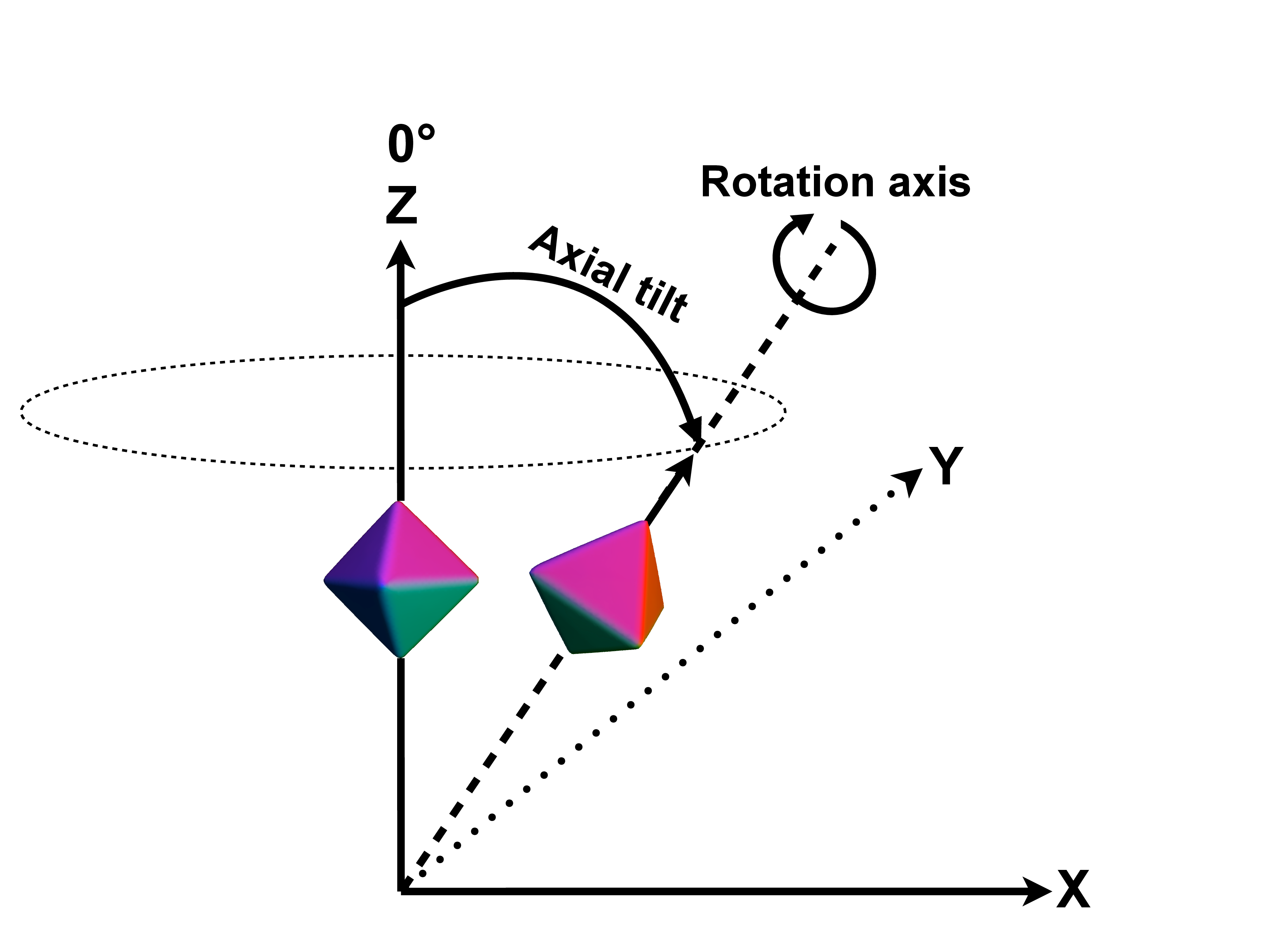}\label{fig:define_3D_orient_cubic}}
\put(-15,48){\includegraphics[scale=0.06]{e.eps}}
\caption{\textbf{Fig. 3 }3D phase-field simulation of cubic crystal growth:
(a) Initial condition with hemispherical nuclei uniformly embedded at the
bottom of the domain and with assigned orientations of A (blue),
B (red) and C (green) degrees. Growth competition leads to
orientation selections according to the orientations (b) 
{A = 15$\degree$, B = 30$\degree$ and C =
0$\degree$ (c) A = 5$\degree$, B = 30$\degree$and C = 45$\degree$
(d) A = 15$\degree$, B = 10$\degree$ and C = 45$\degree$}
{(e) Schematic diagram to
provide orientation definition for 3D cubic crystal
growth simulations. The different colors of crystal 
facets is meant for better visualization 
of orientation components (axial tilt
and rotation axis) 
and is not to be confused with color-coding 
in preceding sub-figures.}}\label{fig:3d_fg}
\end{figure}

In the first set of simulations, we study the effect of
non-neighboring members {(termed as 
"Long distance effect" by \citet{Nollet:2005qo})} on the free-growth regime of cubic
crystals. A 2D simulation is carried out with 10 crystal 
seeds uniformly embedded at the bottom of the simulation 
domain and orientation designated as A, B or C
degrees. In Fig. \ref{fig:Freely_growing_crystals}, the
second grain from left at the bottom grain boundary 
is assigned `A' degrees and sixth grain as `B'
degrees. All other grains are assigned `C' degrees. 
The orientation definitions in 2D simulations are given by
Fig. \ref{fig:orient_def_2D_cubic}. Similarly, we
simulate the free-growth in 3D by embedding 36 crystal seeds
uniformly at the bottom and assigning each orientation 
A and B once in the domain. 
The rest of the grains are assigned with orientation value C as 
shown in Fig. \ref{fig:3d_fg}. 
The numeric value of `A', `B' and `C' are chosen to 
investigate the effect of non-neighboring crystals (if any) 
on orientation selection during free-growth of crystals.
\newpage
\clearpage
\newpage
\clearpage
On the basis of phase-field simulation results as 
shown in Figs. \ref{fig:A_15_B_30_C_45} - 
\ref{fig:A_25_B_min25_C_45}, 
we derive an orientation map Fig. \ref{orient_map} of
crystal orientations `A', `B' and `C'  to establish the following selection 
rules:

\begin{itemize}
\item If every crystal has a different orientation, the crystal
which is most favorably oriented (vertical in this case), 
survives and rest of the crystals are consumed.
\item The consumption or survival of a crystal depends on its
own orientation relative to its neighbors. The non-neighboring
crystals do not effect the growth.
\item Crystals retain {the} equilibrium shape (from Wulff
construction) when in contact with liquid at all times.
\item Crystals having the same growth orientation with respect 
to most preferred orientation (vertical in this case) form 
perpendicular grain boundaries {(relative to initial surface)}
and co-exist in the final 
microstructure, if the other neighboring crystals are not 
more favorably oriented. This is also observed in {the} 3D case 
when two crystals have a similar orientation with 
respect to {the} vertical line but lie in different planes.
\end{itemize}

\begin{figure}[!htbp]
\centering
\subfigure{\includegraphics[scale=0.22]{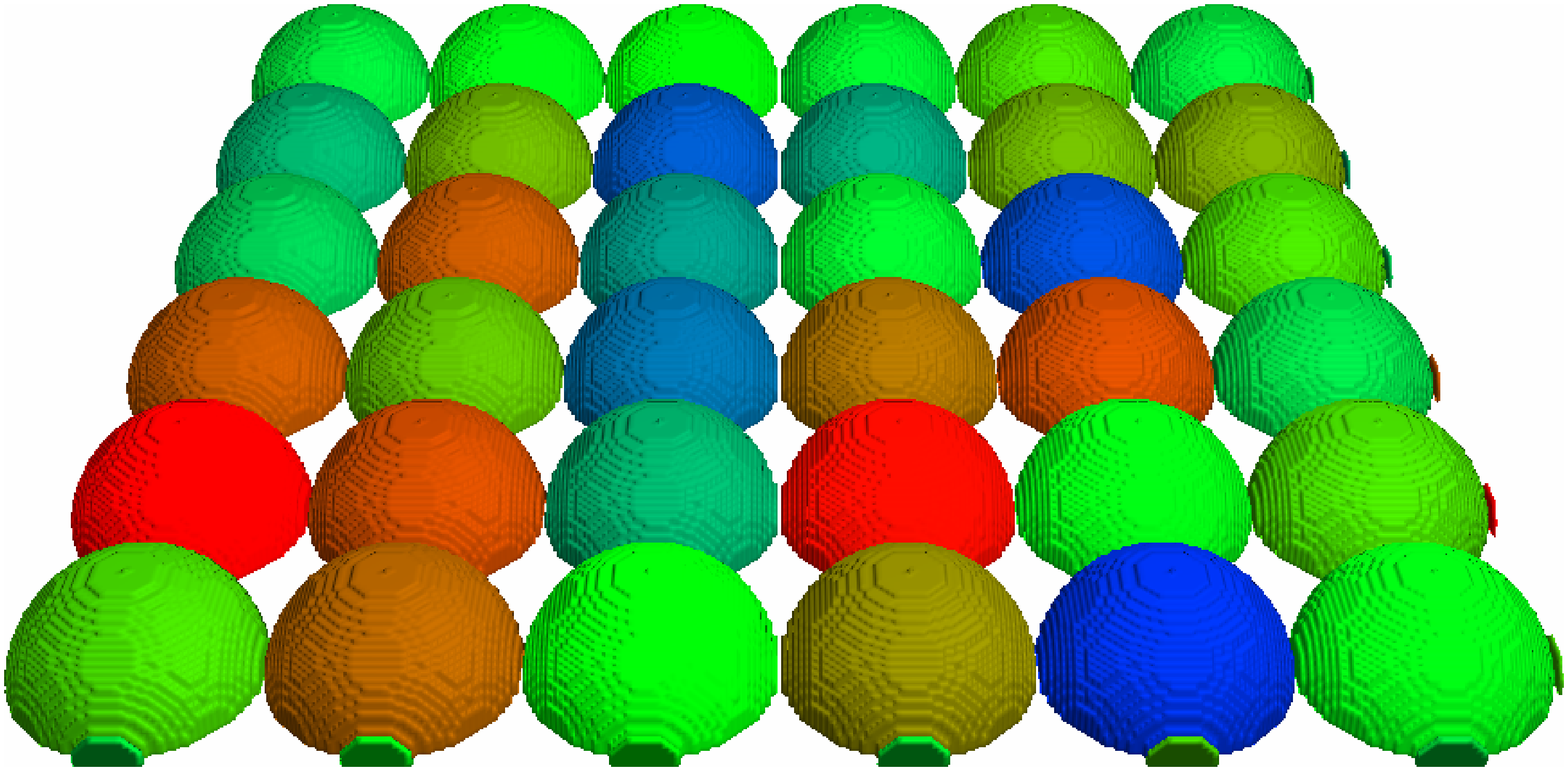}\label{fig:3D_alum_free_time_zero}}
\put(-5,30){\includegraphics[scale=0.06]{a.eps}}\quad
\subfigure{\includegraphics[scale=0.22]{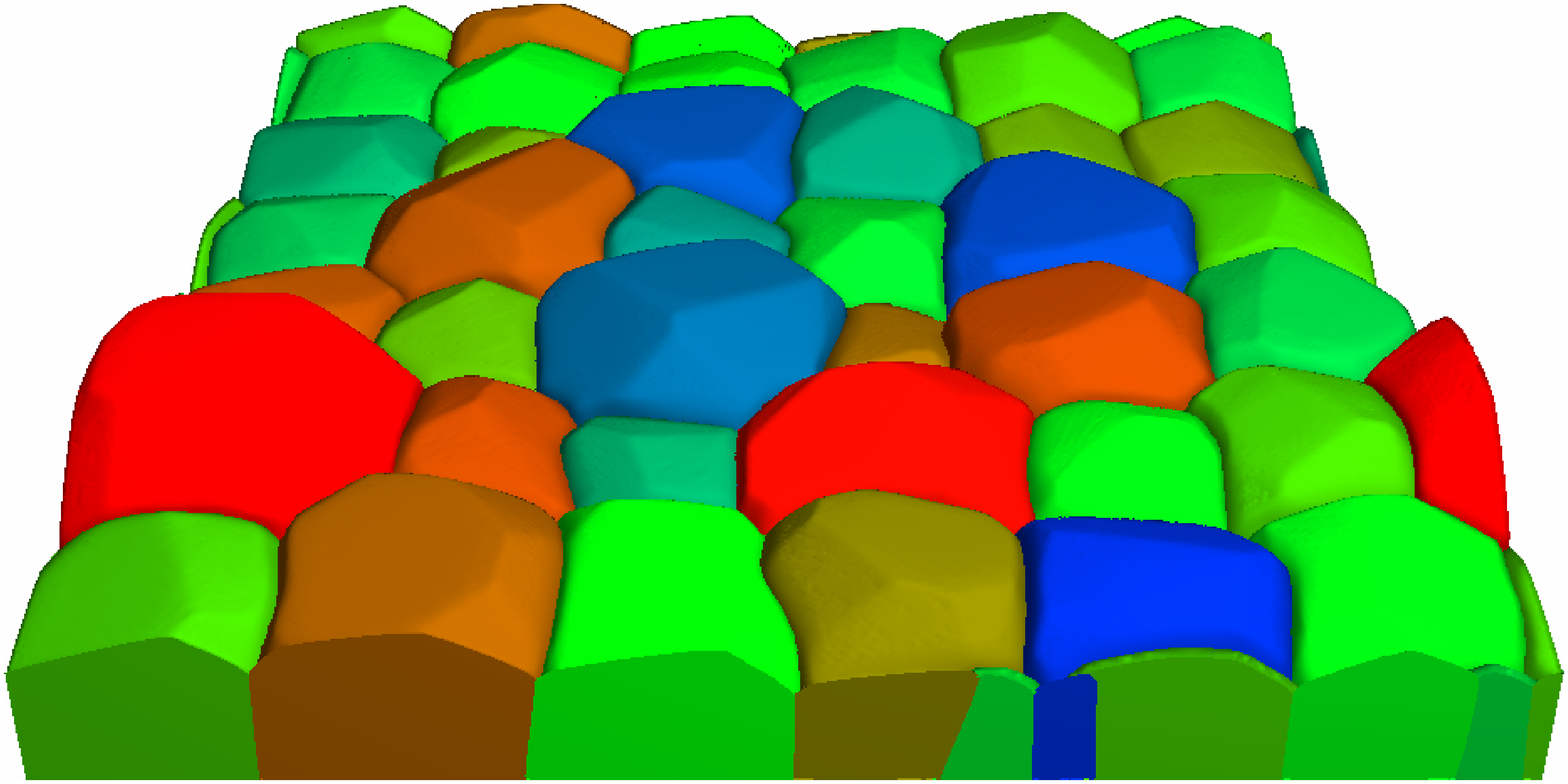}\label{fig:3D_alum_free_time_int}}
\put(-5,30){\includegraphics[scale=0.06]{b.eps}}\\
\subfigure{\includegraphics[scale=0.25]{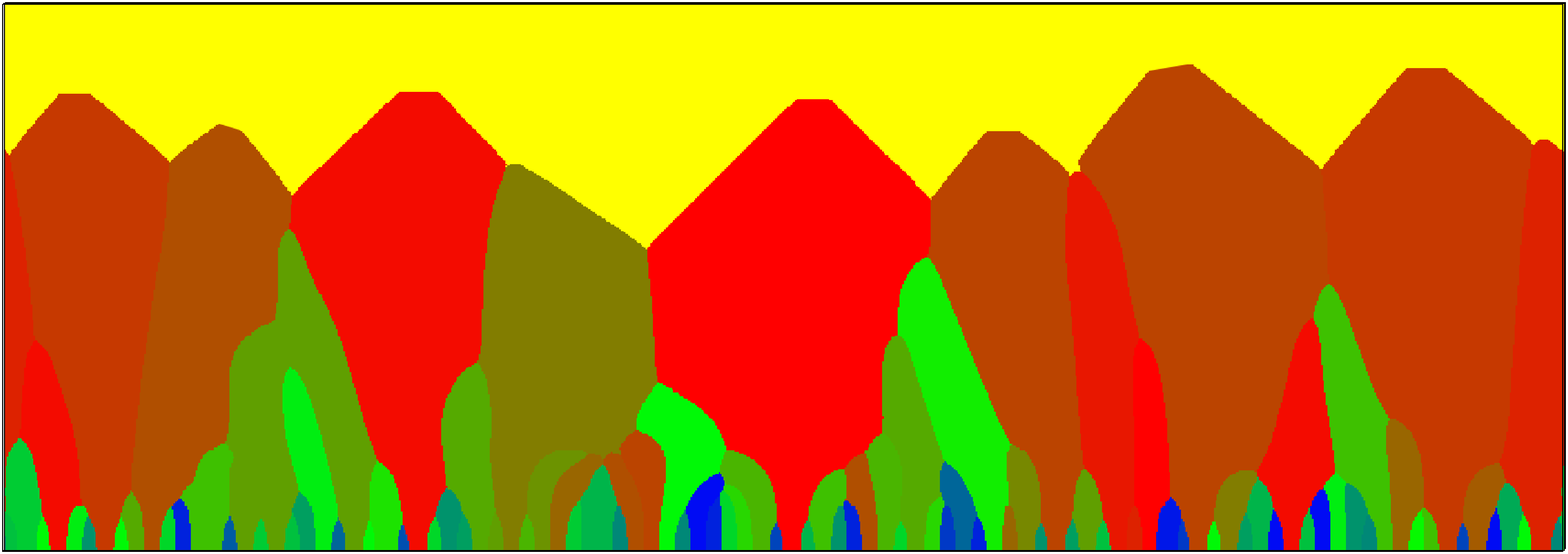}\label{fig:2D_alum_free}}
\put(-4,48){\includegraphics[scale=0.06]{c.eps}}\\
\subfigure{\includegraphics[scale=0.22]{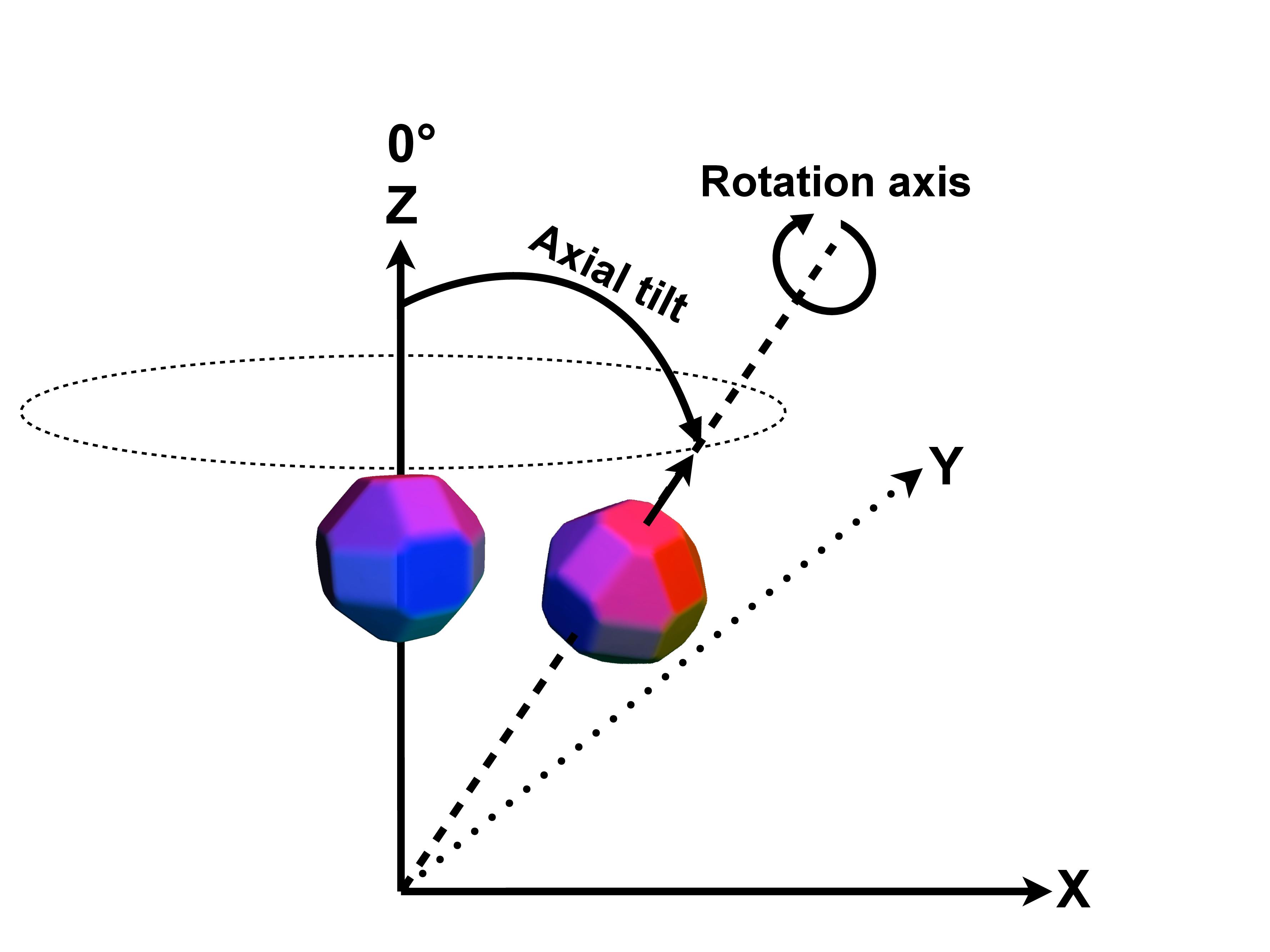}\label{fig:define_3D_orient_alum}}
\put(-15,48){\includegraphics[scale=0.06]{d.eps}}
\subfigure{\includegraphics[scale=0.22]{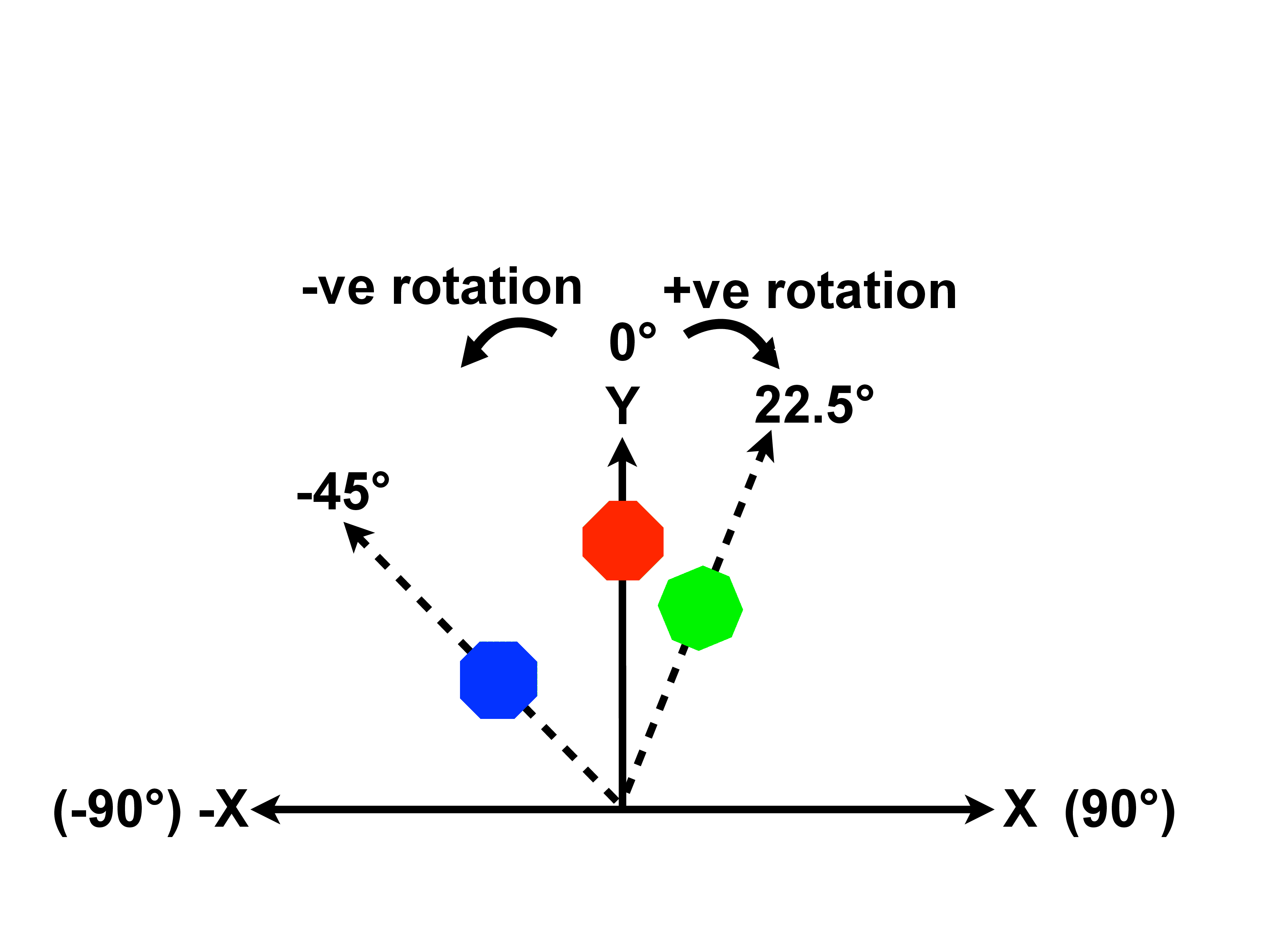}\label{fig:orient_def_2D_alum}}
\put(-10,48){\includegraphics[scale=0.06]{e.eps}}\\
\subfigure{\includegraphics[scale=0.15]{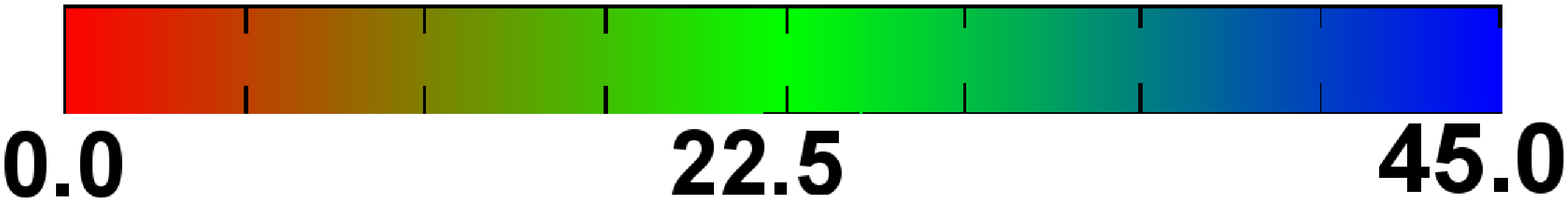}}
\caption{\textbf{Fig. 4 }Free growth of alum crystals with random orientations.
(a) Bottom layer of the 3D phase-field
simulation domain showing the spherical 
crystal nuclei and their assigned orientations
in different colors. 
(b) intermediate stage of alum crystals growing in liquid. 
The facets \{111\}, \{110\} and \{100\} can 
be distinctly identified, (c) 2D 
simulation showing the final 
stage of growth competition. Favorably
oriented crystals (reddish) outgrow their neighbors.
{(d) Orientation definitions for 3D and 
(e) 2D alum growth simulations. The colored crystal facets
in Fig. \ref{fig:define_3D_orient_alum} is provided for a 
better visualization and is \textit{not} related to
colorbar shown below.}}
\label{alum_free_grwth}
\end{figure}

Next, we perform a 3D simulation of alum crystal growth
such that crystals also develop $\left\lbrace110\right\rbrace$
and $\left\lbrace100\right\rbrace$ facets in addition to
$\left\lbrace111\right\rbrace$ as shown in Fig. \ref{alum_free_grwth}.
Hemispherical crystal seeds are uniformly embedded
at the bottom of the domain as shown in 
Fig. \ref{fig:3D_alum_free_time_zero}, 
and every crystal is assigned
a random orientation with respect to the normal direction 
of the growth plane. {The growth competition} results in the
consumption of poorly oriented crystals (greenish and bluish in color) 
and in survival of favorably oriented ones (reddish)
as shown in Fig. \ref{fig:2D_alum_free}. 
\newpage
\clearpage
\section{Crack-seal growth}
\label{sec:crack-seal}
In the following section, we use the phase-field method described
in section \ref{sec:phase_field}, to study the effect of various
crack parameters namely roughness, opening velocity and
trajectory on the crack-seal microstructure. Further, we also 
examine if the number of competing crystal nuclei causes
a variation in the final microstructure.

Numerical simulations are carried out with periodic 
asperities on the crack wall and uniformly distributed
 crystal nuclei placed at the top of the domain. 
 The crystals grow downwards in the direction of
crack opening. All the crystals are assumed to posses a cubic
anisotropy and {are} assigned 
{random} orientation in the range of
$-45\degree$ to $45\degree$ with respect to {the} vertical direction.

\subsection{Effect of crack wall roughness}
\label{subsec:roughness}
\begin{figure}[!htbp]
\centering
\subfigure{\includegraphics[scale=0.4]{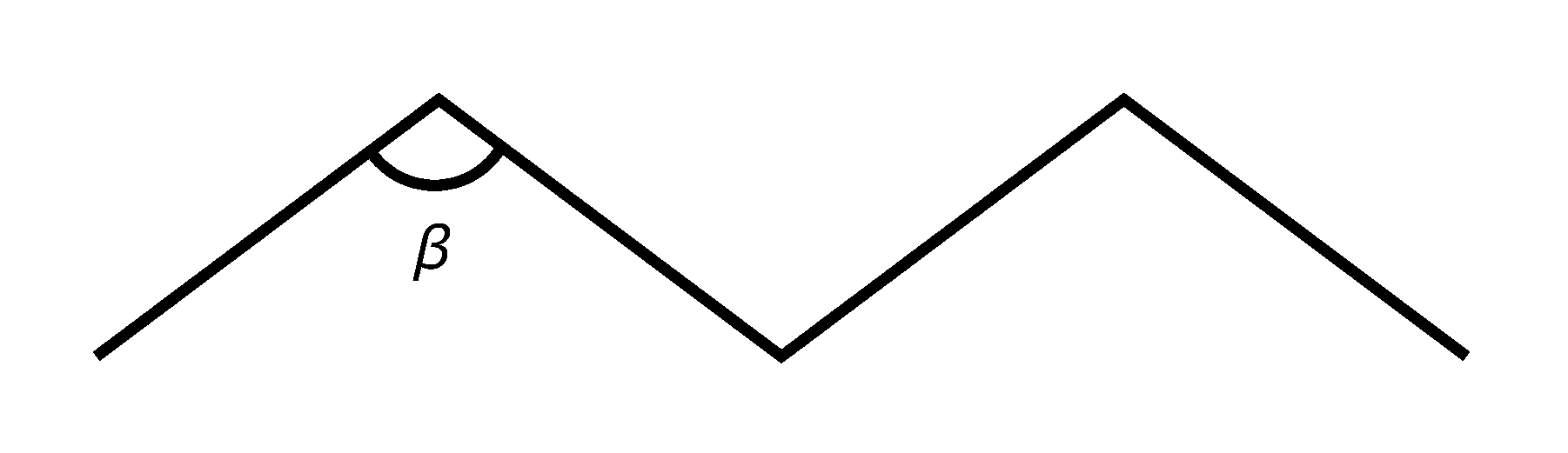}\label{fig:define_beta}}
\put(-71,0){\includegraphics[scale=0.06]{a.eps}}\\
\subfigure{\includegraphics[width=0.3\textwidth,height=0.3\textwidth]{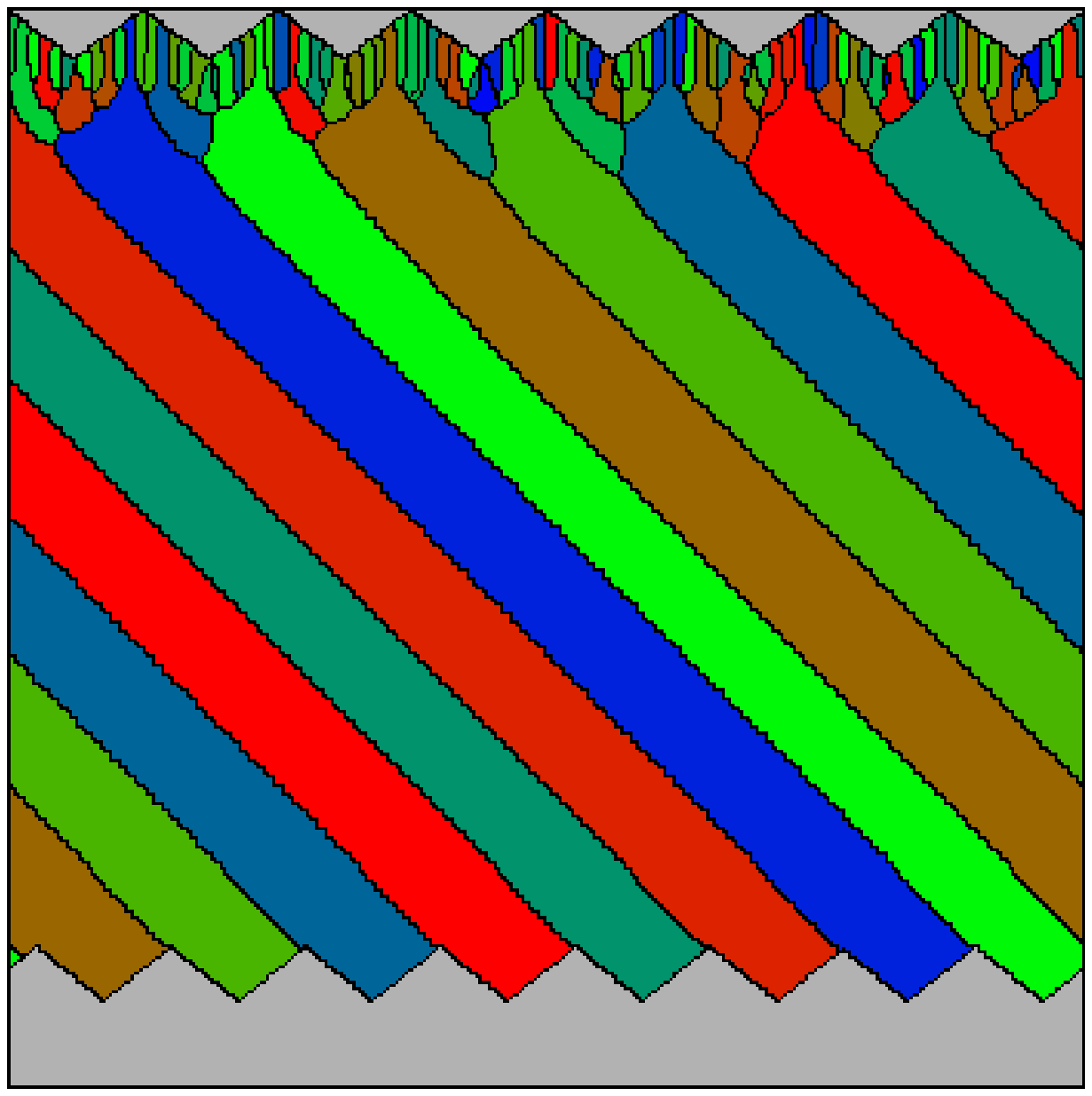}\label{fig:rough4}}
\put(-43,2){\includegraphics[scale=0.06]{b.eps}}\qquad\qquad
\subfigure{\includegraphics[width=0.3\textwidth,height=0.3\textwidth]{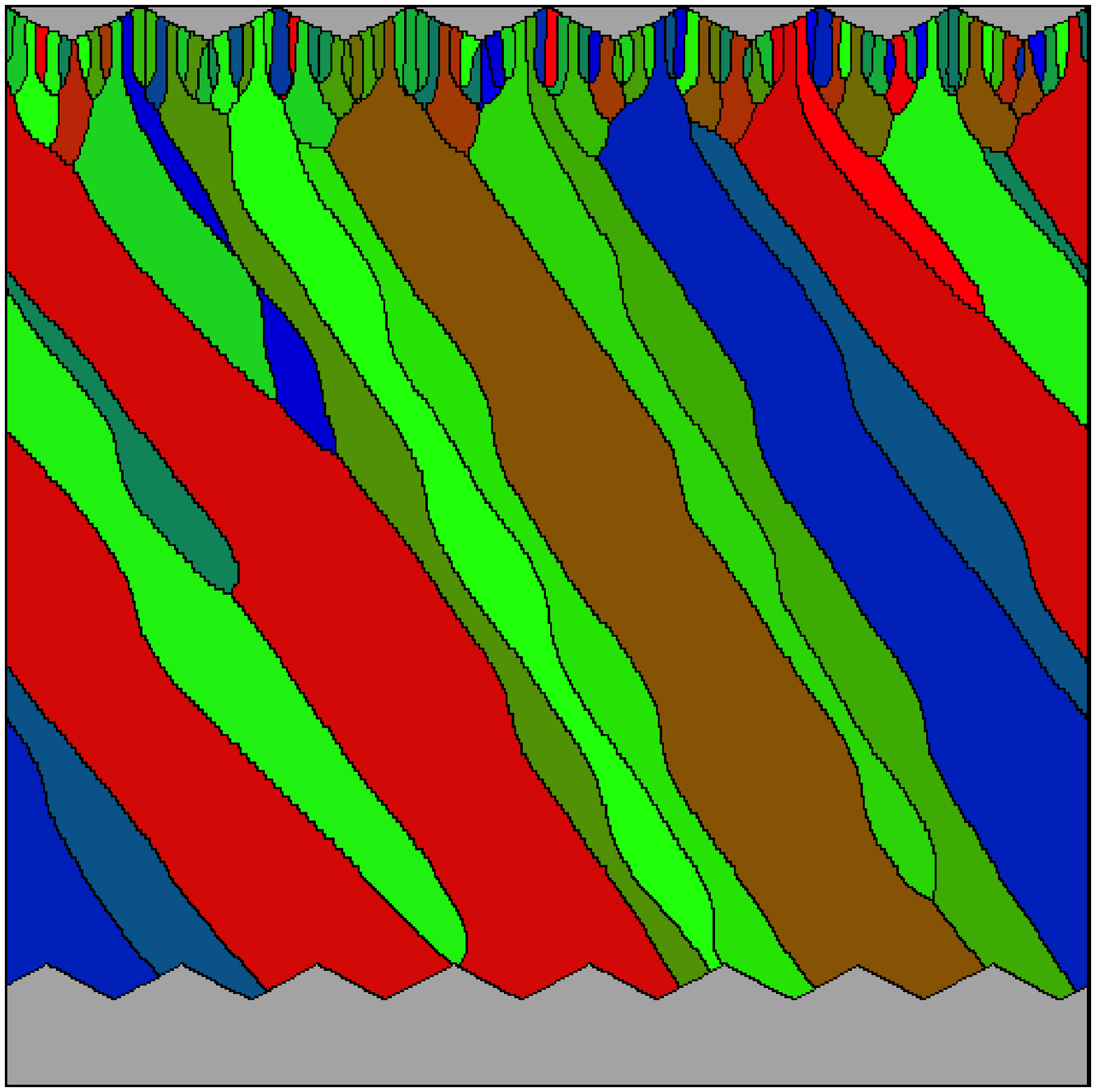}\label{fig:rough6}}
\put(-43,2){\includegraphics[scale=0.06]{c.eps}}\\
\subfigure{\includegraphics[width=0.3\textwidth,height=0.3\textwidth]{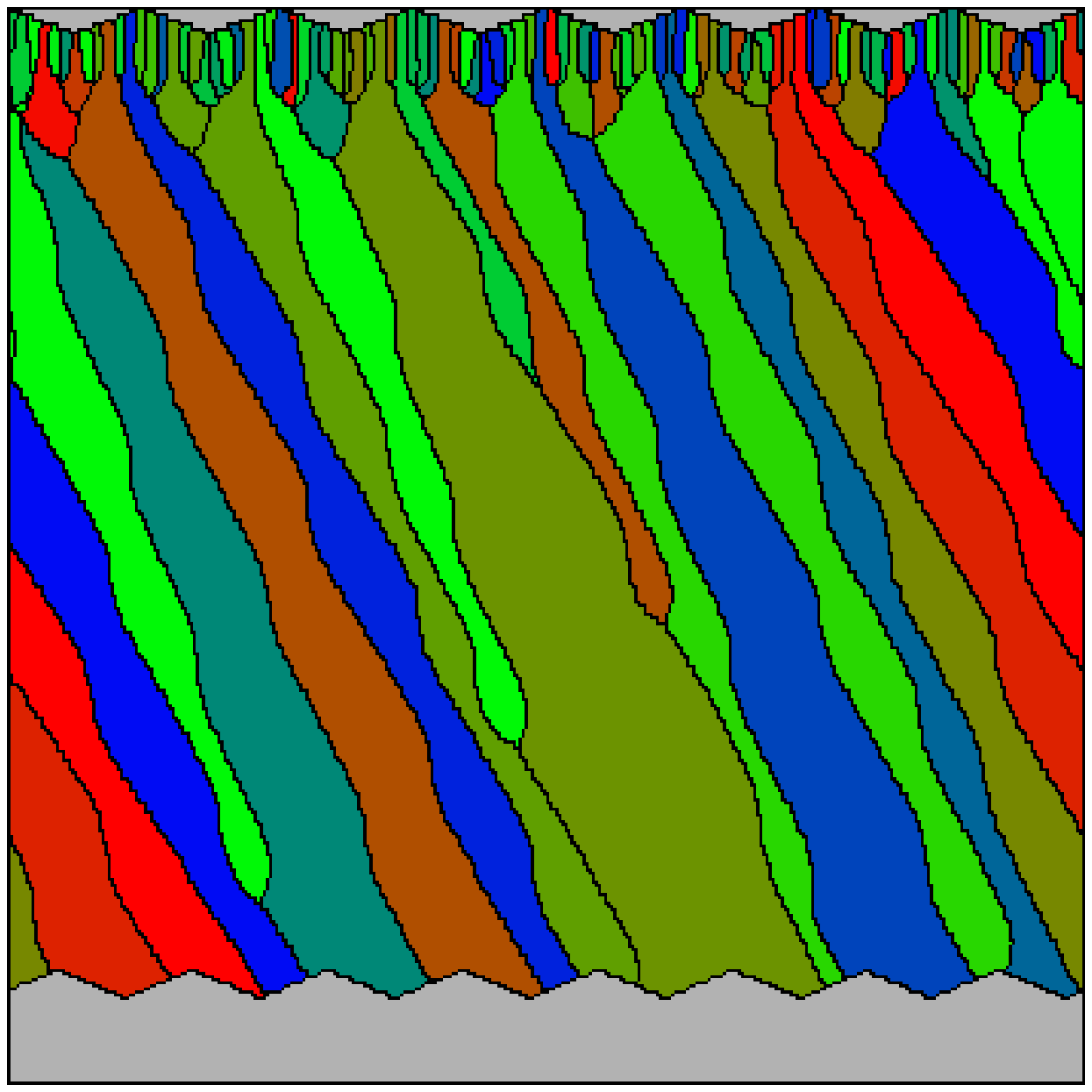}\label{fig:rough8}}
\put(-43,2){\includegraphics[scale=0.06]{d.eps}}\qquad\qquad
\subfigure{\includegraphics[width=0.3\textwidth,height=0.3\textwidth]{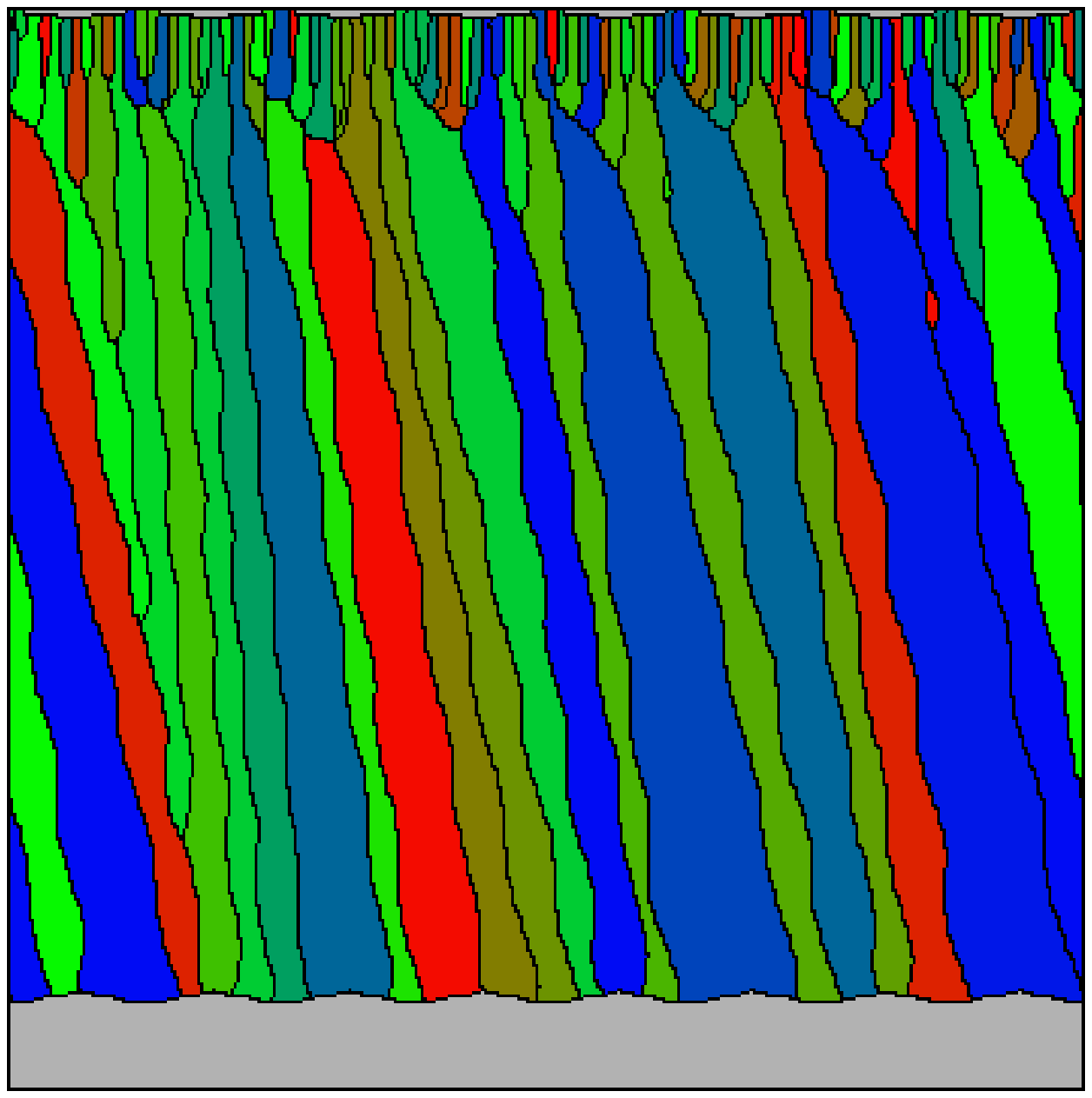}\label{fig:rough16}}
\put(-43,2){\includegraphics[scale=0.06]{e.eps}}\\
\subfigure{\includegraphics[scale=0.22]{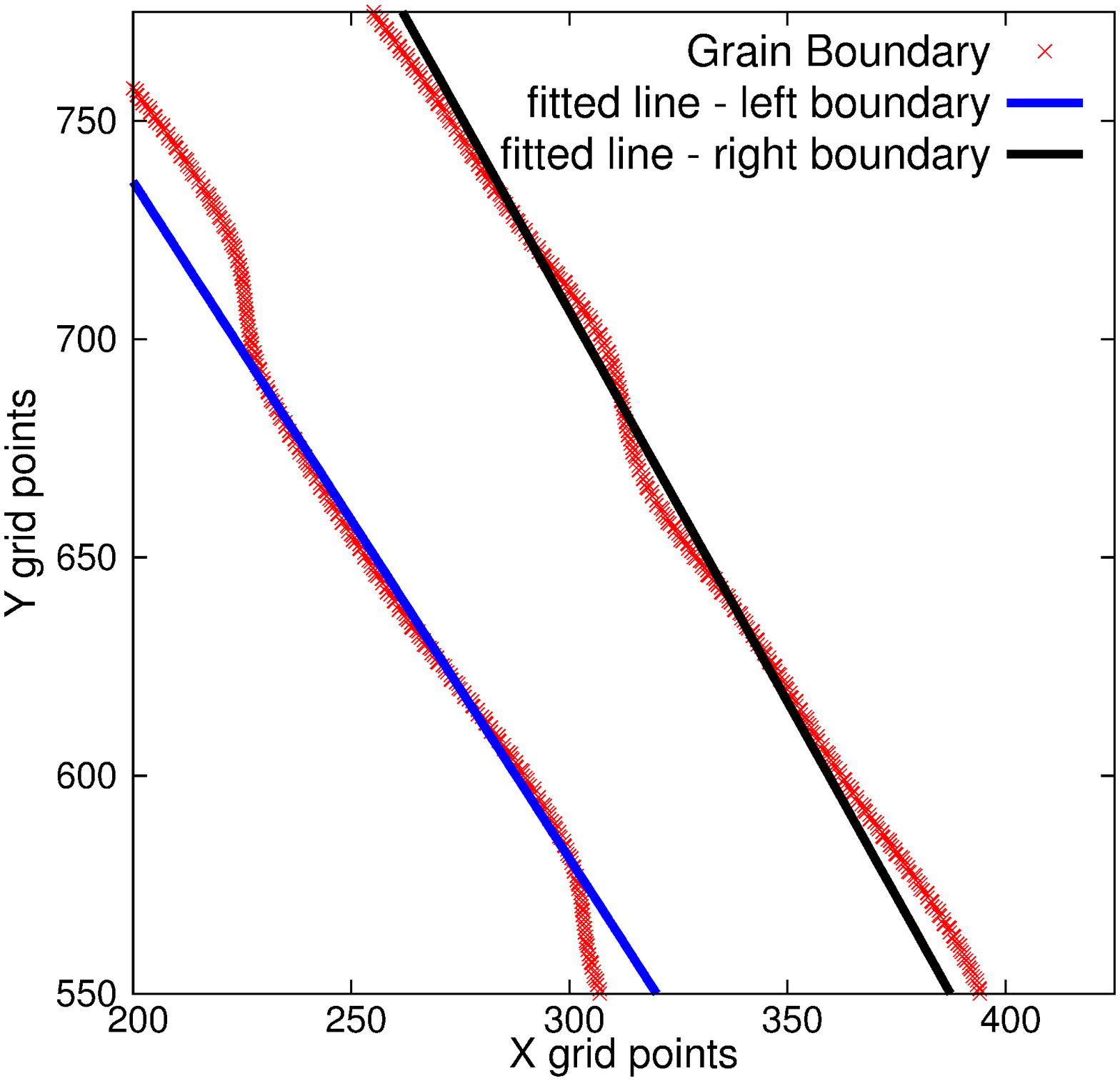}\label{fig:ft_r6}}
\put(-39,5){\includegraphics[scale=0.06]{f.eps}}\qquad
\subfigure{\includegraphics[scale=0.5]{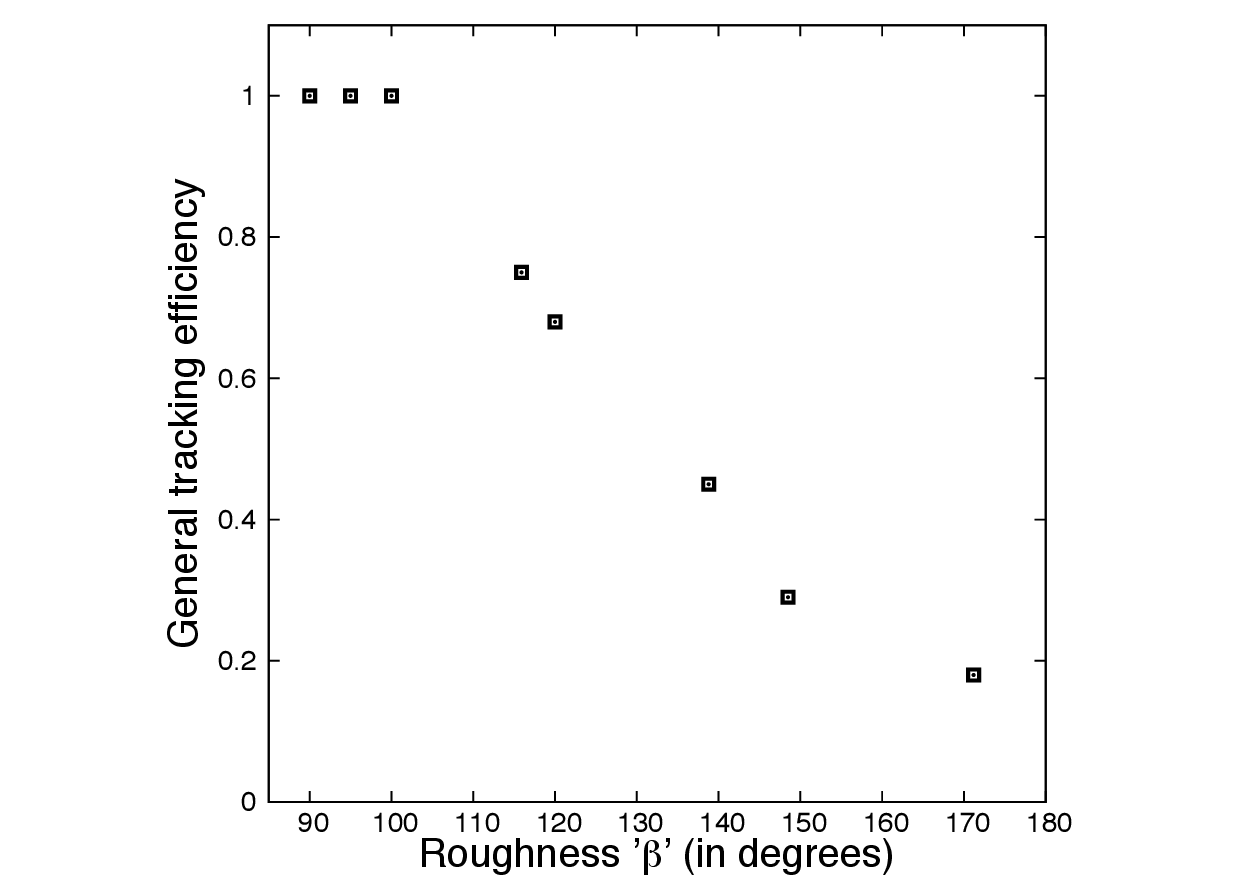}\label{fig:rough_plot}}
\put(-40,6){\includegraphics[scale=0.06]{g.eps}}\\
\subfigure{\includegraphics[scale=0.15]{colormap.eps}}
\caption{\textbf{Fig. 5 }Effect of crack-wall roughness on the grain boundary
tracking behavior. The direction of the crack opening is $\theta_{open}$ =
45\degree to the vertical line. {(a) The roughness of 
the crack is governed by angle $\beta$.} 
The simulated grain structures refer to
wall roughness of (b) $100\degree$,
(c) $120\degree$,
(d) $139\degree$ and
(e) $171\degree$
(f) straight lines are fitted along the grain boundaries 
(g) plot of tracking efficiency as a function of crack roughness.
{The nature of the plot obtained from
2D phase-field simulations is consistent
with the results of \citet{Urai:1991ec}.}
The colors refer to different
crystal orientation with respect 
to the vertical direction (see colormap).}
\label{roughness}
\end{figure}

The crack surface is assumed to be periodic and 
the degree of roughness is controlled by the 
vertical angle $\beta$ 
which is varied between
90$\degree$ and 180$\degree$ while keeping the 
wavelength constant as shown in \ref{fig:define_beta}.
An angle of $\beta = 180\degree$ corresponds to  
a smooth surface. The crack is opened 
periodically at an angle $\theta_{open}$
with respect to the vertical direction. 
In order to study the influence of crack roughness, 
we fix the value of $\theta_{open}$ to be $45\degree$. 
The velocity of crack opening is selected in such a way that 
{a} complete sealing of {the} crack occurs before every 
crack opening event. 

At a higher crack roughness $\left(\beta = 90\degree\right)$ ,
it is observed in Fig. \ref{fig:rough4} that the 
crystals track the crack opening trajectory. 
This is also characterized by the crack peaks
{(referred as 'Grain Boundary Attractor' by 
\citet{Hilgers:2001rw} and \citet{Nollet:2005qo})} 
attracting the grain boundaries and forcing the 
crystals to grow in a fibrous morphology.
As the roughness angle $\beta$ is successively increased
in Figs. \ref{fig:rough6} - \ref{fig:rough16}, 
the tracking behavior decreases and crystals develop 
curved/oscillating grain boundaries. The period
of oscillation reduces with decreasing roughness.

To quantify the grain boundary tracking behavior, 
we plot the contour of the crystal grain boundaries
{(corresponding to one of the tracking grain boundary)} 
and fit straight
lines to the left and right crystal boundaries as
illustrated in Fig. \ref{fig:ft_r6}. The overall tracking 
inclination $\theta_{track}$ is defined as the mean value 
of the slopes of the two lines and is used to 
elaborate the {general tracking efficiency as: 
\begin{eqnarray}
\mathrm{GTE} = \dfrac{\theta_{track}}{\theta_{open}}
 \label{eq:track_eff}
\end{eqnarray}
The general tracking efficiency {(GTE)} is plotted as
a function of roughness angle $\beta$ 
in Fig. \ref{fig:rough_plot}. The resulting
trend indicates that the general tracking efficiency 
decreases for lower crack-surface
roughness, beyond a certain $\beta$ value. 
It is to be noted that the purpose of the prefix 'general' in 
the term 'general tracking efficiency' (GTE) is to highlight the
difference with the definition of 'tracking efficiency'
by \citet{Urai:1991ec}. As evident from the simulations
and plotted results in Fig. \ref{roughness}, GTE converges
 to the definition of 'tracking efficiency' 
when crystal growth is isotropic without any oscillations.
Additionally, the definition of GTE is more general as it remains valid 
even when characterizing the tracking behavior of oscillating grain
boundaries.}
\newpage
\clearpage
\subsection{Effect of crack opening rate}
\label{subsec:open_v}
\begin{figure}[!htbp]
\centering
\subfigure{\includegraphics[width=0.31\textwidth,height=0.31\textwidth]{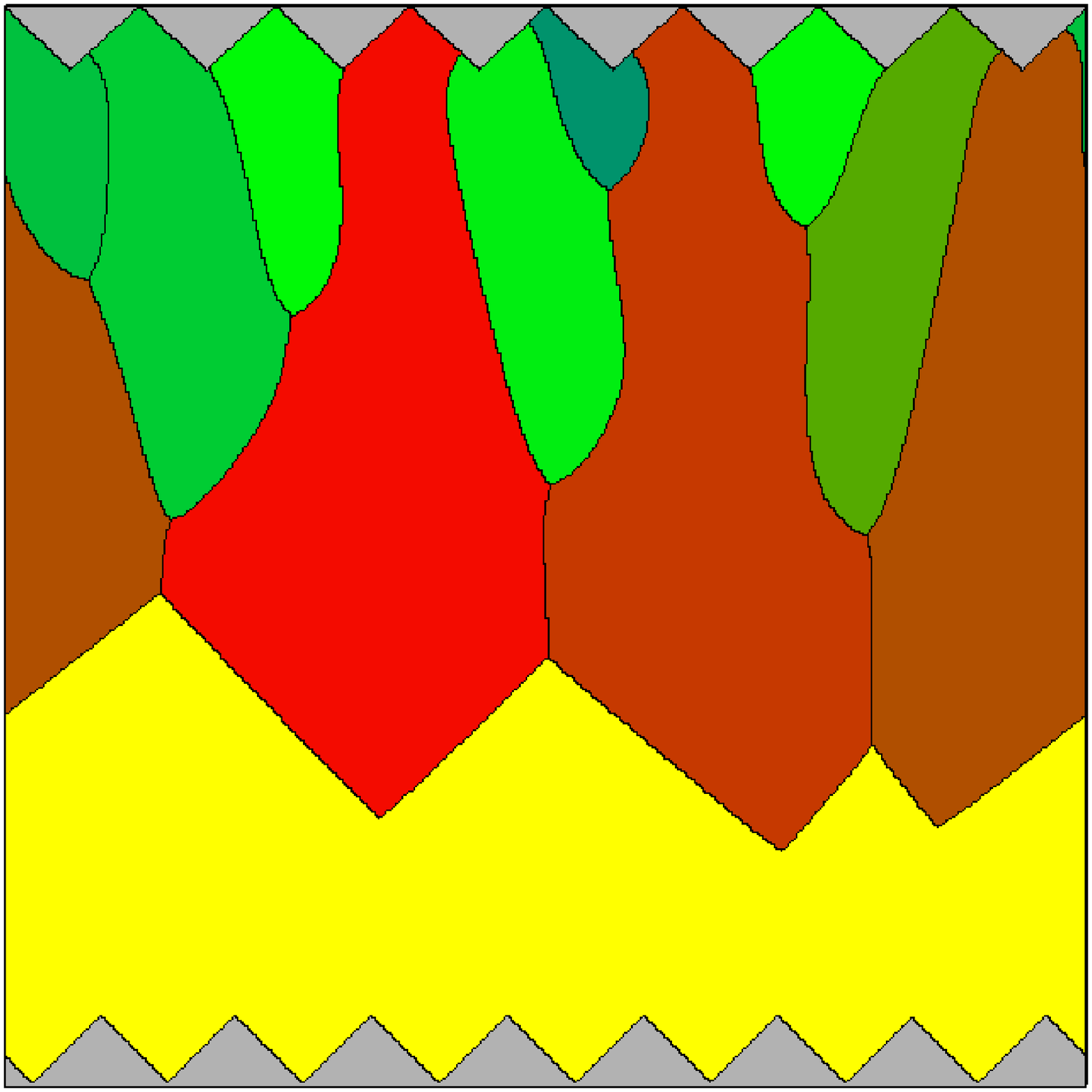}\label{fig:open_v_200}}
\put(-45,41.5){\includegraphics[scale=0.06]{a.eps}}\quad
\subfigure{\includegraphics[width=0.31\textwidth,height=0.31\textwidth]{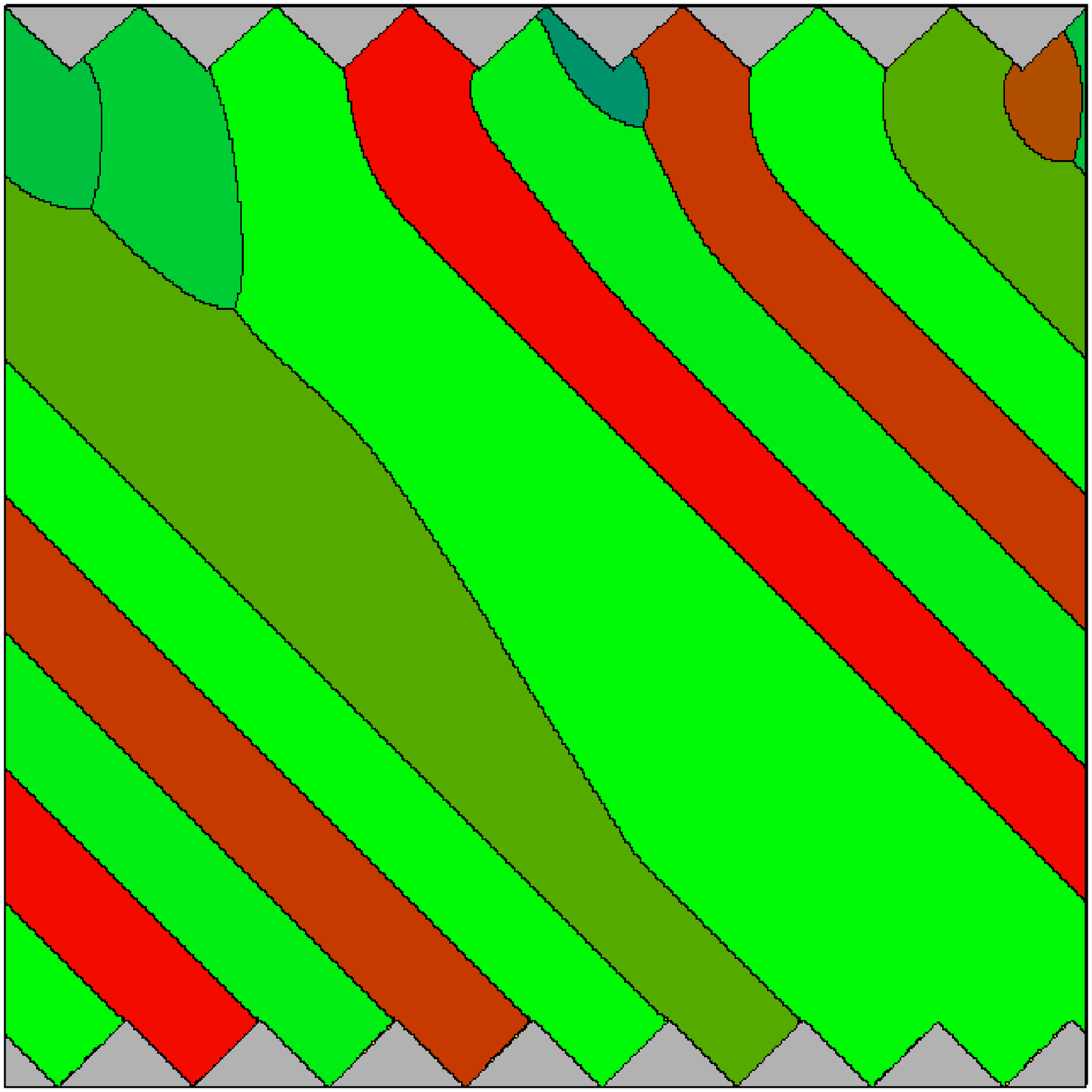}\label{fig:open_v_500}}
\put(-45,41.5){\includegraphics[scale=0.06]{b.eps}}\quad
\subfigure{\includegraphics[width=0.31\textwidth,height=0.31\textwidth]{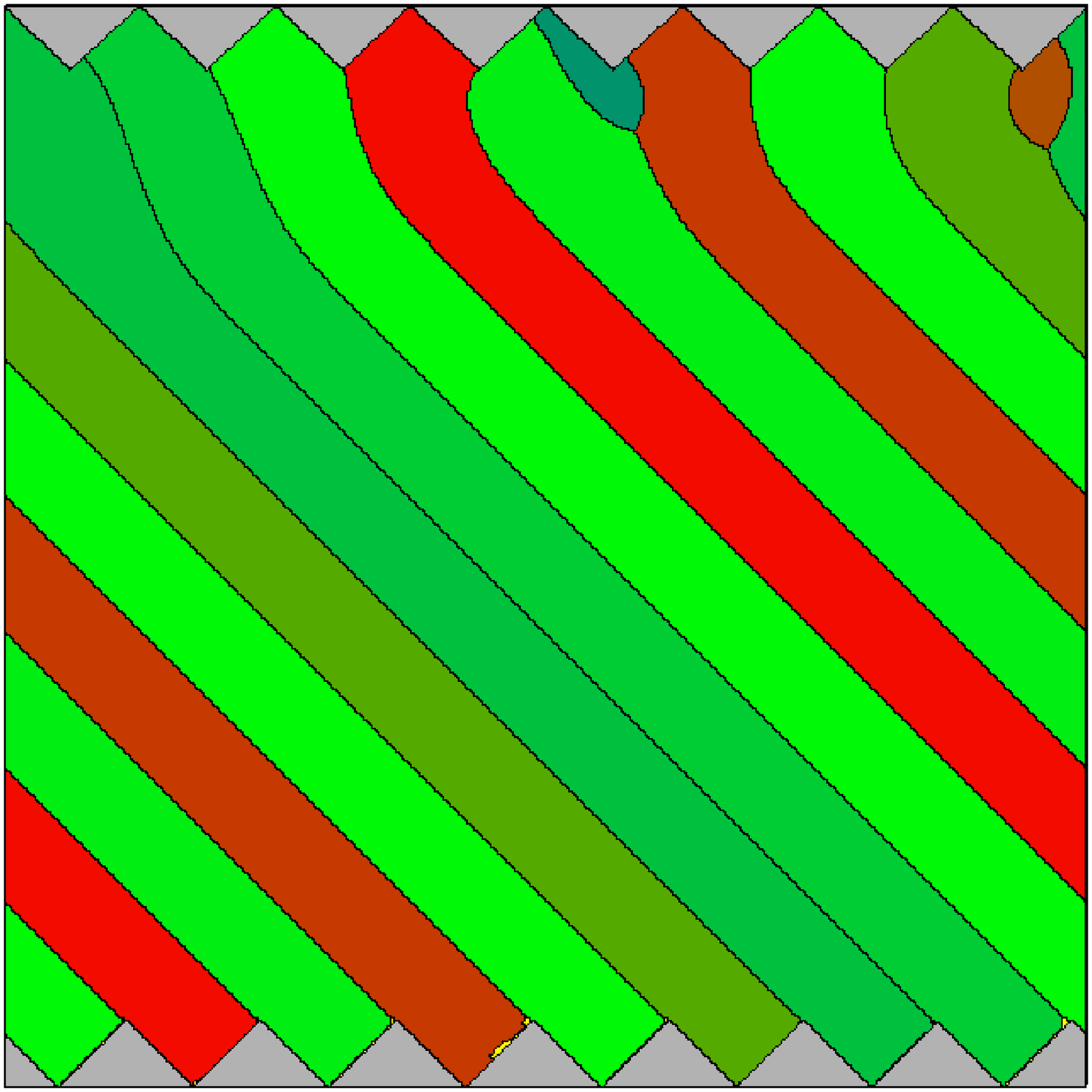}\label{fig:open_v_700}}
\put(-45,41.5){\includegraphics[scale=0.06]{c.eps}}\\
\subfigure{\includegraphics[scale=0.15]{colormap.eps}}
\caption{\textbf{Fig. 6 }Effect of crack-wall opening velocity on crystal growth
morphology and final microstructure. The growth is elongate-blocky
at higher crack opening rate in contrast to fibrous when the
crack opening rate is slower. Further, within the complete crack seal
regime, a morphological transition from curved to straight boundaries 
is observed. The direction of the crack opening
is 45\degree with respect to vertical. The crack opens after every {%
				(a) 200*$\Delta t$, %
				(b) 500*$\Delta t$ and %
     			(c) 700*$\Delta t$ seconds where $\Delta t=0.12$ second refers to time-step width.}
     			The colors refer to different
     			crystal orientation with respect 
     			to vertical (see colormap).}\label{open_vel}
\end{figure}

To study the effect of crack opening velocity on vein growth,
a simulation setup similar to section \ref{subsec:roughness} 
is considered with 10 crystal nuclei.  
The roughness angle $\beta=100\degree$ is selected 
to ensure a high level of tracking efficiency,
if complete sealing occurs before every opening event.
The crack opening angle is kept constant $\left(\theta_{open} = 45\degree\right)$
and rate of opening is varied as shown in Fig. \ref{open_vel}.
It is observed in Fig. \ref{fig:open_v_200} that at 
higher crack opening rate (as compared
to crystal growth rate), crystals lose contact with the crack surface.
This leads to the formation of {an} elongate-blocky growth morphology,
since crystals now grow more or less anisotropically
and follow the orientation rule as observed for 
free-growth in Fig. \ref{orient_map}.
At lower crack opening rate as in Figs. 
\ref{fig:open_v_500} and \ref{fig:open_v_700}, 
crystals grow isotropically
in fibrous morphology, since the facet formation is
suppressed. Further, within the complete crack-seal regime,
at a higher opening velocity, the crystal boundaries have a
tendency to curve in contrast to the case when 
opening velocity is smaller. The simulation results for
the influence of crack opening velocity on crystal growth
morphology are presented in Fig. \ref{open_vel}.
A definitive change in the crystal growth 
morphology is observed if crack opening rate
is varied.
\newpage
\clearpage
\subsection{Effect of crack opening trajectory}
\label{subsec:Crack_traj}
\begin{figure}[!htbp]
\centering
\subfigure{\includegraphics[width=0.31\textwidth,height=0.31\textwidth]{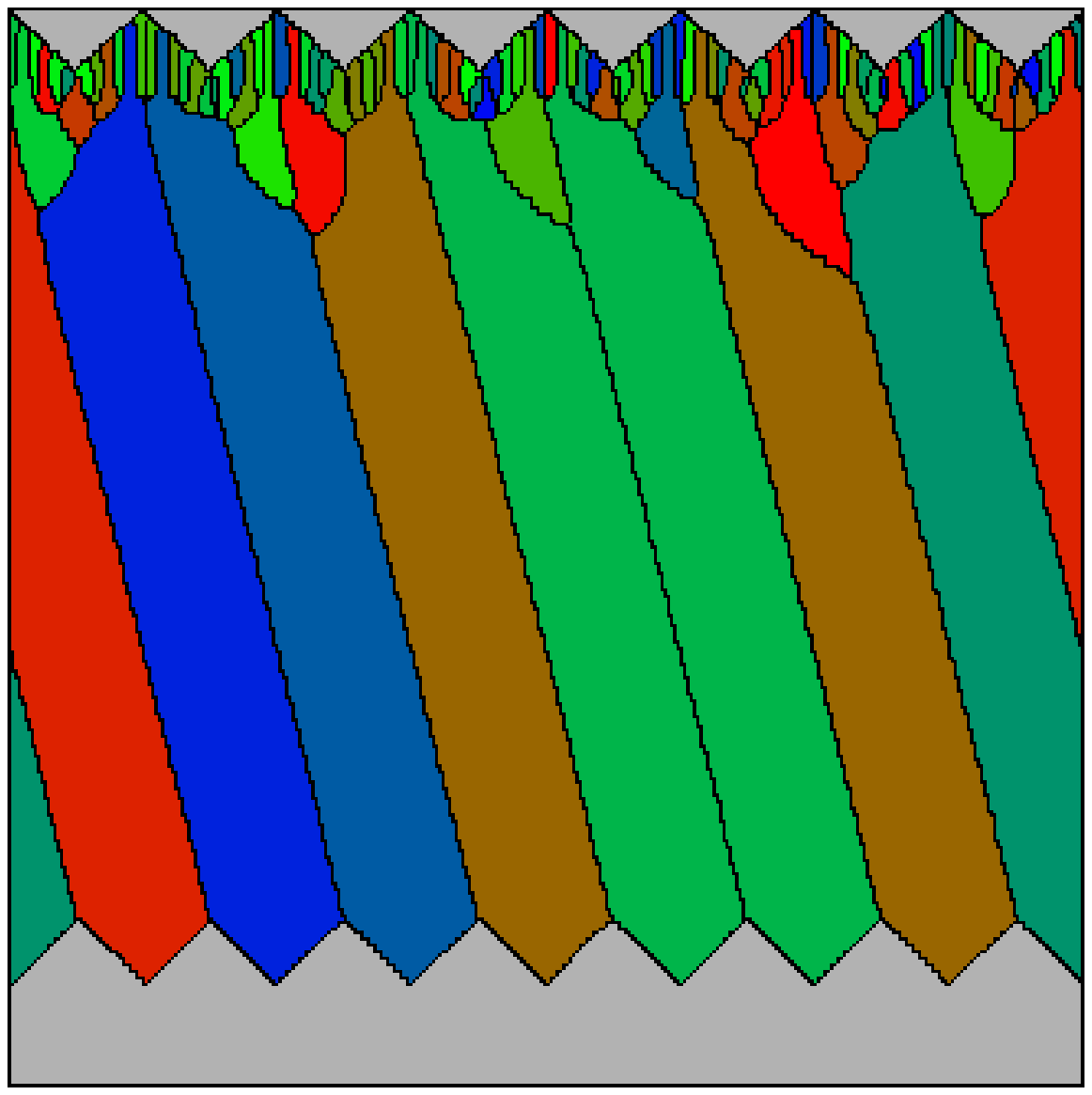}\label{fig:theta_14p0}}
\put(-45,41.5){\includegraphics[scale=0.06]{a.eps}}\quad
\subfigure{\includegraphics[width=0.31\textwidth,height=0.31\textwidth]{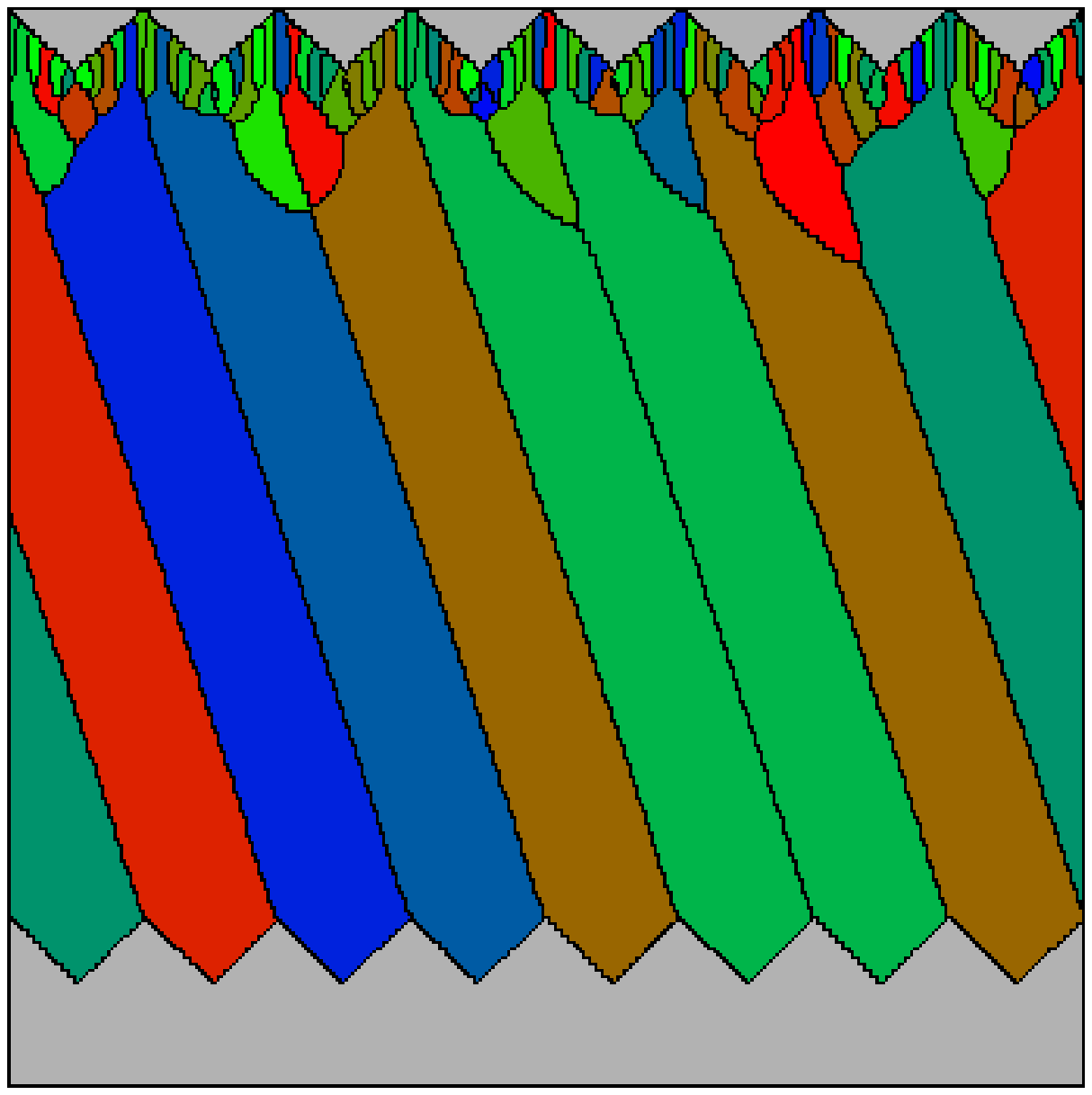}\label{fig:theta_18p4}}
\put(-45,41.5){\includegraphics[scale=0.06]{b.eps}}\quad
\subfigure{\includegraphics[width=0.31\textwidth,height=0.31\textwidth]{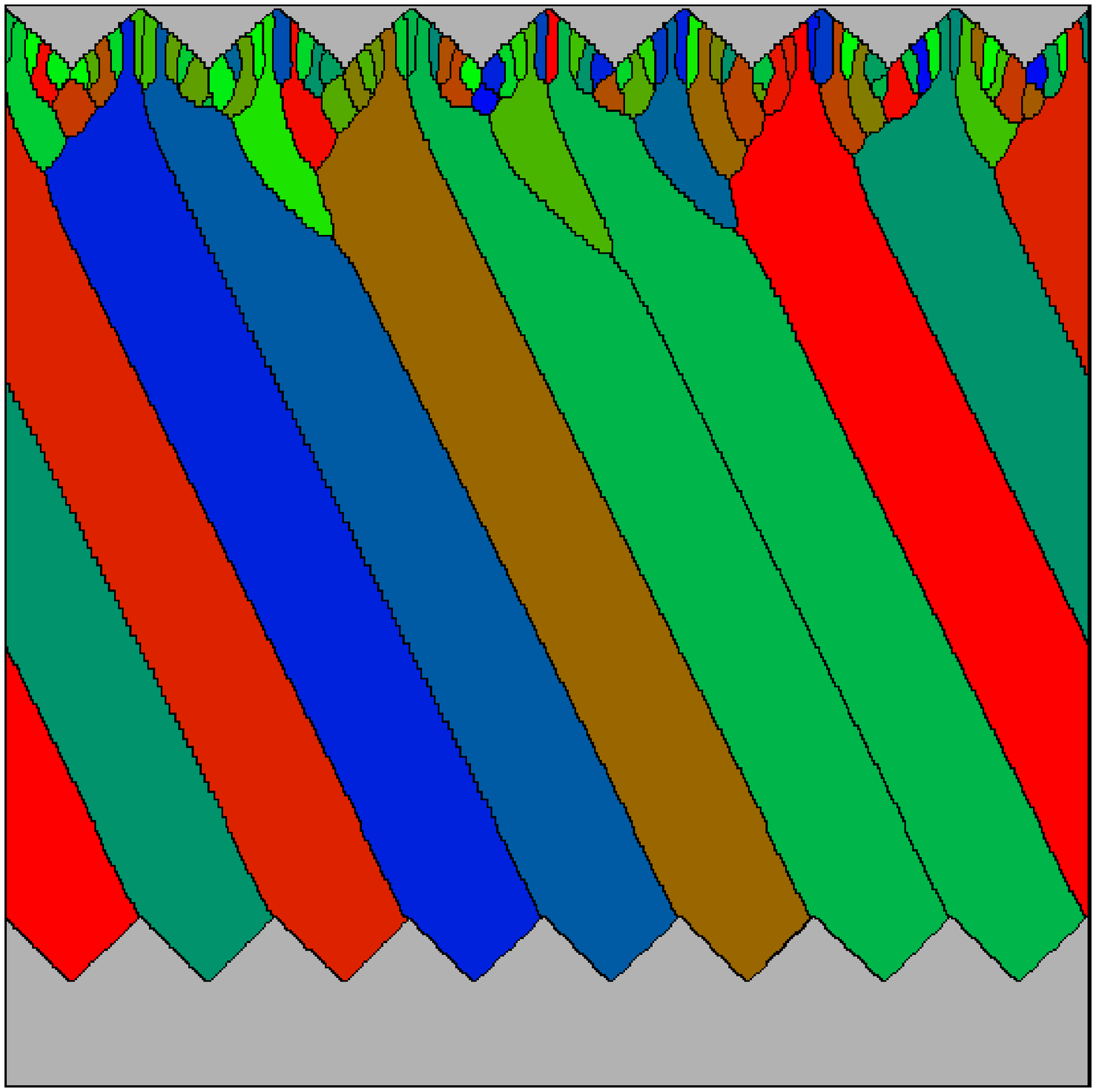}\label{fig:theta_26p5}}
\put(-45,41.5){\includegraphics[scale=0.06]{c.eps}}\\
\subfigure{\includegraphics[width=0.31\textwidth,height=0.31\textwidth]{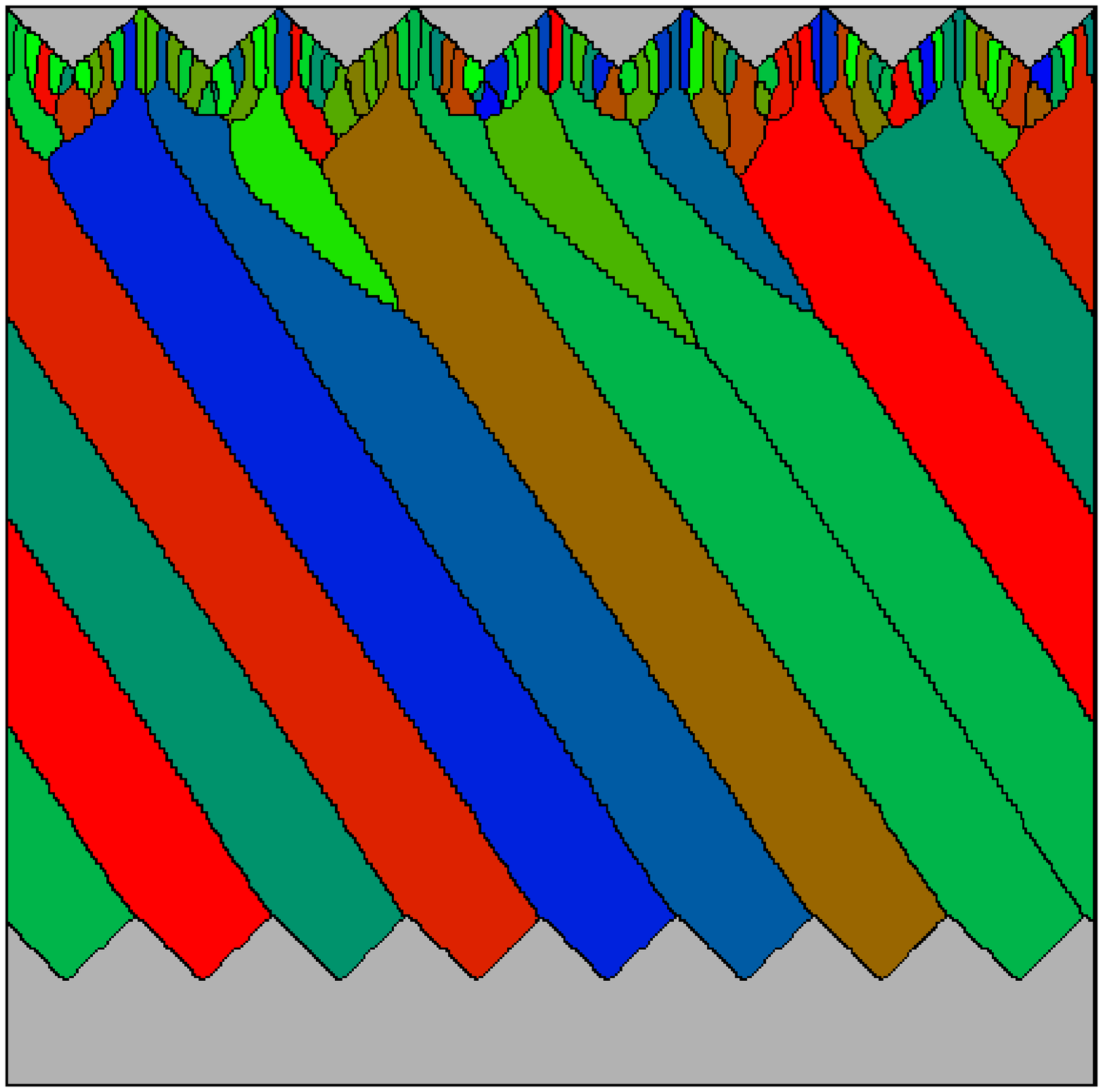}\label{fig:theta_33p6}}
\put(-45,41.5){\includegraphics[scale=0.06]{d.eps}}\quad
\subfigure{\includegraphics[width=0.31\textwidth,height=0.31\textwidth]{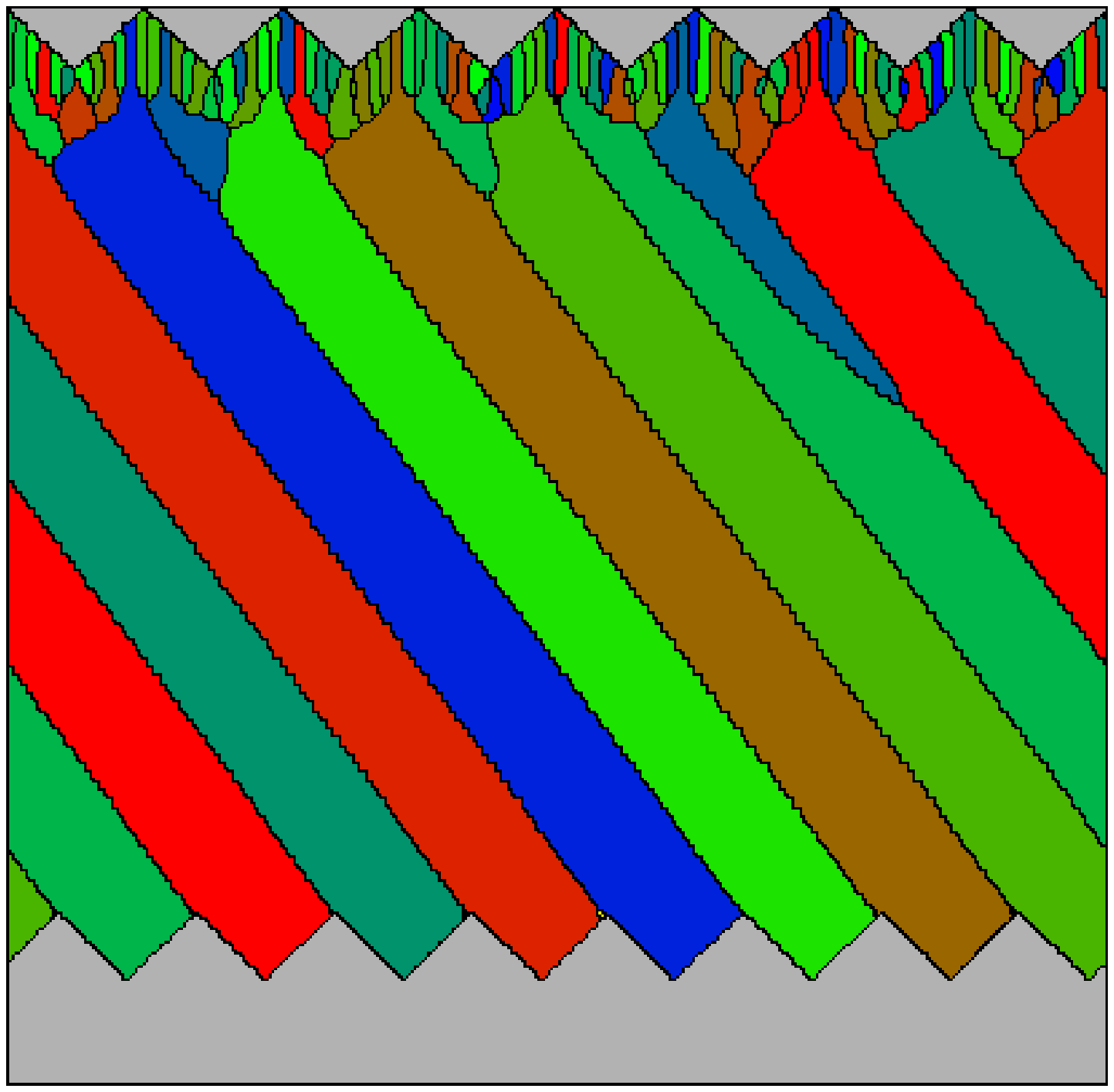}\label{fig:theta_36p8}}
\put(-45,41.5){\includegraphics[scale=0.06]{e.eps}}\quad
\subfigure{\includegraphics[scale=0.15]{colormap.eps}}
\caption{\textbf{Fig. 7 }Effect of crack opening trajectory on crack-seal
mictrostructure. The angles of opening $\theta_{open}$ are %
		   (a) 14.0\degree,%
		   (b) 18.4\degree,%
	       (c) 26.5\degree,%
	       (d) 33.6\degree and%
	       (e) 36.8\degree %
with respect to vertical in anti-clockwise direction.
The crystals track the crack opening 
irrespective of the trajectory, if opening
increments are small and crack surface
is sufficiently rough. The colors refer to different
crystal orientation with respect 
to the vertical direction (see colormap).}\label{open_angle}
\end{figure}

\begin{figure}[!htbp]
\centering
\subfigure{\includegraphics[width=0.31\textwidth,height=0.31\textwidth]{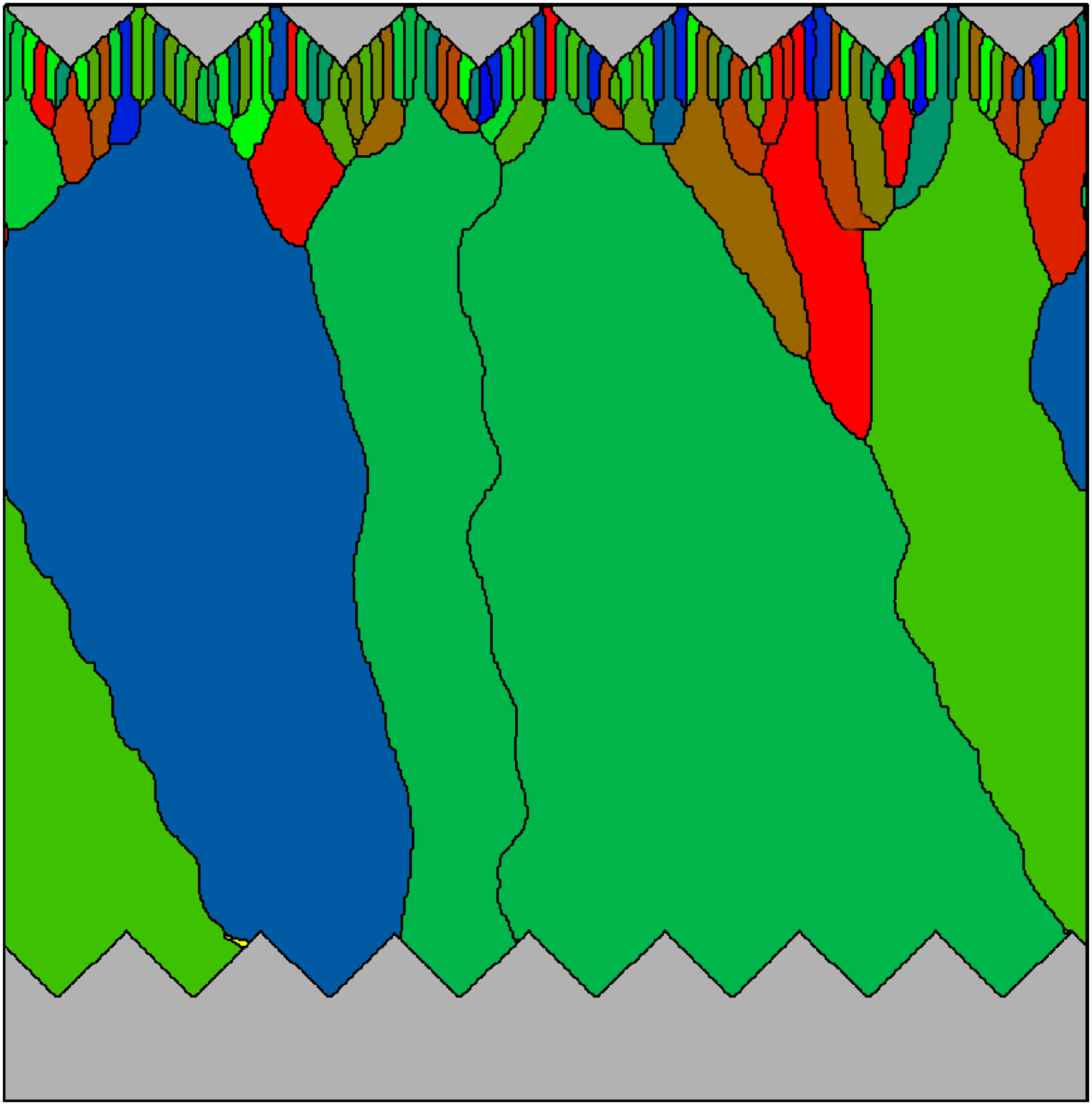}\label{fig:path_20_2020}
\put(-45,41.5){\includegraphics[scale=0.06]{a.eps}}
\put(10,0){\includegraphics[height=0.2\textwidth]{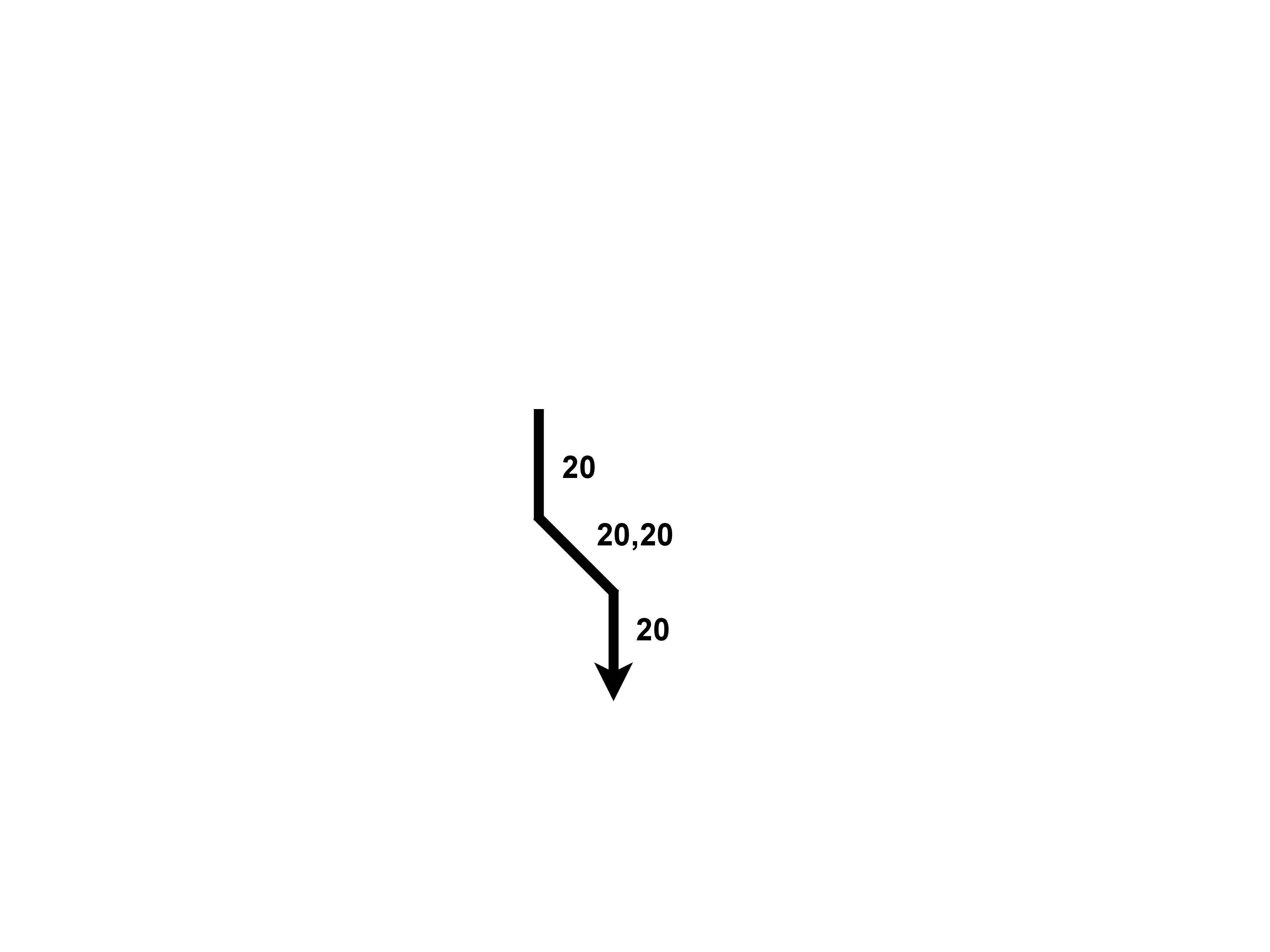}\label{fig:open_20}}}\\
\subfigure{\includegraphics[width=0.31\textwidth,height=0.31\textwidth]{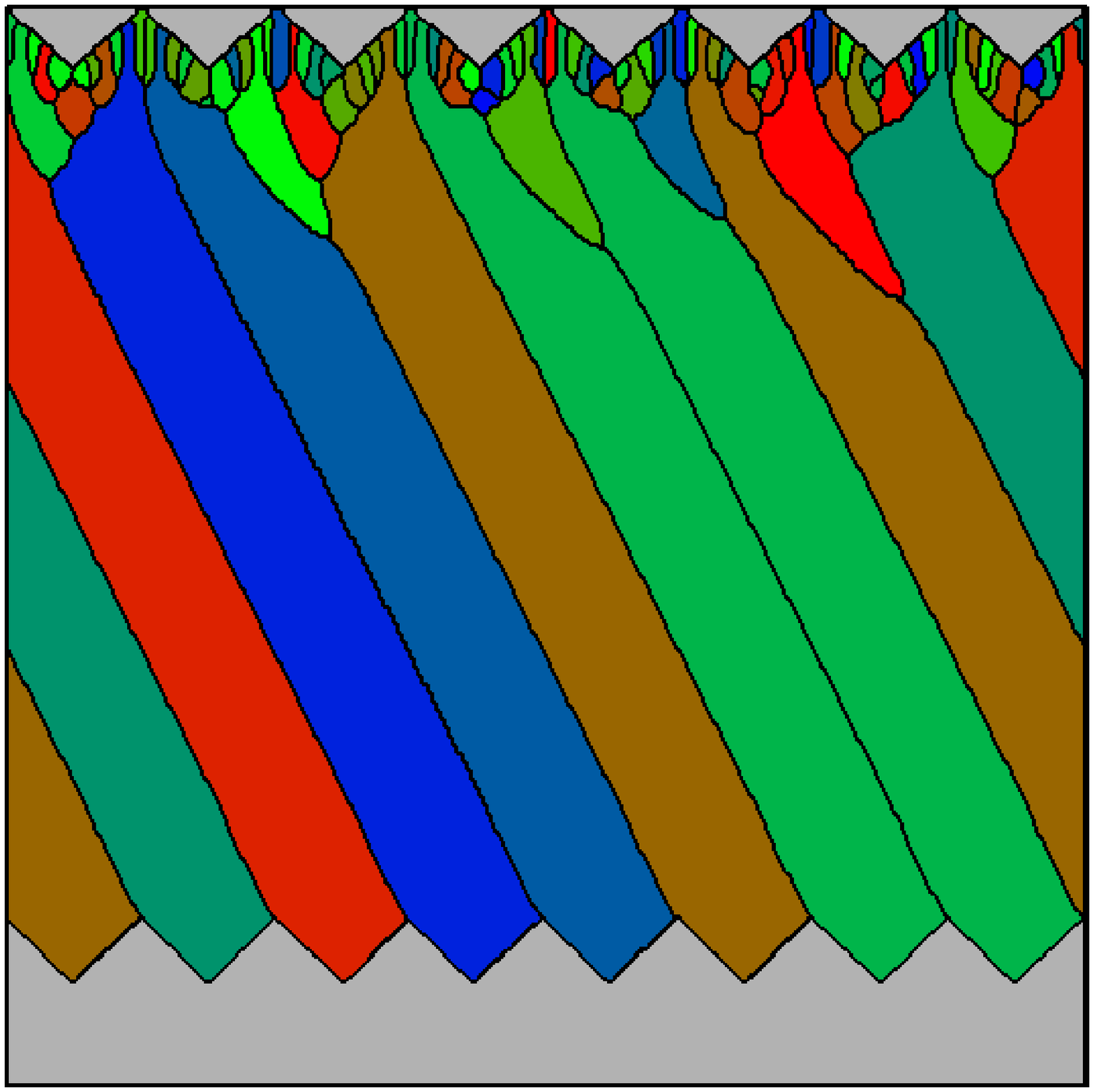}\label{fig:path_2_22}
\put(-10,41.5){\includegraphics[scale=0.08]{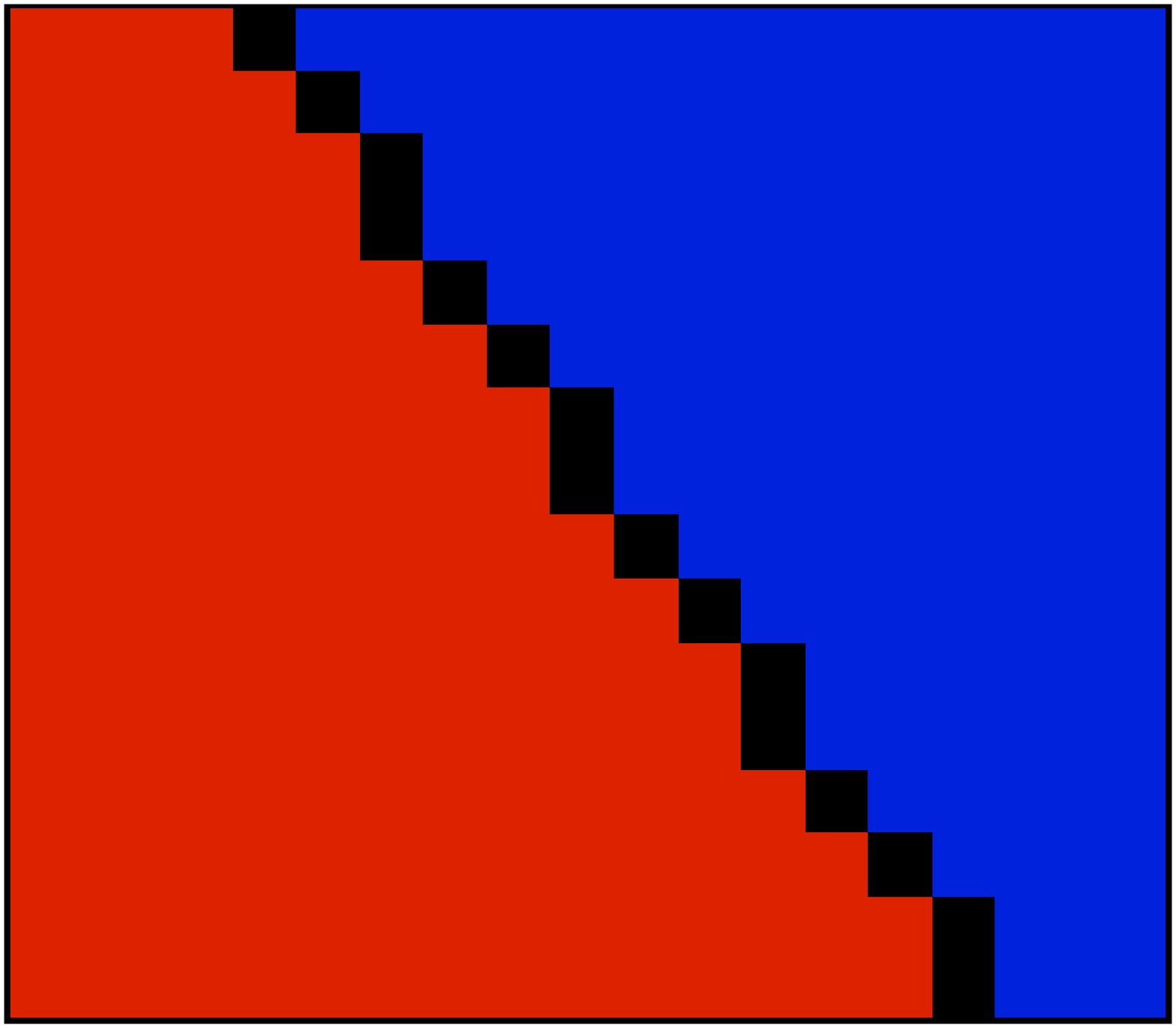}}
\put(-45,41.5){\includegraphics[scale=0.06]{b.eps}}\qquad\qquad
\put(10,0){\includegraphics[height=0.17\textwidth]{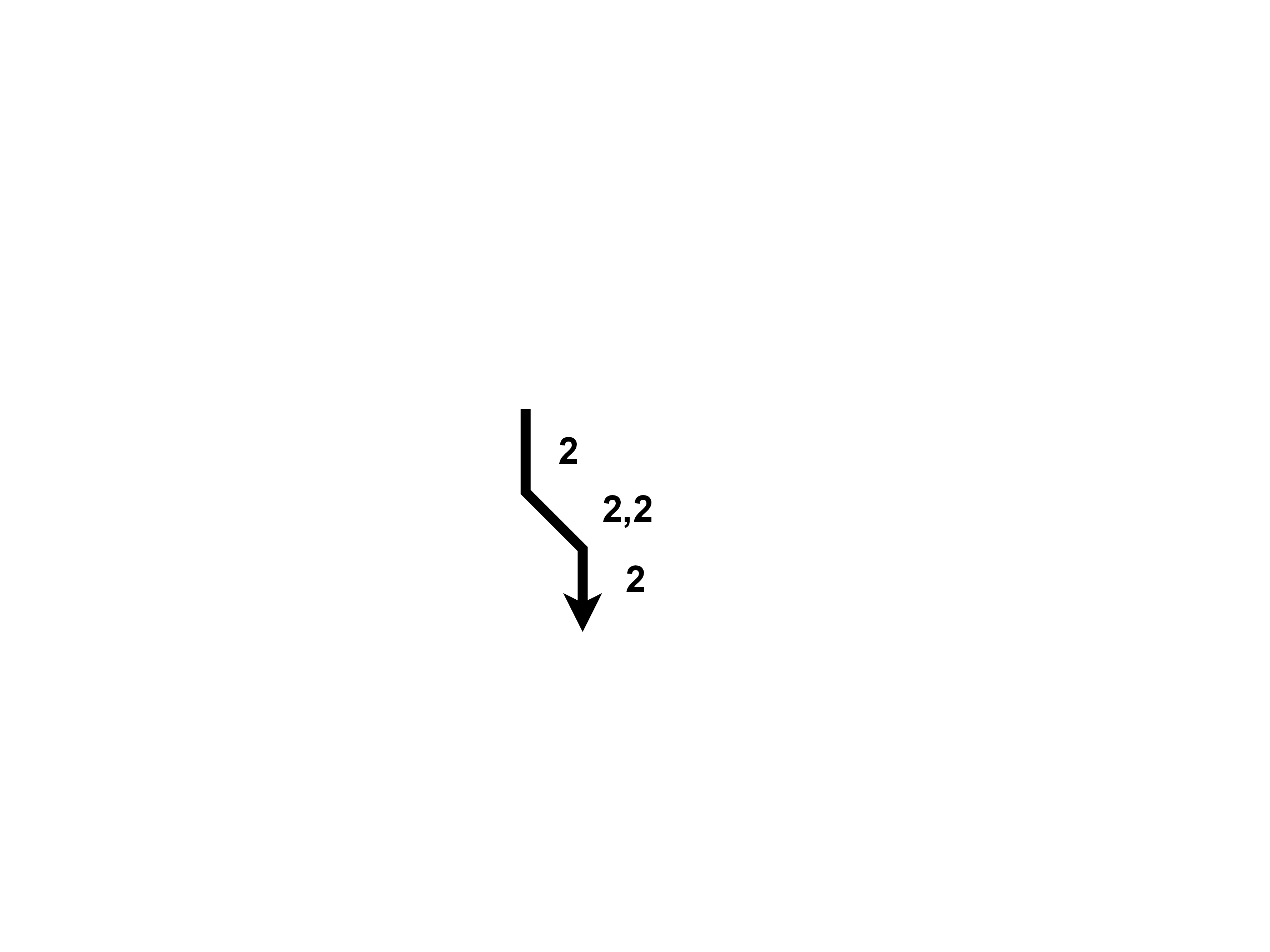}\label{fig:open_2}}}\\
\subfigure{\includegraphics[scale=0.15]{colormap.eps}}
\caption{\textbf{Fig. 8 }Effect of crack opening increment on the 
crack-seal mictrostructure. The same number of crystal
nuclei, crystal anisotropy
and wall morphology are used as in 
Fig. \ref{open_angle}. (a)  Loss of tracking behavior for linear and oblique 
opening in large increments (20 grid-points lateral offset). 
(b) Grain boundary tracking for linear and oblique opening in smaller 
increments (2 grid-points lateral offset). 
Crystals track the trajectory
only when opening increments are 
sufficiently small. 
{The picture on 
upper right hand side
(plot of $\phi=0.5$) 
shows the grain boundary 
morphology provided to 
compare with crack-opening path.}
The colors refer to different
crystal orientation with respect
to the vertical direction (see colormap).}\label{open_traj}
\end{figure}

In the next series of simulations,  
100 crystal nuclei with different orientation
are uniformly embedded at the top of the domain and 
the crack opening angle $\theta_{open}$
is varied. The crack roughness angle $\beta$ 
is selected to be $100\degree$ and a complete
sealing is ensured at every opening event.
The phase-field simulation results show 
that crystals track the crack opening trajectory, 
irrespective of $\theta_{open}$.  Thus,
the grain boundary tracking efficiency is 
not affected by the crack opening angle.
The results of the simulation are
summarized in Fig. \ref{open_angle}. 
On further increasing the magnitude of the crack opening
increment in Fig. \ref{open_traj} while ensuring complete sealing before
every opening event, we observe that the 
tracking behavior is lost and crystals grow 
in a random morphology (Fig. \ref{fig:path_20_2020}). 
However, when the 
crack opening increment is reduced ten-folds, 
while keeping $\theta_{open}$ unchanged, 
crystals re-establish a tracking of the crack 
opening trajectory as 
shown in Fig. \ref{fig:path_2_22}.
\newpage
\clearpage
\newpage
\clearpage
\subsection{Effect of number of crystal nuclei}
\label{subsec:num_nuc}
\begin{figure}[!htbp]
\centering
\subfigure{\includegraphics[width=0.31\textwidth,height=0.31\textwidth]{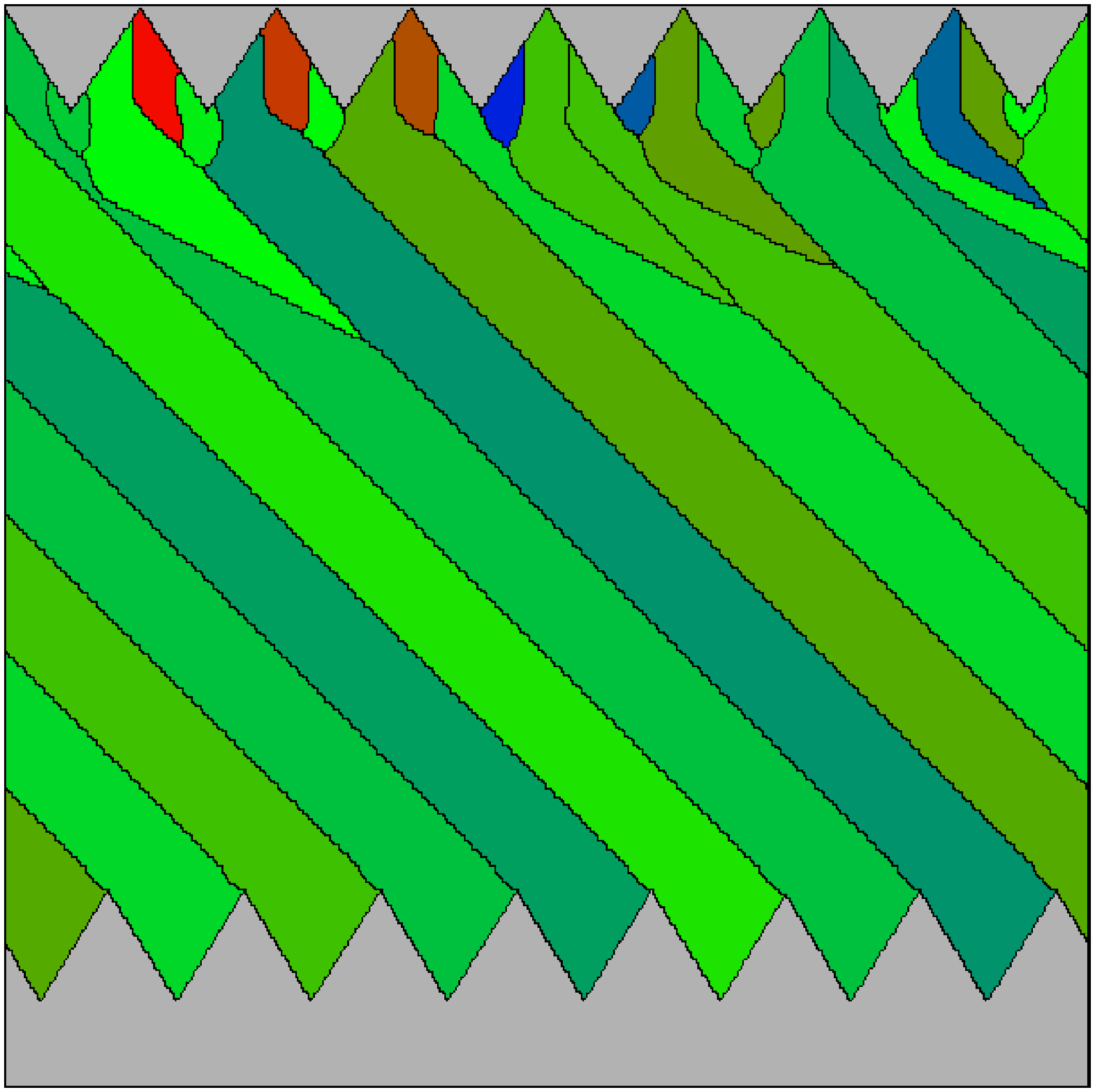}\label{fig:num_nuc_25}}
\put(-45,41.5){\includegraphics[scale=0.06]{a.eps}}\qquad
\subfigure{\includegraphics[width=0.31\textwidth,height=0.31\textwidth]{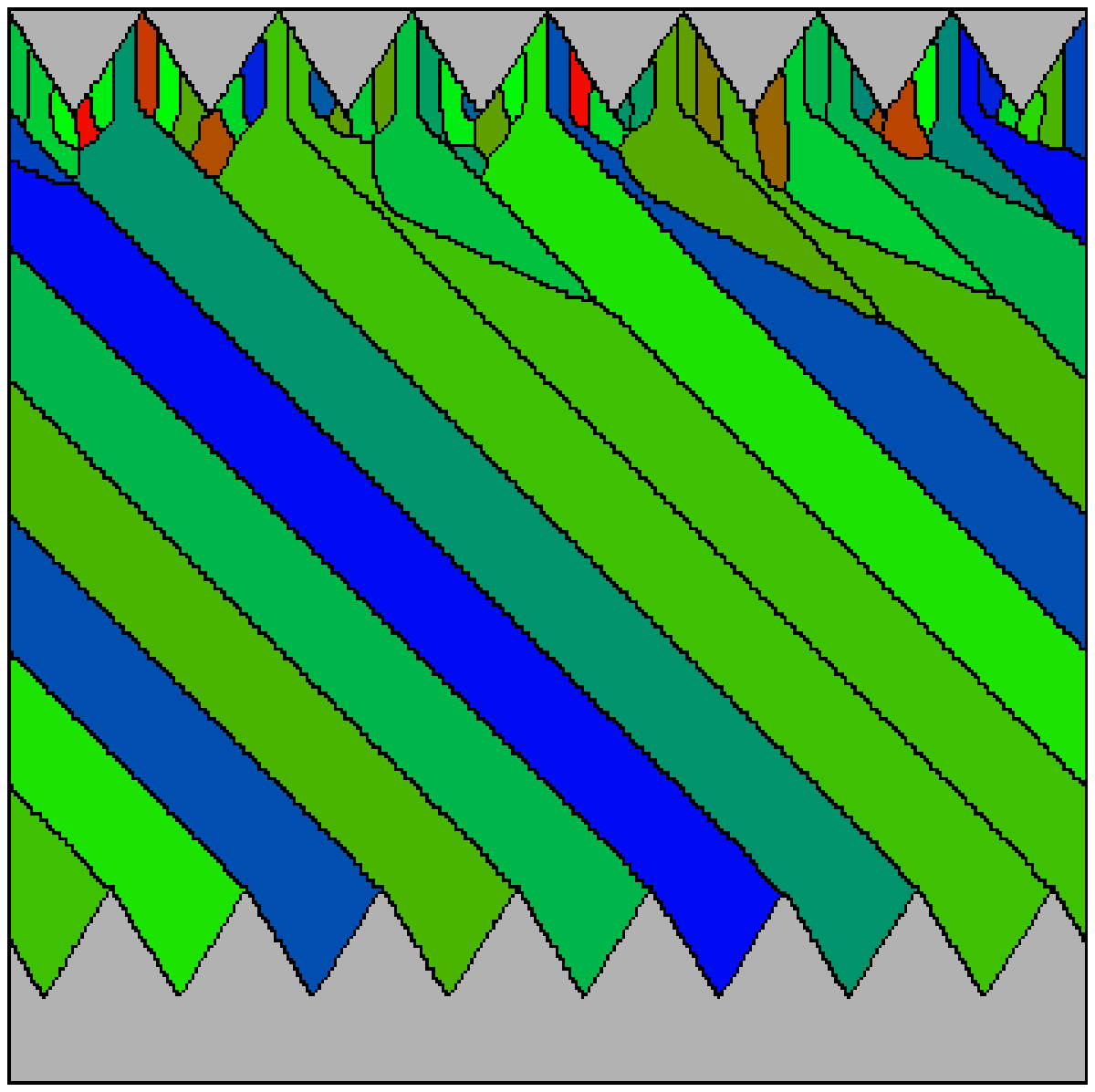}\label{fig:num_nuc_50}}
\put(-45,41.5){\includegraphics[scale=0.06]{b.eps}}\\
\subfigure{\includegraphics[width=0.31\textwidth,height=0.31\textwidth]{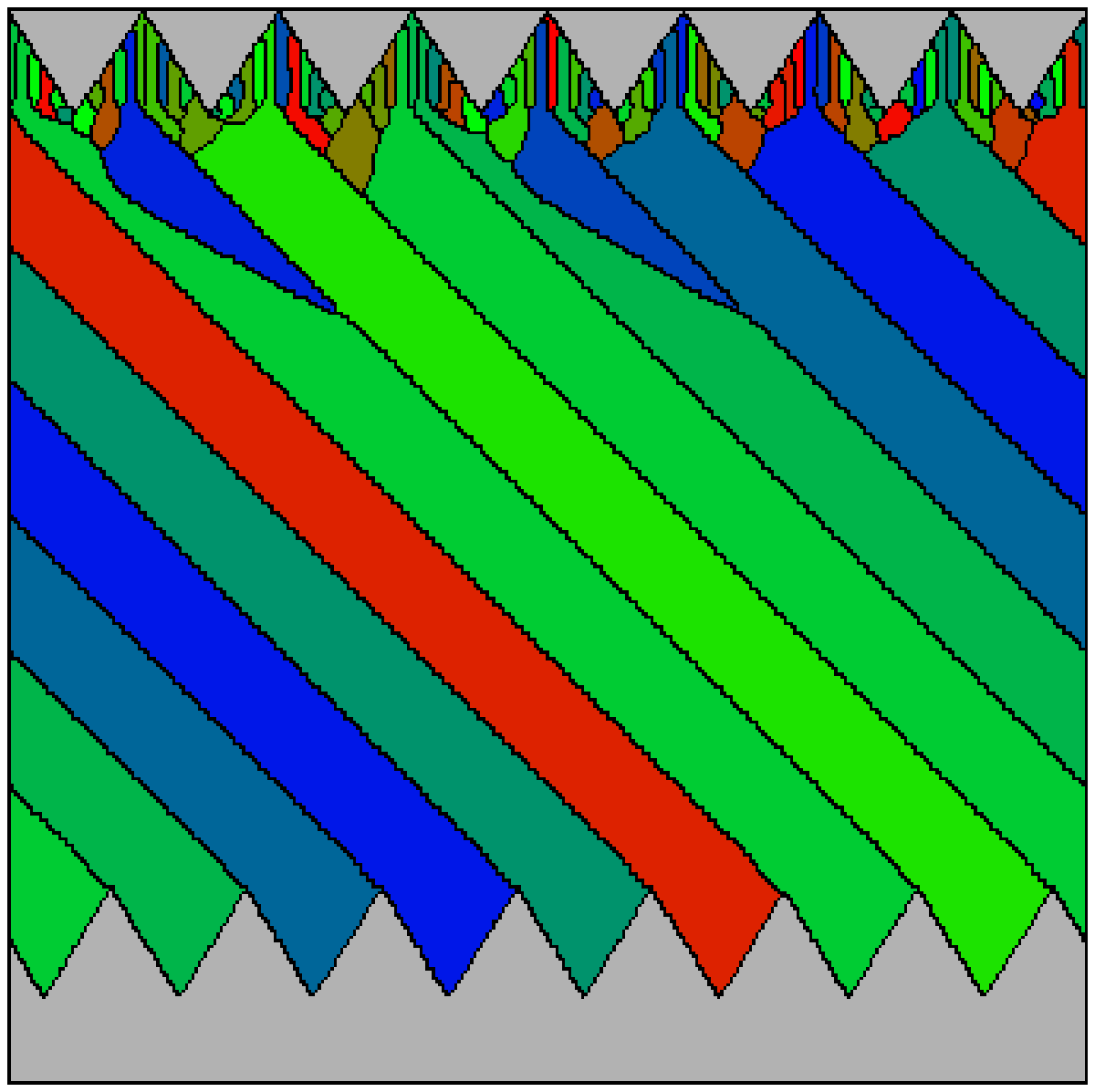}\label{fig:num_nuc_100}}
\put(-45,41.5){\includegraphics[scale=0.06]{c.eps}}\qquad
\subfigure{\includegraphics[width=0.31\textwidth,height=0.31\textwidth]{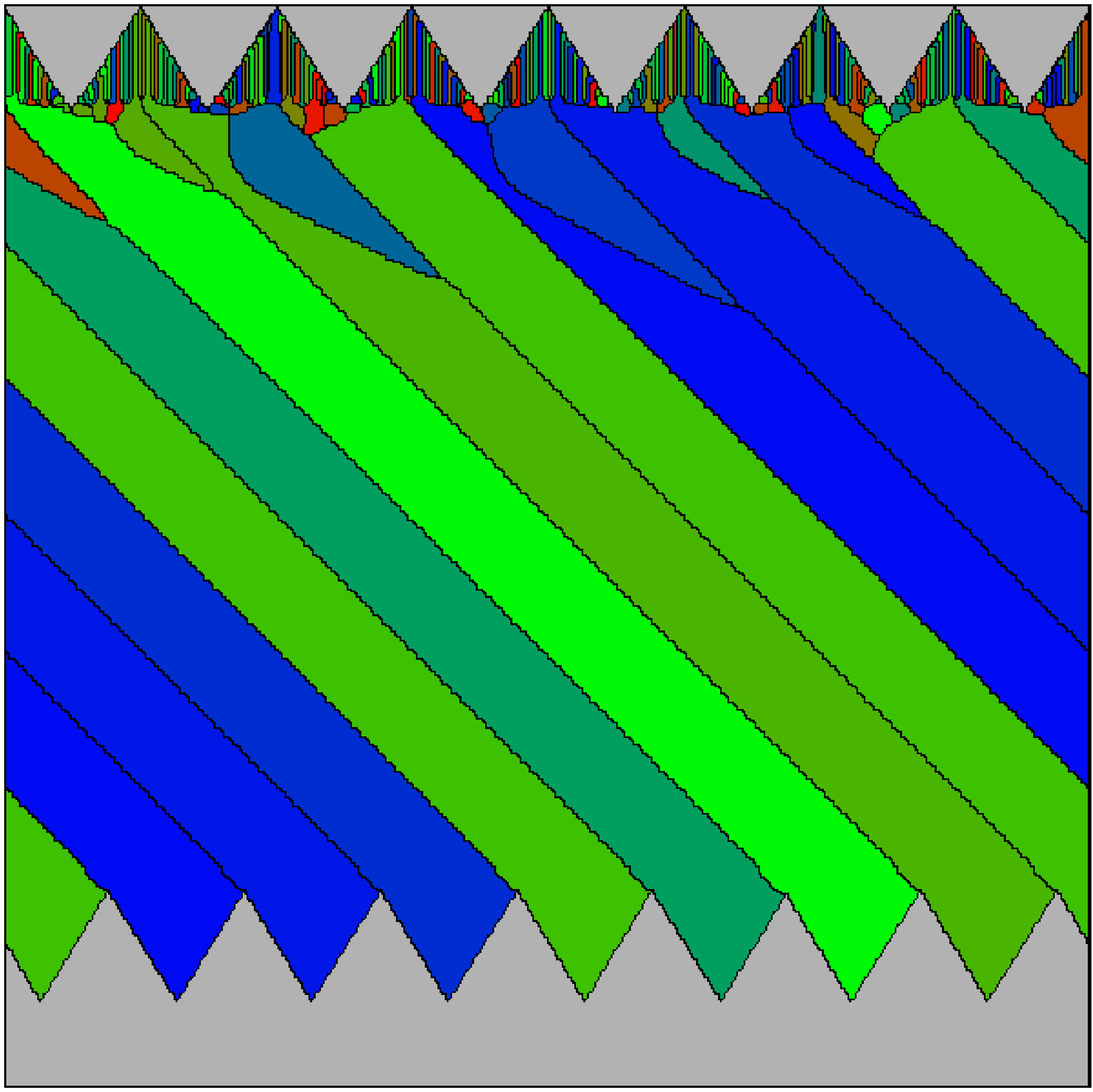}\label{fig:num_nuc_250}}
\put(-45,41.5){\includegraphics[scale=0.06]{d.eps}}\\
\subfigure{\includegraphics[scale=0.15]{colormap.eps}}
\caption{\textbf{Fig. 9 }Effect of initial number of nuclei on crystal growth
morphology and final microstructure. The crack opening
increments are small in the direction 45\degree 
with respect to vertical. The crack surface facing the crystal
growth front has 8 peaks. The number of crystal nuclei are
varied as in %
				(a) 25, %
				(b) 50, %
				(c) 100 and %
				(d) 250 %
to observe the effect on final microstructure. 
The number of crystals that end up tracking
the crack openings is equal to the number of 
peaks on the advancing crack. Colors refer 
to different crystal orientation with respect 
to vertical (see colormap).}\label{num_nuc}
\end{figure}

The number of nuclei is varied to study its 
influence on the vein growth morphology.
The crack is assumed to be sufficiently rough 
for 100\% tracking of crystal boundaries
and complete sealing is ensured. 
The phase-field simulations
show that varying the initial number of nuclei 
does not influence the tracking 
behavior. It is noteworthy that the number of 
surviving crystals remains constant and is numerically
equal to the number of peaks in the crack surface
which faces the crystal growth front. The crystal 
boundaries stabilize at the crack peaks which
suggest that these attract the grain
boundaries. Further, we do not
observe any change in growth morphology once the
number of crystals surviving equals the crack peaks,
irrespective of the initial number of nuclei 
as shown in Fig. \ref{num_nuc}.

\begin{figure}[!htbp]
\centering
\subfigure{\includegraphics[scale=0.2]{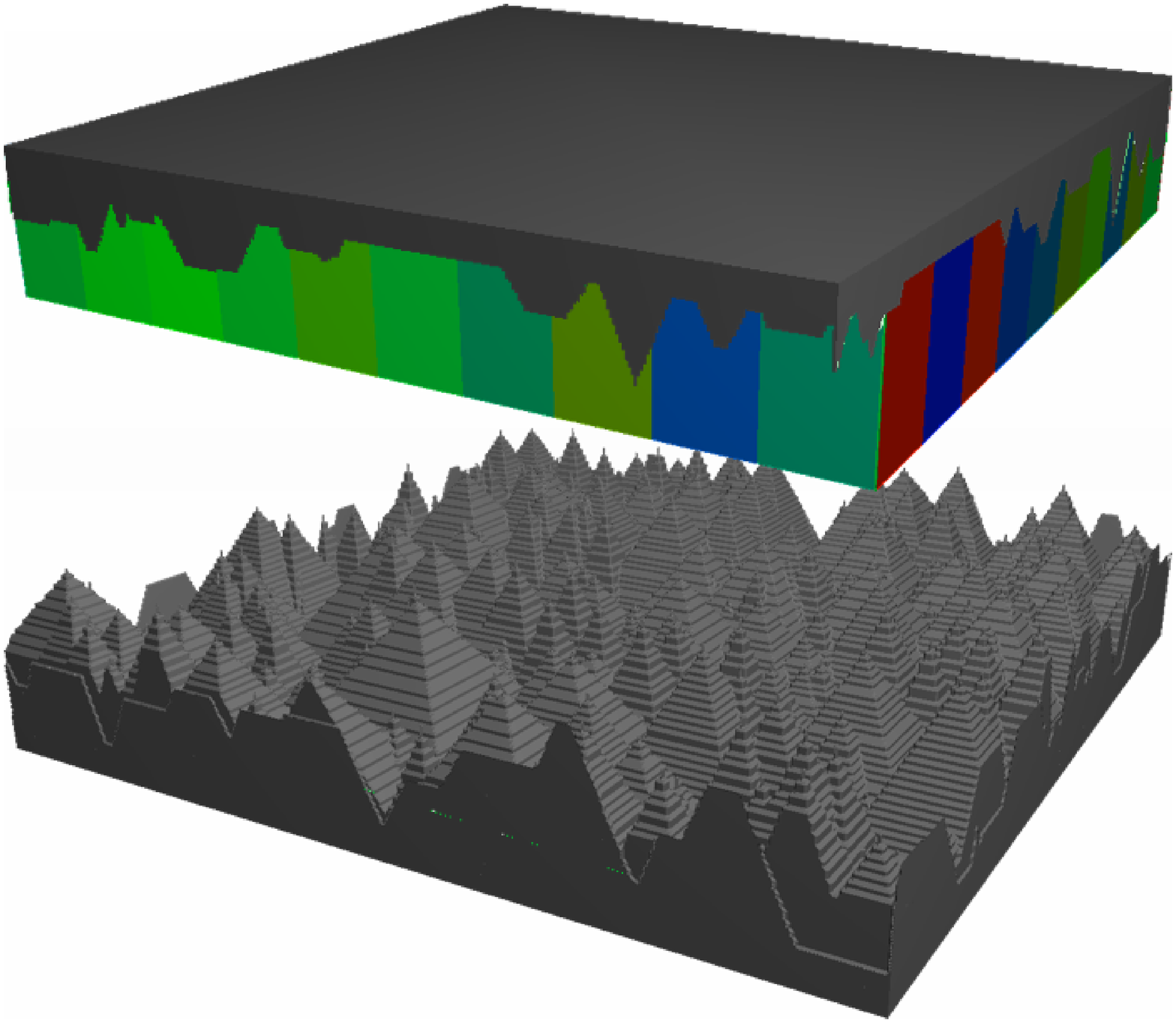}\label{fig:initial_CS}}
\put(-3,42){\includegraphics[scale=0.06]{a.eps}}\qquad\qquad
\subfigure{\includegraphics[scale=0.2]{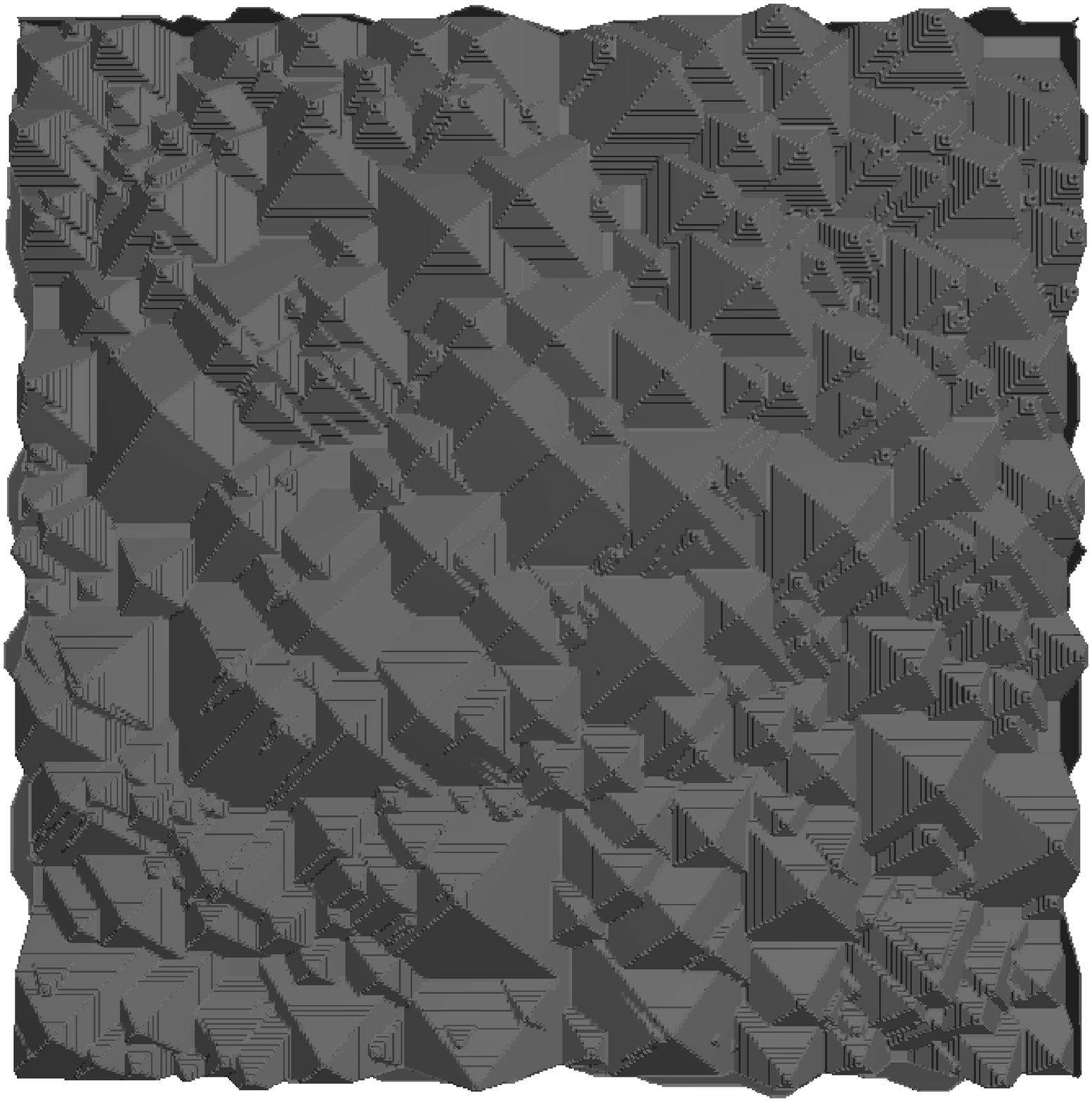}\label{fig:3D_crack_plan}}
\put(-5,42){\includegraphics[scale=0.06]{b.eps}}\\
\subfigure{\includegraphics[scale=0.25]{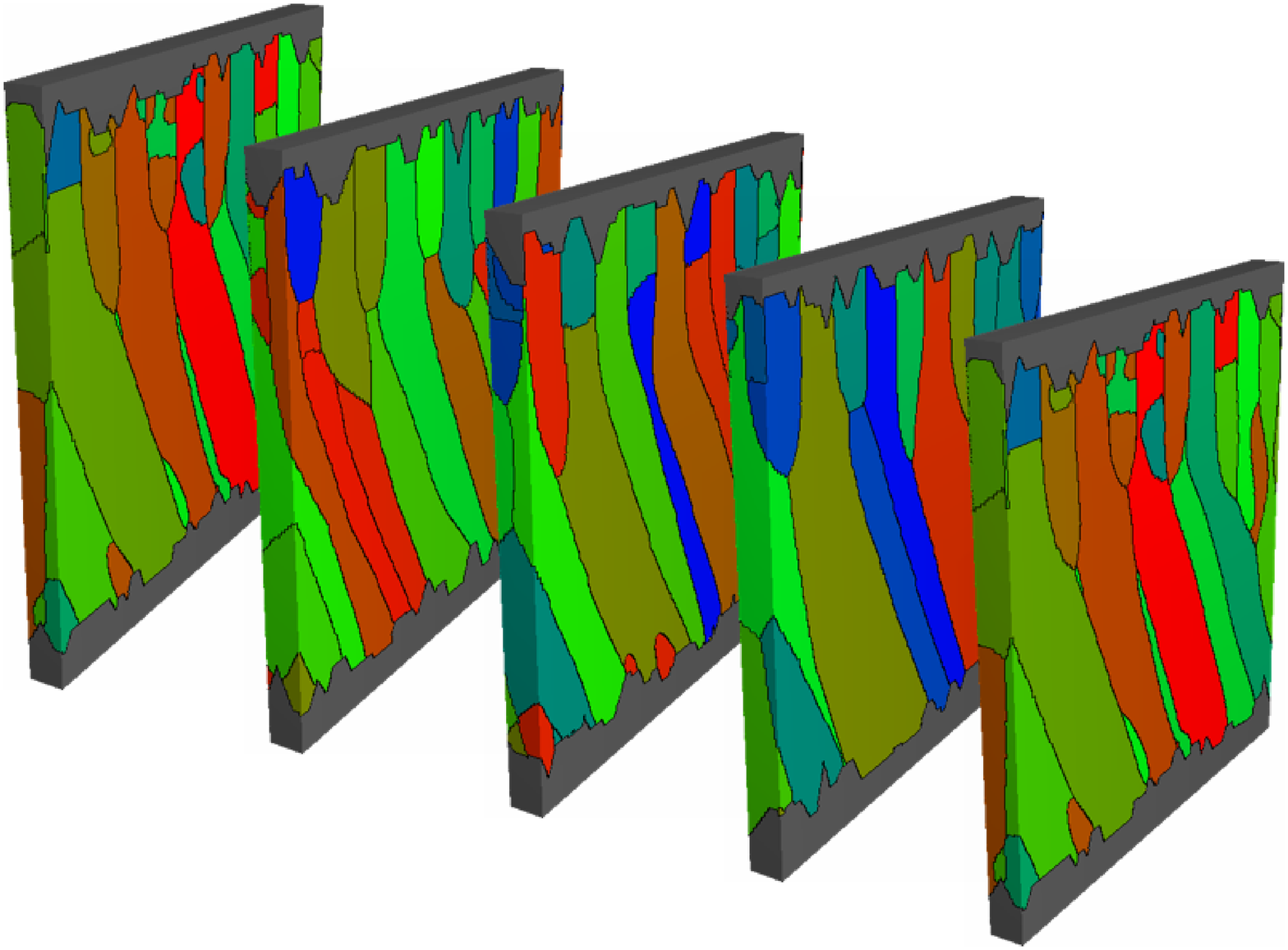}\label{fig:3D_cs_r6}}
\put(-3,50){\includegraphics[scale=0.06]{c.eps}}\qquad\qquad
\subfigure{\includegraphics[scale=0.23]{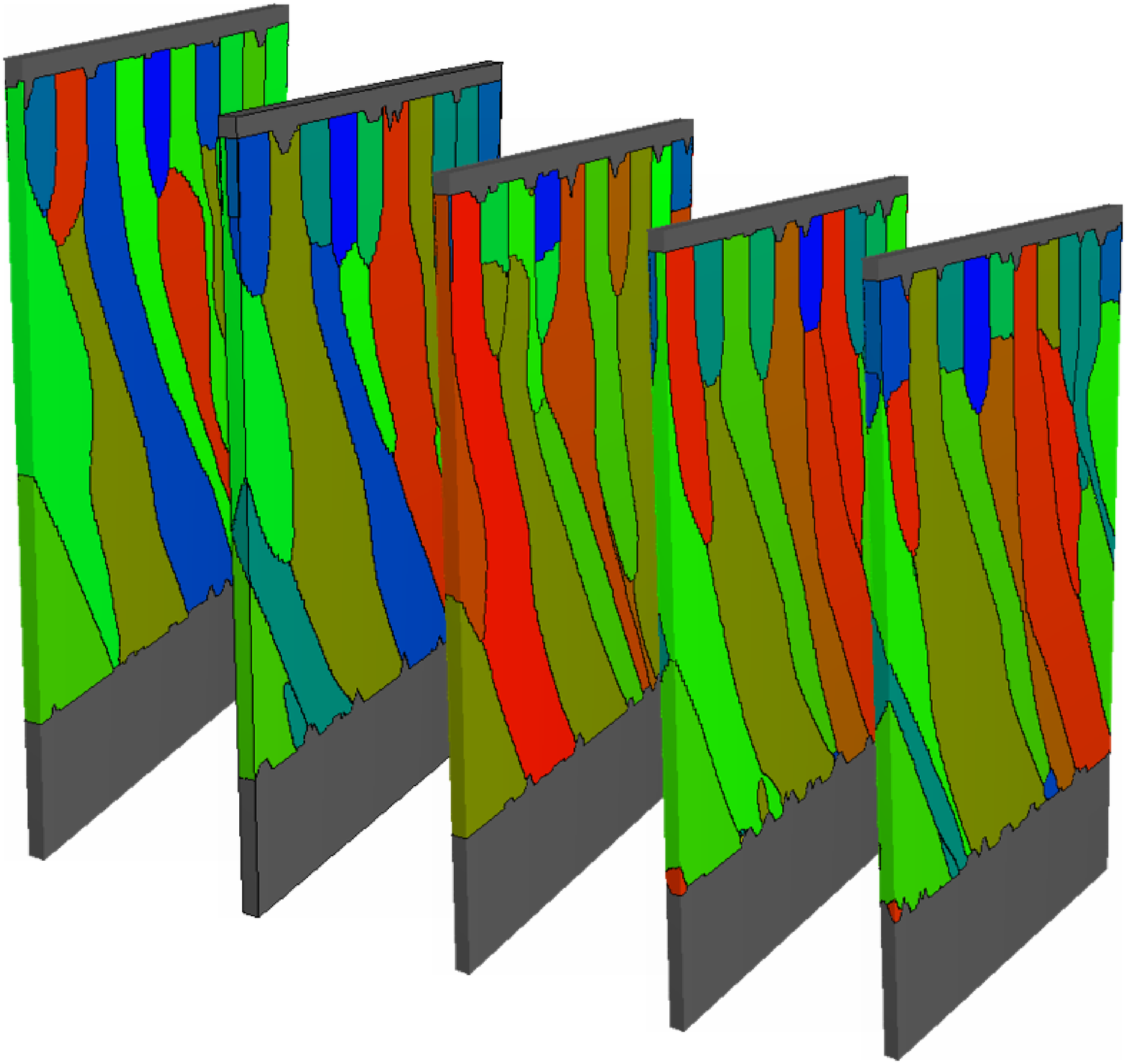}\label{fig:3D_cs_r10}}
\put(-3,52){\includegraphics[scale=0.06]{d.eps}}\\
\subfigure{\includegraphics[scale=0.15]{colormap.eps}}
\caption{\textbf{Fig. 10 }Crystals are embedded on an algorithmically irregular 
surface for simulating crack-sealing process in 3D. The crack surface 
roughness is defined by the amplitude of peaks. To reduce the 
surface roughness, the amplitude of every peak is reduced by a numeric factor. 
(a) The initial simulation domain setting. (b) Top view of the crack surface.
(c) 3D layer plot of the phase-field simulation of crack-sealing process
showing (c) grain boundary tracking along with occasional occurrence of
curved boundaries for the roughness factor 6 and for a crack opening velocity of 
2 grid points lateral offset, in (d), the number of curved/inflected grain 
boundaries increases for the roughness factor 10 and for a crack opening velocity 
of 4 grid points lateral offset. The crack opening trajectory is similar
to the one shown in Fig. \ref{fig:path_2_22}. Colors refer to different
crystal orientation with respect to vertical direction (see colormap).}
\label{3D_cs}
\end{figure}

Finally, 3D phase-field simulations of
crack-sealing are carried out by embedding 100
{alum} crystal nuclei on computationally
generated rough crack surface in
Figs. \ref{fig:initial_CS} and \ref{fig:3D_crack_plan}. 
The roughness of the 3D crack surface is controlled by 
varying the range of amplitude of peaks.
{Higher roughness relates to a wider 
range of maximum and minimum height of peak 
which is permissible (chosen randomly at different spatial locations).}
The crack is opened slowly to ensure 
ensuring complete sealing before every opening event. 
The 3D simulation results are displayed in Fig. \ref{3D_cs}
which shows a layer-by-layer plot of the simulation domain.
The simulation results reveal that a crack surface with 
higher roughness forces the crystals to track the opening 
trajectory. As the surface roughness is reduced, the resulting grain 
boundaries partially form curved/oscillatory morphologies 
(Fig. \ref{fig:3D_cs_r10}). It can be 
observed, that crystals grow in a mixed regime when
the wall roughness is not sufficiently high.
\newpage

\section{Discussion of results}
\label{sec:discussion}
In the present article, the multiphase-field model for grain-growth
has been extended to study the crystal growth in veins in two different
growth conditions namely, free-growth in fluid and crack-sealing. 
{In spite of considering the role of surface energy anisotropy
which is responsible for crystal facet formation 
and growth competition, the previous algorithm 
for modeling vein growth \citep{Bons:2001ve} 
suffer from model artifacts namely `Crystal
terminations' and `Long distance effect''.}
The phase-field model presented in this article 
explains the role of surface energy anisotropy 
and additionally account {for} the interface 
thermodynamics and force balances
at multi-grain junctions (crystal triple/quadruple points
in current context). Thus, the selection
dynamics of faceted crystal growth observed in 
{the} present simulations are more reliable as they 
compare well with the
experimental findings \citep{Hilgers:2002uq}.
  
It is observed that the consumption 
or the survival of a crystal
depends on its own orientation 
relative to neighbors, when
allowed to grow freely. The
non-neighboring crystals do not effect the growth of the crystal
in consideration directly. Additionally, the inter-facet angle 
defined by the equilibrium shapes 
are always preserved for the crystals
growing freely in liquid
Figs. \ref{fig:ECS_octa} and 
\ref{fig:ECS_alum}. Moreover, we do not
observe any exceptions in the deduced crystal growth selection
rule contrary to the findings of \citet{Nollet:2005qo} for cubic 
crystal symmetry. {In particular, these
exceptions refer to
the survival of crystals 
which have different orientations as well
as equilibrium angle between the facets
described as model artifacts.
The present method is able to rectify
such problems, evident from the 
simulations as well as orientation
map and shown in Fig. \ref{fig:A_15_B_30_C_45} 
-  Fig. \ref{fig:A_25_B_min25_C_45} and  Fig. \ref{orient_map}
respectively.}
The phase-field simulation results in
Figs. \ref{fig:A_10_B_min15_C_45}
and \ref{fig:A_11_B_min15_C_45} 
corroborate the findings of \citet{Bons:2003fk} 
by the observation that
the interface boundary between 
crystal `A' and `B' is found to be inflected.
{This highlights the reproducibility
of previous simulation results when phase-field
method is utilized.}
Apparently, if sufficient 
time is available for the development of 
facets, the favorably oriented crystals always
out-grow the poorly oriented neighbors. 
We are aware of the fact that 
crystal growth results from an interplay of 
surface energy anisotropy and growth 
kinetics. Typically, a crystal nucleates 
in its equilibrium (Wulff) shape and grows 
asymptotically towards its "kinetic Wulff shape"
if long-range transport of energy and/or 
mass is assumed to be extremely fast
\citep{Sekerka:2005fk}.
Additionally, we do not rule out the 
possibility of different 
contact angles at 
liquid–solid triple junctions 
as a result of different force balances 
corresponding to kinetic or surface 
energy anisotropy. Therefore, the values of 
the kinetic coefficient $\tau$ and of interfacial
energies need to be chosen appropriately to
simulate correct selection dynamics.

The presence of a barrier for example,
a rigid and rough crack surface, 
obstructs the freely growing crystals, 
forcing them to grow into a
morphology unrelated to the 
equilibrium crystal shape. It is found that
the crack parameters such as the roughness, 
opening velocity and trajectory determine the final
crack-seal microstructure. The phase-field simulation
results of crack-sealing process suggest
that if the wall rock is sufficiently rough and 
opening velocity is small, the crystals grow isotropically, 
in a fibrous morphology and track the 
crack-opening trajectory. In this case, anisotropy in surface energy
does not effect the growth morphology 
since facet formation
is suppressed due to the additional boundary condition, 
which corroborates the previous simulation and 
experimental studies on crack-sealing process
\citep{Urai:1991ec,Hilgers:2001rw,Nollet:2005qo}.
{Moreover, higher roughness
of the facing wall rock} also increases the fineness of fibrous microstructure.
This is inferred from Fig. \ref{num_nuc} which shows that if {the}
roughness is sufficiently high, the number of fibers formed is numerically
consistent with the peaks on the facing crack surface.
At a lower crack surface roughness, the grain boundary 
tracking efficiency decreases and the crystal boundaries
have a curved/oscillating morphology as presented in Figs. 
\ref{fig:rough6}, \ref{fig:rough8} and \ref{fig:rough16}.
The anisotropy of the surface energy 
provides an explanation for the curvature observed in 
grain boundaries in this case, as the
crack opening events occur along an oblique $45\degree$  
line with respect to {the} vertical direction.
It is noteworthy, that grain boundaries alone do 
not provide much information about the crack-opening trajectory
in such cases. The simulation results also reveal that 
straight grain boundaries are formed when 
surface energy of the growing crystals
is assumed to be isotropic. Further, within the complete
crack-seal regime, when the opening increment is 
relatively {faster}, curved grain boundaries 
are formed. The 3D phase-field 
study of {the} crack-sealing process 
accentuate these findings; if the wall rock 
roughness is low and crack opening
velocity is increased while still ensuring 
complete sealing  before every 
opening event, the crystals
growth occurs in a mixed regime, characterized
by a decrease in tracking behavior with a
propensity to form curved grain boundaries.
{A systematic study to establish the correlation 
of crack roughness (by choosing more realistic
boundary condition for e.g. fractal surface) and opening rate 
(in 3 dimensions for more relevant vein forming crystals like quartz or calcite)
with the extent/amplitude of grain boundary curvature/oscillation 
and influence of hydrodynamic convection, is a part of 
on-going effort.\\
At this point, we would like to clarify that the 
objective of present work is not to 
question the numerical algorithm 
or the simulation results of
\citet{Urai:1991ec}, \citet{Bons:2001ve}, \citet{Hilgers:2001rw} and \citet{Nollet:2005qo}.
Rather, it is aimed to advance 
the numerical studies to 
3D by adopting a 
thermodynamically consistent approach
and to discuss the applicability of phase-field
method to study polycrystal growth in veins.
In this scope, the foremost intention is to
discuss the reproducibility
of previous results (in free-growth as well
as crack-sealing 2D simulations) as well as advantages
of using the present phase-field model (3D numerical studies
for crystal of any shape, large-scale simulations and provision to 
implement transport).}

\section{Conclusion and outlook}
\label{sec:conclusion}
In the present article, the parameters 
controlling the vein  microstructure are
studied using the phase-field method. 
The role of surface energy 
anisotropy during free growth of crystals
and its influence on the 
grain orientation selection is addressed. 
The 3D phase-field
simulation of free crystal 
growth process provides
further insight on
how poorly 
oriented crystals are consumed
by more favorably oriented ones
and previous ambiguities
reported in literature \citep{Nollet:2005qo}
are ruled out.
Further, the influence of crack
parameters on the final
microstructure is discussed
in detail. One of the intriguing
finding of the current work 
is the role of surface energy anisotropy
in the formation of crack-seal 
microstructures. 
Of particular importance, is the
appearance of curved/oscillating
crystal boundaries which draws similarities
with the natural vein microstructures.
The current phase-field study of crack-sealing 
process (in 2D and 3D) indicate that 
anisotropy in surface energy of the growing 
crystals cause the grain boundaries to curve/oscillate,
if wall rock is not sufficiently
rough and crack opening rate is increased gradually 
while still ensuring complete crack-sealing before every 
opening event. The fineness of the fibrous microstructure
is directly related to the number of peaks
on the facing wall surface.    
Further, the transition from fibrous to
elongate-blocky morphology in vein microstructure
caused by varying the crack opening rate
is demonstrated in the simulation results.
The present work also
establishes the phase-field method
as comprehensive, {artifact-free} and standout 
approach to simulate the geological processes
occurring during vein formation.
The simulation results 
also demonstrate the general
capability of  multiphase-field
method in dealing with anisotropic 3D vein-growth
problem. {It is noteworthy 
that the numerical model 
presented in the current work
provides a general framework 
to simulate crystals of any shape.} Since the underlying
model equations are based on 
continuum mechanical 
and thermodynamical concepts, 
several extensions of 
the present vein-growth model 
are possible.
This includes the studies
related to diffusion
driven grain evolution and 
hydrodynamic convection. Including
such effects in the present 
phase-field model is imperative for the
complete understanding of the vein
growth problem. Once implemented,
the model can be further utilized to 
study the precise effect of hydrodynamic
convection on the morphology of the 
vein front and crystal boundaries.
Further, this also helps in debating 
the question posed by 
\citet{Barker:2006fk} whether vein 
formation involves advective fluid 
flow, or occurs by local diffusion 
of material from the surrounding 
wall rock. Though, this requires 
significant computational resources,
the efficient parallelization 
of the phase-field solver \citep{Nestler:2008kx}
(utilized for phase-field simulations 
presented in this article) makes it feasible.

\begin{acknowledgements}
Drs. Abhik Choudhury (Ecole Polytechnique, Palaiseau, France) 
and Frank Wendler (Institute of Applied Geosciences, Karlsruhe
Institute of Technology) are thanked for many insightful discussions.
KA and BN acknowledge the financial support by Graduate school 
1483 of German Research Foundation
and by the project CCMSE of the European Union (EFRE) 
together with the state Baden-Wuerttemberg.
KA also thank former co-workers Drs. Denis Pilipenko and Michael Fleck 
(Materials and Process Simulations, University of Bayreuth) for preliminary discussions concerning the model and
Center for Computing and Communication at RWTH Aachen University (HPC Cluster) for computational resources.
\end{acknowledgements}





\bibliographystyle{spbasic} 







\end{document}